\documentclass{aa}
\usepackage{graphicx,txfonts,natbib}
\bibpunct{(}{)}{;}{a}{}{,}

\begin{document}

\title{Modelling the formation of double white dwarfs}
\author{M.V. van der Sluys \inst{1} \and F. Verbunt \inst{1,2} \and O.R. Pols \inst{1}}
\offprints {M.V. van der Sluys, \email{sluys@astro.uu.nl}}

\institute{ Astronomical Institute, P.O. Box 80000, NL-3508 TA Utrecht, the Netherlands, 
            {\tt(sluys@astro.uu.nl)}, {\tt(verbunt@astro.uu.nl)} and {\tt(pols@astro.uu.nl)}
            \and
            SRON Laboratory for Space Research, Sorbonnelaan 2, NL-3584 CA
            Utrecht, the Netherlands
          }

\date{22 February 2006 / 25 August 2006}

\abstract{We investigate the formation of the ten double-lined double white dwarfs
that have been observed so far.  A detailed stellar evolution code is used to calculate
grids of single-star and binary models and we use these to reconstruct possible 
evolutionary scenarios.  
We apply various criteria to select the acceptable solutions from these scenarios.
We confirm the conclusion of \citet{2000A&A...360.1011N} 
that formation via conservative mass transfer and a common envelope with spiral-in 
based on energy balance or via two such spiral-ins cannot explain the formation 
of all observed systems.  We investigate three different prescriptions of envelope ejection due
to dynamical mass loss with angular-momentum balance and show that they can explain 
the observed masses and orbital periods well.  Next, we demand that the age difference 
of our model is comparable to the observed cooling-age difference and show that this puts a 
strong constraint on the model solutions.  However, the scenario in which the primary 
loses its envelope in an isotropic wind and the secondary transfers its envelope, which
is then re-emitted isotropically, can explain the observed age differences as well.
One of these solutions explains the DB-nature 
of the oldest white dwarf in PG\,1115+116 along the evolutionary scenario proposed by 
\citet{2002MNRAS.334..833M}, in which the helium core of the primary becomes exposed due 
to envelope ejection, evolves into a giant phase and loses its hydrogen-rich outer layers.  


}

\maketitle


\section{Introduction}
\label{sec:intro}

Ten double-lined spectroscopic binaries with two white-dwarf components are currently known.
These binaries have been systematically searched for to find possible progenitor systems for Type Ia supernovae,
for instance by the SPY (ESO SN Ia Progenitor surveY) project ({\it e.g.} Napiwotzki et al. 
2001, 2002\nocite{2001A&A...378L..17N,2002A&A...386..957N}).  Short-period double white dwarfs can lose orbital 
angular momentum by emitting gravitational radiation and if the total mass of the binary exceeds the Chandrasekhar 
limit, their eventual merger might produce a supernova of type Ia \citep{1984ApJS...54..335I}.

The observed binary systems all have short orbital periods that, with one exception, range from an hour and a half 
to a day or two (see Table\,\ref{tab:obswds}), corresponding to orbital separations between
0.6\,$R_\odot$ and 7\,$R_\odot$.  The white-dwarf masses of 0.3\,$M_\odot$ or more indicate that their 
progenitors were (sub)giants with radii of a few tens to a few hundred solar radii. This 
makes a significant orbital shrinkage (spiral-in) during the last mass-transfer phase necessary and fixes the
mechanism for the last mass transfer to common-envelope evolution.  In such an event the envelope of the
secondary engulfs the oldest white dwarf due to dynamically-unstable mass transfer. 
Friction then causes the two white dwarfs to spiral in towards each other while the envelope is expelled. 
The orbital energy that is freed due to the spiral-in provides for the necessary energy for the expulsion \citep{1984ApJ...277..355W}.

The first mass transfer phase is usually thought to be either another spiral-in or stable and conservative mass transfer.  
The first scenario predicts that the orbit shrinks appreciably during the mass transfer whereas the second
suggests a widening orbit.  Combined with a core-mass\,--\,radius relation ({\it e.g.} Refsdal \& Weigert 1970\nocite{1970A&A.....6..426R}) 
these scenarios suggest that the mass ratio $q_2 \equiv M_2/M_1$ of the double white dwarfs is much smaller than unity 
in the first scenario and larger than unity in the second scenario.  The observed systems all have mass ratios 
between 0.70 and 1.28 (Table\,\ref{tab:obswds}), which led \citet{2000A&A...360.1011N} to conclude that a 
third prescription is necessary to explain the evolution of these systems.  They suggested envelope ejection due to
dynamical mass loss based on angular-momentum balance, in which little orbital shrinkage takes place.  
They used analytical approximations to reconstruct the evolution of three double white dwarfs and concluded 
that these three systems can only be modelled if this angular-momentum prescription is included.

In this article we will use the same method as \citet{2000A&A...360.1011N}, to see if a stable-mass-transfer episode
followed by a common envelope with spiral-in can explain the observed double white dwarfs.  We will improve on their calculations
in several respects.  First, we extend the set of observed binaries from 3 to 10 systems.  Second, we take
into account progenitor masses for the white dwarf that was formed last up to $10\,M_\odot$ and allow them
to evolve beyond core helium burning to the asymptotic giant branch.  \citet{2000A&A...360.1011N} 
restricted themselves to progenitor masses of $2.3\,M_\odot$ or less and did not allow these stars to evolve 
past the helium flash.  This was justified because the maximum white-dwarf mass that should be created by these 
progenitors was $0.47\,M_\odot$, the maximum helium-core mass of a low-mass star and less than the minimum mass
for a CO white dwarf formed in a spiral-in (see Fig.\,\ref{fig:basic_mc-r}).  The most massive white
dwarf in our sample is $0.71\,M_\odot$ and cannot have been created by a low-mass star on the red-giant branch.
Third, we use more sophisticated stellar models to reconstruct the evolution of the observed systems.  This
means that the radius of our model stars does not depend on the helium-core mass only, but also on total mass
of the star (see Fig.\,\ref{fig:basic_mc-r}).  Furthermore, we can calculate the binding energy of the hydrogen
envelope of our models so that we do not need the envelope-structure parameter $\lambda_\mathrm{env}$ and can 
calculate the common-envelope parameter $\alpha_\mathrm{ce}$ directly.  Last, because we use a full 
binary-evolution code, we can accurately model the stable mass transfer rather than estimate the upper limit for
the orbital period after such a mass-transfer phase.  This places a strong constraint on the possible 
stable-mass-transfer solutions.  The evolution code also takes into account the fact that the core mass of
a donor star can grow appreciably during stable mass transfer, a fact that alters the relation between 
the white-dwarf mass and the radius of the progenitor mentioned earlier for the case of stable mass transfer.

Our research follows the lines of \citet{2000A&A...360.1011N}, calculating the evolution of the systems in
reverse order, from double white dwarf, via some intermediate system with one white dwarf, to the initial ZAMS binary.
In Sect.\,\ref{sec:observed} we list the observed systems that we try to model.
The stellar evolution code that we use to calculate stellar models is described in Sect.\,\ref{sec:code}.
In Sect.\,\ref{sec:gbmodels} we present several grids of single-star models from which we will use the helium-core 
mass, stellar radius and envelope binding energy to calculate the evolution during a spiral-in.  We show a grid of `basic' 
models with standard parameters and describe the effect of chemical enrichment due to accretion 
and the wind mass loss.  We find that these two effects may be neglected for our purpose.
In Sect.\,\ref{sec:mtp2} we use the single-star models to calculate spiral-in evolution for each observed binary
and each model star in our grid and thus produce a set of progenitor binaries.  Many of these systems can be 
rejected based on the values for the common-envelope parameter or orbital period.
The remainder is a series of binaries consisting of a white dwarf and a giant star that would cause a common 
envelope with spiral-in and produce one of the observed double white dwarfs.  
In Sect.\,\ref{sec:mtp1} we model the first mass-transfer scenario that produces the systems found in Sect.\,\ref{sec:mtp2} to complete the evolution.  
We consider three possible prescriptions: stable and conservative mass transfer, a common envelope with spiral-in based on 
energy balance and envelope ejection based on angular-momentum balance.  We introduce two variations
in the latter prescription and show that they can explain the observed binaries.  In addition, we show that
the envelope-ejection scenario based on angular-momentum balance can also explain the second mass-transfer
episode.  In Sect.\,\ref{sec:age_difference}
we include the observed age difference in the list of parameters our models should explain and find that this
places a strong constraint on our selection criteria.  In Sect.\,\ref{sec:discussion} we compare this study
to earlier work and discuss an alternative formation scenario for PG\,1115+116.  Our conclusions  
are summed up in Sect.\,\ref{sec:conclusions}.


\section{Observed double white dwarfs}
\label{sec:observed}

At present, ten double-lined spectroscopic binaries consisting of two white dwarfs have 
been observed.  The orbital periods of these systems are well determined.  The fact that 
both components are detected makes it possible to constrain the mass ratio of the system 
from the radial-velocity amplitudes.  The masses of the components are usually 
determined by fitting white-dwarf atmosphere models to the observed effective temperature 
and surface gravity, using mass--radius relations for white dwarfs.  
The values thus obtained are clearly better for the brightest
white dwarf but less well-constrained than the values for the period or mass ratio.
It is also harder to estimate the errors on the derived mass.
In the publications of these observations, the brightest white dwarf is usually denoted
as `star\,1' or `star\,A'.  Age determinations suggest in most 
cases that the brightest component of these systems is the youngest white dwarf.
These systems must have evolved through two mass-transfer episodes and the brightest
white dwarf is likely to have formed from the originally less massive component of
the initial binary (consisting of two ZAMS stars).  We will call this
star the secondary or `star 2' throughout this paper, whereas the primary or `star 1'
is the component that was the initially more massive star in the binary.  The two components
will carry these labels throughout their evolution, and therefore white dwarf 1 will be
the oldest and usually the faintest and coldest of the two observed components.  The properties of the
ten double-lined white-dwarf systems are listed in Table\,\ref{tab:obswds}.
\begin{table*}
\centering
\begin{tabular}{lllllllll}
\hline \hline                                                                                                    
Name             & $P_\mathrm{orb}$ (d)  & $a_\mathrm{orb}$ ($R_\odot$)    & $M_1$ ($M_\odot$) & $M_2$ ($M_\odot$) & $q_2$ = $M_2/M_1$ & $\tau_2$ (Myr) & $\Delta \tau$ (Myr) & Ref/Note \\
\hline
\object{WD\,0135--052}     & 1.556      & 5.63             & 0.52 $\pm$ 0.05   & 0.47 $\pm$ 0.05   & 0.90 $\pm$ 0.04  &  950 & 350 &  1,2  \\ 
\object{WD\,0136+768}      & 1.407      & 4.98             & 0.37		     & 0.47 	         & 1.26 $\pm$ 0.03  &  150 & 450 &  3,10  \\ 
\object{WD\,0957--666}     & 0.061      & 0.58             & 0.32		     & 0.37 	         & 1.13 $\pm$ 0.02  &   25 & 325 &  3,5,6,10  \\ 
\object{WD\,1101+364}      & 0.145      & 0.99             & 0.33		     & 0.29 	         & 0.87 $\pm$ 0.03  &  135 & 215 &  4,(10)  \\ 
\object{PG\,1115+116}      & 30.09      & 40.0             & 0.7		     & 0.7               & 0.84 $\pm$ 0.21  &   60 & 160 &  8,9   \\
\multicolumn{9}{c}{}\\
\object{WD\,1204+450}      & 1.603      & 5.72             & 0.52		     & 0.46 	         & 0.87 $\pm$ 0.03  &   40 & 80  &  6,10  \\ 
\object{WD\,1349+144}      & 2.209      & 6.65             & 0.44		     & 0.44 	         & 1.26 $\pm$ 0.05  &   -- & --  &  12  \\ 
\object{HE\,1414--0848}    & 0.518      & 2.93             & 0.55 $\pm$ 0.03   & 0.71 $\pm$ 0.03   & 1.28 $\pm$ 0.03  & 1000 & 200 &  11   \\ 
\object{WD\,1704+481a}     & 0.145      & 1.13             & 0.56 $\pm$ 0.07   & 0.39 $\pm$ 0.05   & 0.70 $\pm$ 0.03  &  725 & -20 &  7,a \\ 
\object{HE\,2209--1444}    & 0.277      & 1.89             & 0.58 $\pm$ 0.08   & 0.58 $\pm$ 0.03   & 1.00 $\pm$ 0.12  &  900 & 500 &  13   \\
\hline
\end{tabular}
\caption{Observed double white dwarfs discussed in this paper. The table shows for each system the orbital period $P_\mathrm{orb}$,
the orbital separation $a_\mathrm{orb}$, the masses $M_1$ and $M_2$, the mass ratio $q_2$ = $M_2/M_1$, 
the estimated cooling age of the youngest white dwarf $\tau_2$ and the difference between the cooling ages of the 
components $\Delta\tau$. $M_1$ is the mass of the oldest white dwarf and 
thus presumably the original primary. The errors on the periods are smaller than the last digit.  The values for $a_\mathrm{orb}$ are 
calculated by the authors and meant to give an indication.  
References: 
(1) \citet{1988ApJ...334..947S},
(2) \citet{1989ApJ...345L..91B}, 
(3) \citet{1990ApJ...365L..13B}, 
(4) \citet{1995MNRAS.275L...1M}, 
(5) \citet{1997MNRAS.288..538M}, 
(6) \citet{1999ASPC..169..275M}, 
(7) \citet{2000MNRAS.314..334M}, 
(8) \citet{2002ApJ...566.1091B}, 
(9) \citet{2002MNRAS.334..833M}, 
(10) \citet{2002MNRAS.332..745M}, 
(11) \citet{2002A&A...386..957N}, 
(12) \citet{2003whdw.conf...43K}, 
(13) \citet{2003A&A...410..663K}.
Note: (a) WD\,1704+481a is the close pair of a hierarchical triple.  It seems unclear which of the two stars in this pair is the youngest 
(see the text).
\label{tab:obswds}  }
\end{table*}
For our calculations we will use the parameters that are best determined from the Table:
$P_\mathrm{orb}$, $q_2$ and $M_2$.  For $M_1$ we will {\em not} use the value listed in 
Table\,\ref{tab:obswds}, but the value $M_2/q_2$ instead. We hereby ignore the observational 
uncertainties in $q_2$, because they are small with respect to the uncertainties in the mass.  
In Sects.\,\ref{sec:mtp2} and \ref{sec:mtp1} we will use a typical value of $0.05\,M_\odot$ \citep{2002MNRAS.332..745M}
for the uncertainties in the estimate of the secondary mass.

Although the cooling-age determinations are strongly dependent on the cooling model used, the 
thickness of the hydrogen layer on the surface and the occurrence of shell flashes, the cooling-age {\it difference}
is thought to suffer less from systematic errors.  The values for $\Delta\tau$ in Table\,\ref{tab:obswds} 
have an estimated uncertainty of 50\%\ \citep{2002MNRAS.332..745M}.
The age determinations of the components of WD\,1704+481a suggest that star\,2 may be 
the oldest white dwarf, although the age difference is small in both absolute (20\,Myr) 
and relative ($\approx$\,3\%) sense \citep{2002MNRAS.332..745M}.  
Because of this uncertainty we will introduce an eleventh system with a reversed mass ratio. This new system
will be referred to as WD\,1704+481b or 1704b and since we assume that the 
value for $M_2$ is better determined, we will use the following values for this system: 
$M_1 = 0.39\,M_\odot$, $q_2 = 1.43 \pm 0.06$ and $M_2 \equiv q_2 M_1 = 0.56\,M_\odot$.


\section{The stellar evolution code}
\label{sec:code}

We calculate our models using the \textsc{STARS} binary stellar evolution
code, originally developed by \citet{1971MNRAS.151..351E,1972MNRAS.156..361E} 
and with updated input physics as described in \citet{1995MNRAS.274..964P}. 
Opacity tables are taken from \textsc{OPAL}~\citep{1992ApJ...397..717I}, 
complemented with low-temperature opacities  from \citet{1994ApJ...437..879A}.

The equations for stellar
structure and composition are solved implicitly and simultaneously, along
with an adaptive mesh-spacing equation. Because of this, the code is quite
stable numerically and relatively large timesteps can be taken.  As a result of
the large timesteps and because hydrostatic equilibrium is assumed, the code does
not easily pick up short-time-scale instabilities such as thermal pulses.  We can
thus quickly evolve our models up the asymptotic giant branch (AGB), without having to
calculate a number of pulses in detail. We thus assume that such a model is a 
good representation of an AGB star.

Convective mixing is modelled by a
diffusion equation for each of the composition variables, and we assume a
mixing-length to scale-height ratio $l/H_\mathrm{p} = 2.0$.  Convective overshooting is taken 
into account as in \citet{1997MNRAS.285..696S}, with a parameter
$\delta_\mathrm{ov}=0.12$ which corresponds to overshooting lengths of about 0.3 
pressure scale heights ($H_\mathrm{p}$) and is calibrated against accurate stellar data from
non-interacting binaries \citep{1997MNRAS.285..696S,1997MNRAS.289..869P}.
The code circumvents the helium flash in the degenerate core of a low-mass star 
by replacing the model at which the flash occurs by a model with the same total 
mass and core mass but a non-degenerate helium core in which helium was just ignited. 
The masses of the helium and carbon-oxygen cores are defined as the mass coordinates 
where the abundances of hydrogen and helium respectively become less than 10\%.
The binding energy of the hydrogen envelope of a model is calculated by integrating
the sum of the internal and gravitational energy over the mass coordinate, from
the helium-core mass $M_\mathrm{c}$ to the surface of the star $M_\mathrm{s}$:
\begin{equation}
  U_\mathrm{b,e} = \int_{M_\mathrm{c}}^{M_\mathrm{s}} \left(U_\mathrm{int}(m) - \frac{G m}{r(m)}\right)\, \mathrm{d}m
  \label{eq:be}
\end{equation}
The term $U_\mathrm{int}$ is the internal energy per unit of mass, that contains 
terms such as the thermal energy and recombination energy of hydrogen and helium.
It has been argued that the binding energy of the envelope depends
strongly on the definition of its inner boundary, {\it i.e.}\ on the definition
for $M_\mathrm{c}$ \citep{2000A&A...360.1043D}. This is true
for stars with relatively high masses, whose cores and 
envelope binding energies are ill-defined. In our calculations,
however, we deal with relatively low-mass stars on the giant branch:
these have steep density and composition gradients at the edge
of the core, and as a result the mass of the core and binding energy
of the envelope in fact are only weakly dependent on the exact
definition of the core.

We use a version of the code (see Eggleton \& Kiseleva-Eggleton 2002\nocite{2002ApJ...575..461E}) 
that allows for non-conservative binary evolution.  We use the code to calculate
the evolution of both single stars and binaries in which both components are calculated
in full detail. With the adaptive mesh, mass loss by stellar winds or by Roche-lobe
overflow (RLOF) in a binary is simply accounted for in the boundary condition for
the mass.  The spin of the stars is neglected in the calculations and the spin-orbit 
interaction by tides is switched off.  The initial composition of our model stars 
is similar to solar composition: $X=0.70, Y=0.28$ and $Z = 0.02$.


\section{Giant branch models}
\label{sec:gbmodels}

As we have seen in Sect.\,\ref{sec:intro}, each of the double white dwarfs that are observed
today must have formed in a common-envelope event that caused a spiral-in of
the two degenerate stars and expelled the envelope of the secondary.  The intermediate 
binary system that existed before this event, but after the first mass-transfer episode, 
consisted of the first white dwarf (formed from the original primary) and a giant-branch 
star (the secondary).  This giant is thus the star that caused the common envelope and 
in order to determine the properties of the spiral-in that formed each of the observed systems, 
we need a series of giant-branch models.  In this section we present a grid of models
for single stars that evolve from the ZAMS to high up the asymptotic giant branch (AGB).
For each time step we saved the total mass of the star, the radius, the helium-core mass
and the binding energy of the hydrogen envelope of the star.

In an attempt to cover all possibilities, we need to take into account the
effects that can change the quantities mentioned above.  We consider the chemical 
enrichment of the secondary by accretion in a first mass-transfer phase and the effect 
a stellar-wind mass loss may have.  For each of these changes, we compare the results 
to a grid of `basic' models with default parameters.  We keep the overshooting parameter 
$\delta_\mathrm{ov}$ constant for all these grids, because this effect is unimportant for
low-mass stars ($M\!\mathrel{\hbox{\rlap{\lower.55ex \hbox {$\sim$}} \kern-.3em \raise.4ex \hbox{$<$}}}\!2.0\,M_\odot$) and its value is well calibrated for 
intermediate-mass stars (see Sect.\,\ref{sec:code}).

\subsection{Basic models}
\label{sec:basic_models}

In order to find the influence of the effects mentioned above, we want to compare the
models including these effects to a standard.  We therefore calculated a grid of
stellar models, from the zero-age main sequence to high up the asymptotic giant branch
(AGB), with default values for all parameters.  
These models have solar composition and no wind mass loss.  
We calculated a grid of $199$ single-star models with these 
parameters with masses between 0.80 and 10.0\,$M_\odot$, with the logarithm of their 
masses evenly distributed. Model stars with masses lower than about 2.05\,$M_\odot$
experience a degenerate core helium flash and are at that point replaced by a post-helium-flash 
model as described in Sect.\,\ref{sec:code}. Because of the large timesteps the code can take, 
the models evolve beyond the point on the AGB where the carbon-oxygen core (CO-core) mass has 
caught up with the helium-core mass and the first thermal pulse should occur. 

\begin{figure}
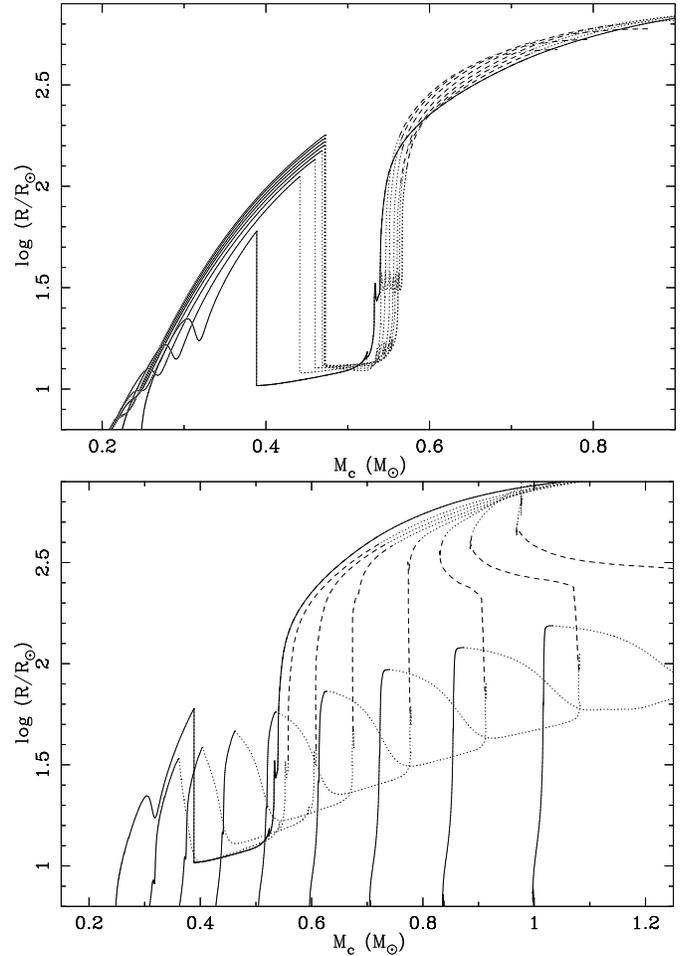

\resizebox{\hsize}{!}{\rotatebox{-90}{\includegraphics{5066f01a.eps}}} 
\resizebox{\hsize}{!}{\rotatebox{-90}{\includegraphics{5066f01b.eps}}} 
\caption{
 Core-mass\,--\,radius relations for the `basic' models, as described in the text. 
 The lines show the logarithm of the radius of the stars as a function of the helium-core mass.  
 {\it Upper panel} ({\bf a}): low-mass grid models with masses of 0.91, 1.01, 1.14, 
 1.30, 1.48, 1.63, 1.81 and 2.00\,$M_\odot$. {\it Lower panel} ({\bf b}): high-mass grid models
 with masses of 2.00, 2.46, 2.79, 3.17, 3.60, 4.09, 4.65, 5.28 and 6.00\,$M_\odot$.
 The 2\,$M_\odot$ model is plotted in both panels throughout as a solid line for easier comparison.
 The other models are shown as solid lines on the first giant branch (FGB), where they could cause a common envelope
 resulting in a spiral-in and creating a helium white dwarf.  The dashed lines show the asymptotic giant branch (AGB), where a
 common envelope would lead to the formation of a carbon-oxygen white dwarf. Dotted lines are parts
 of the evolution where the stars either are smaller than at the tip of the FGB (at lower radii) or 
 where their envelope binding energies become positive on the AGB (at large radii).
 \label{fig:basic_mc-r}  }
\end{figure}
Figure\,\ref{fig:basic_mc-r} shows the radii of a selection of our grid models as a function 
of their helium-core masses.  
We used different line styles to mark different phases in the evolution of these stars,
depending on their ability to fill their Roche lobes or cause a spiral-in and the type of star
a common envelope would result in.  The solid lines show the evolution up the first giant branch (FGB),
where especially the low-mass stars expand much and could cause a common envelope with spiral-in,
in which a helium white dwarf would be formed.  Fig.\,\ref{fig:basic_mc-r}a shows that low-mass stars
briefly contract for core masses around 0.3\,$M_\odot$.  This is due to the first dredge-up,
where the convective envelope deepens down to just above the hydrogen burning shell and increases the hydrogen
abundance there.  The contraction happens when the hydrogen-burning shell catches
up with this composition discontinuity.  After ignition of helium in the core, all stars
shrink and during core helium burning and the first phase of helium fusion in a shell, their radii
are smaller than at the tip of the FGB.  This means that these stars could never start filling their
Roche lobes in this stage.  These parts of the evolution are plotted with dotted lines.  
Once a CO core is established, the stars evolve up the AGB and eventually get a radius that is larger
than that on the FGB.  The stars are now capable of filling their Roche lobes again and cause a 
common envelope with spiral-in.  In such a case 
we assume that the whole helium core survives the spiral-in and that the helium burning shell will
convert most of the helium to carbon and oxygen, eventually resulting in a CO white dwarf, probably with 
an atmosphere that consists of a mixture of hydrogen and helium.  This part of the evolution is marked with dashed lines.  
Fig.\,\ref{fig:basic_mc-r}b shows that the most massive
models in our grid have a decreasing helium-core mass at some point on the AGB.  This happens
at the so-called second dredge-up, where the convective mantle extends inward, into the helium core
and mixes some of the helium from the core into the mantle, thereby reducing the mass of the core.  
Models with masses between about 1.2 and 5.6\,$M_\odot$ expand to such large radii that the binding
energy of their hydrogen envelopes become positive.  In Sect.\,\ref{sec:mtp2} we are looking for 
models that can cause a spiral-in based on energy balance in the second mass-transfer phase, 
for which purpose we require stars that have hydrogen envelopes with a negative binding energy.  
A positive binding energy means that there is no orbital energy 
needed for the expulsion of the envelope and thus the orbit will not shrink during a common envelope 
caused by such a star.  We have hereby implicitly assumed that the recombination energy is 
available during common-envelope ejection.

\begin{figure}
\resizebox{\hsize}{!}{\rotatebox{-90}{\includegraphics{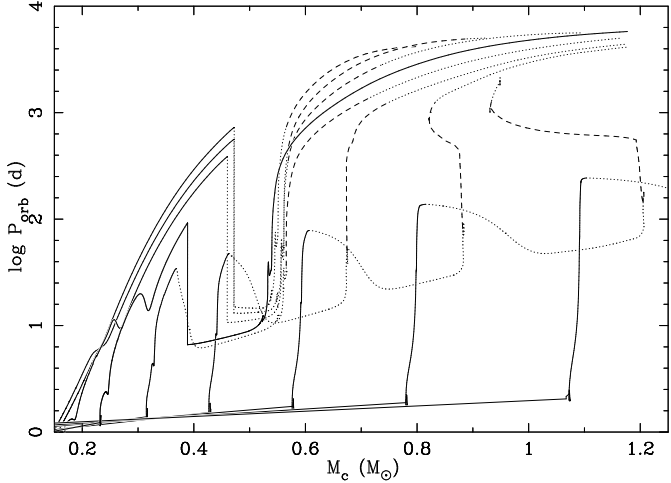}}} \caption{
Helium-core-mass\,--\,orbital period tracks for the `basic' models.  
The lines show the logarithm of the orbital period at which the Roche lobe is filled for grid 
models with masses of 1.01, 1.27, 1.59, 2.00, 2.52, 3.17, 3.99, 5.02 and 6.32\,$M_\odot$.
The period was obtained from the radius of the model star, under the assumption that it fills
its Roche lobe and the companion has a mass equal to the helium-core mass of the model.  This way, the
system would undergo a spiral-in that would lead to a binary with mass ratio $q = 1$.  The line 
styles have the same meaning as in Fig.\,\ref{fig:basic_mc-r}.
 \label{fig:basic_mc-p} }
\end{figure}

To give some idea what kind of binaries can cause a spiral-in and could be the progenitors 
of the observed double white dwarfs, we converted the radii of the stars displayed in 
Fig.\,\ref{fig:basic_mc-r} into orbital periods of the pre-common-envelope systems.  To do this, 
we assumed that the Roche-lobe radius
is equal to the radius of the model star, and that the mass of the companion is
equal to the mass of the helium core of the model.  This is justified by Table\,\ref{tab:obswds}, 
where the geometric mean of the mass ratios is equal to 1.03.  The result is shown in 
Fig.\,\ref{fig:basic_mc-p}.

\begin{figure}
\resizebox{\hsize}{!}{\rotatebox{-90}{\includegraphics{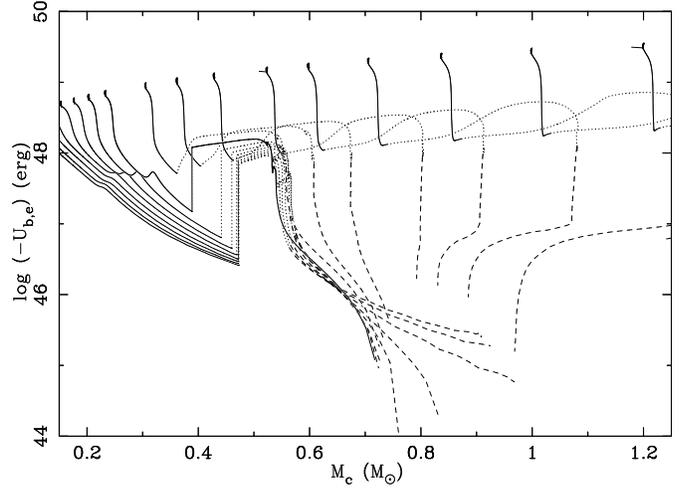}}} \caption{
The logarithm of the binding energy of the `basic' model stars as a function 
of the helium-core mass.  The grid models with masses of 0.91, 1.01, 1.14, 1.30, 1.48, 1.63, 1.81, 2.00, 
2.46, 2.79, 3.17, 3.70, 4.09, 4.65, 5.28, 6.00 and 6.82\,$M_\odot$ are shown.  The 2.00\,$M_\odot$ model
is drawn as a solid line, the line styles for the other models have the same meaning as in Fig.\,\ref{fig:basic_mc-r}.
The parts where the envelope binding energy is zero (before a helium core develops) or positive are not shown.
 \label{fig:basic_mc-eb}  }    
\end{figure}

In Sect.\,\ref{sec:mtp2} we will need the efficiency parameter $\alpha_\mathrm{ce}$ of each 
common-envelope model to judge whether that model is acceptable or not.  In order to calculate
this parameter we must know the binding energy of the hydrogen envelope of the
progenitor star (see Eq.\,\ref{eq:forwardce}), that is provided by the evolution code as shown in Eq.\,\ref{eq:be}.
The envelope binding energy is therefore an important parameter and we show it for a selection of models in 
Fig.\,\ref{fig:basic_mc-eb}, again as a function of the helium-core mass.
Because the binding energy is usually negative, we plot the logarithm of $-U_\mathrm{b,e}$.  The 
phases where the envelope binding energy is non-negative are irrelevant for our calculations of $\alpha_\mathrm{ce}$ and therefore 
not shown in the Figure.

\begin{figure}
\resizebox{\hsize}{!}{\rotatebox{-90}{\includegraphics{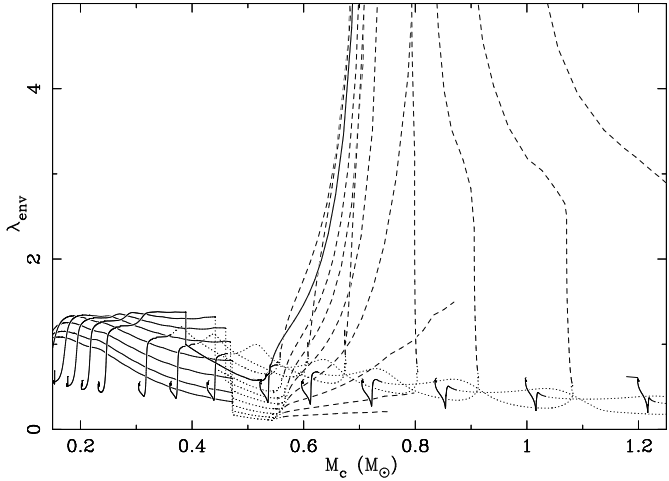}}} \caption{
The envelope-structure parameter $\lambda_\mathrm{env}$ for the `basic' models, as a 
function of the helium-core mass.  The same grid models are shown as in Fig.~\ref{fig:basic_mc-eb}. 
The meaning of the line styles is explained in the caption of Fig.\,\ref{fig:basic_mc-r}.
 \label{fig:basic_mc-lam}  }
\end{figure}

Many common-envelope calculations in the literature use the so-called envelope-structure 
parameter $\lambda_\mathrm{env}$ to estimate the envelope binding energy from basic
stellar parameters in case a detailed model is not available
\begin{equation}
  U_\mathrm{b,e} \, = \, - \frac{G\, M_*\, M_\mathrm{env}}{\lambda_\mathrm{env}\, R_*}.
\label{eq:lambda}
\end{equation}
De Kool et al. (1987)\nocite{1987A&A...183...47D} suggest that $\lambda_\mathrm{env} \approx 0.5$.  
Since we calculate the binding energy of the stellar envelope, we can invert Eq.\,\ref{eq:lambda} 
and calculate $\lambda_\mathrm{env}$ 
(because we mostly consider low-mass stars, the binding energy and hence $\lambda_\mathrm{env}$
does not depend strongly on the definition of the core mass, see the discussion below Eq.\,\ref{eq:be} 
and see also Dewi \& Tauris 2000\nocite{2000A&A...360.1043D}).
Figure\,\ref{fig:basic_mc-lam} shows the results of these calculations as a function of the helium 
core mass, for the same selection of models as in Fig.\,\ref{fig:basic_mc-eb}.  We see that a value of 
$\lambda_\mathrm{env}=0.5$ is a good approximation for the lower FGB of a low-mass star, or the FGB of
a higher-mass star.  A low-mass star near the tip of the first giant branch has a structure parameter 
between 0.5 and 1.5 and for most stars $\lambda_\mathrm{env}$ increases to more than unity rather quickly,
especially when the stars expand to large radii and the binding energies come close to zero.

\subsection{Chemical enrichment by accretion}
\label{sec:enrichment}

The secondary that causes the common envelope may have gained mass by accretion during the first 
mass-transfer phase.  If this mass transfer was stable, the secondary has probably accreted much 
of the envelope of the primary star.  The deepest layers of the envelope of the donor are usually 
enriched with nuclear burning products, brought up from the core by a dredge-up process.  This way, 
the secondary may have been enriched with especially helium which, in sufficiently large quantities, 
can have an appreciable effect on the opacity in the envelope of the star and thus its radius.
This would change the core-mass\,--\,radius relation of the star and the common envelope it causes.

To see whether this effect is significant, we considered a number of binary models that evolved through 
stable mass transfer to produce a white dwarf and a main-sequence secondary.  The latter had a mass between 
2 and $5\,M_\odot$ in the cases considered, of which 50--60\%\ was accreted.  We then took this secondary 
out of the binary and let it evolve up the asymptotic giant branch, to the point where the code picks up 
a shell instability and terminates.  We then compared this final model to 
a model of a single star with the same mass, but with solar composition, that was evolved to
the same stage.  In all cases the core-mass\,--\,radius relations coincide with those in 
Fig.\,\ref{fig:basic_mc-r}.  When we compared the surface helium abundances of these 
models, after one or two dredge-ups, we found that although the abundances were enhanced 
appreciably since the ZAMS, they were enhanced with approximately the same amount and the relative 
difference of the helium abundance at the surface between the different models was always less than 1.5\%.  
In some cases the model that had accreted from a companion had the lower surface helium abundance.

The small amount of helium enrichment due to accretion gives rise to such small changes in the
core mass\,--\,radius relation, that we conclude that this effect can be ignored in our common-envelope
calculations in Sect.\,\ref{sec:mtp2}.

\subsection{Wind mass loss}
\label{sec:wind}

\begin{figure}
\resizebox{\hsize}{!}{\rotatebox{-90}{\includegraphics{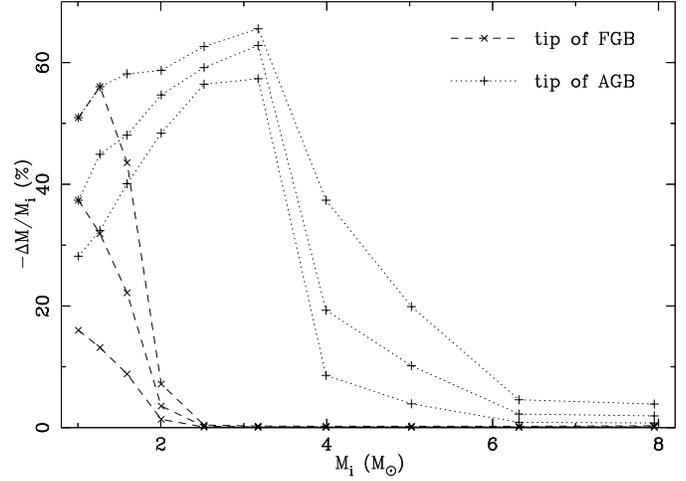}}} 
\caption{
 The fraction of mass lost at two moments in the evolution of a star as a function of its initial mass, for the 
 three different wind strengths ($C_\mathrm{rml}$ = 0.2, 0.5 and 1.0) used in the grid.  This fraction 
 is shown for the tip of the FGB (dashed lines and crosses), and the `tip of the AGB' (dotted lines and 
 plusses). See the text for details.
 \label{fig:wind_mi-mf}  }
\end{figure}

The mass loss of a star by stellar wind can change the mass of a star appreciably before 
the onset of Roche-lobe overflow, and the mass loss can influence the relation between the core mass
and the radius of a star.  From Fig.\,\ref{fig:basic_mc-r} it is already clear that this relation 
depends on the total mass of the star.  In this section, we would therefore like to find out whether
a conservative model star of a certain total mass and core mass has the same radius and envelope
binding energy as a model with the same total mass and core mass, but that started out as a more 
massive star, has a strong stellar wind and just passes by this mass on its evolution down to 
even lower masses.  We calculated a small grid of models with ten different initial masses between 
1.0\,$M_\odot$ and 8.0\,M$_\odot$, evenly spread in $\log M$ and included a Reimers type mass loss 
\citep{1975MSRSL...8..369R} of variable strength:
\begin{equation}
  \dot{M}_\mathrm{rml} ~=~ -4\times 10^{-13}\, M_\odot\, \mathrm{yr}^{-1} ~C_\mathrm{rml} ~\left(\frac{L}{L_\odot}\right) \left(\frac{R}{R_\odot}\right) \left(\frac{M}{M_\odot}\right)^{-1},
  \label{eq:rml}
\end{equation}
where we have used the values $C_\mathrm{rml}$ = 0.2, 0.5 and 1.0.  The basic models of Sect.\,\ref{sec:basic_models} 
are conservative and therefore have $C_\mathrm{rml} = 0$.  The effect of these winds
on the total mass of the model stars in our grid is displayed in Fig.\,\ref{fig:wind_mi-mf}.
 It shows the fraction of mass lost at the tip of the first giant branch (FGB) and the `tip of the
 asymptotic giant branch' (AGB).  The first moment is defined as the point where the star reaches 
 its largest radius before helium ignites in the core, the second as the point where the radius of 
 the star reaches its maximum value while the envelope binding energy is still negative.  Values for
 both moments are plotted in Fig.\,\ref{fig:wind_mi-mf} for each non-zero value of $C_\mathrm{rml}$ in the grid.
 For the two models with the lowest masses the highest mass-loss rates are so high that the 
 total mass is reduced sufficiently on the FGB to keep the star from igniting helium in the core, and the
 lines in the plot coincide.  Stars more massive than 2\,$M_\odot$ have negligible mass loss on the FGB,
 because they have non-degenerate helium cores so that they do not ascend the FGB as far as stars of lower mass. Their
 radii and luminosities stay relatively small, so that Eq.\,\ref{eq:rml} gives a low mass loss rate.
 For stars of 4\,$M_\odot$ or more, the mass loss is diminutive and happens only shortly before the 
 envelope binding energy becomes positive.  We can conclude that for these stars the wind mass loss has
 little effect on the core-mass\,--\,radius relation.

\begin{figure}
\resizebox{\hsize}{!}{\rotatebox{-90}{\includegraphics{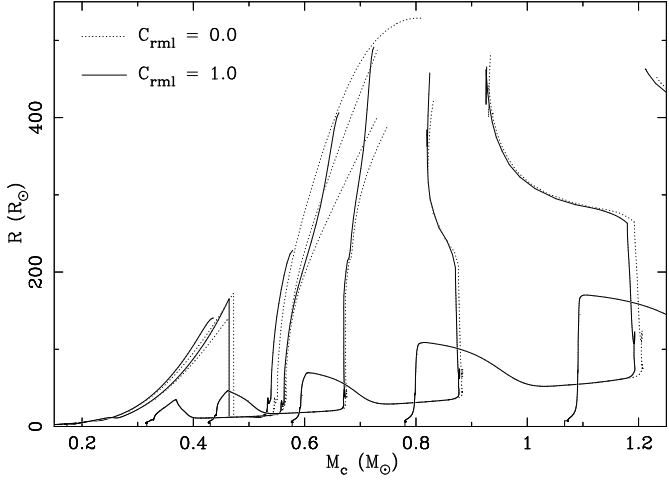}}} 
\caption{
 Comparison of a selection from the small grid of models with a stellar wind.  The models displayed have
 masses of 1.0, 1.6, 2.5, 3.2, 4.0, 5.0 and 6.3\,$M_\odot$.  The wind strength parameters are  
 $C_\mathrm{rml}$ = 0.0 (dotted lines) and $C_\mathrm{rml}$ = 1.0 (solid lines, the strongest mass loss 
 in the grid).  Stars with mass loss are usually larger, but for models of 4.0\,$M_\odot$ or more this
 effect becomes negligible.  The 1.0\,$M_\odot$ model loses so much mass that it never ignites helium in the core.
 \label{fig:wind_mc-r}  }
\end{figure}

 The core-mass\,--\,radius relations for a selection of the models from our wind grid are shown in 
 Fig.\,\ref{fig:wind_mc-r}.
The Figure compares models without stellar wind with models that have the strongest stellar wind in our 
grid ($C_\mathrm{rml} = 1.0$).  Models with the other wind strengths would lie between those shown, but
are not plotted for clarity.  The greatest difference in Fig.\,\ref{fig:wind_mc-r} is in the 1.0\,$M_\odot$ 
model.  The heavy mass loss reduces the total mass of the star to 0.49\,$M_\odot$ on the first giant
branch, so that the star is not massive enough to ignite helium in the core.  
Fig.\,\ref{fig:wind_mc-r} shows that models with mass loss are larger than conservative models for the same 
core mass, as one would expect from Fig.\,\ref{fig:basic_mc-r}.  This becomes clear on the FGB for stars
that have degenerate helium cores, because they have large radii and luminosities and lose large amounts of
mass there.  For stars more massive than about $2\,M_\odot$ the mass loss
becomes noticeable on the AGB.  Stars of $4\,M_\odot$ or more show little difference in 
Fig.\,\ref{fig:wind_mc-r}.  The envelope binding energies have similar differences in the same mass regions.

\begin{figure}
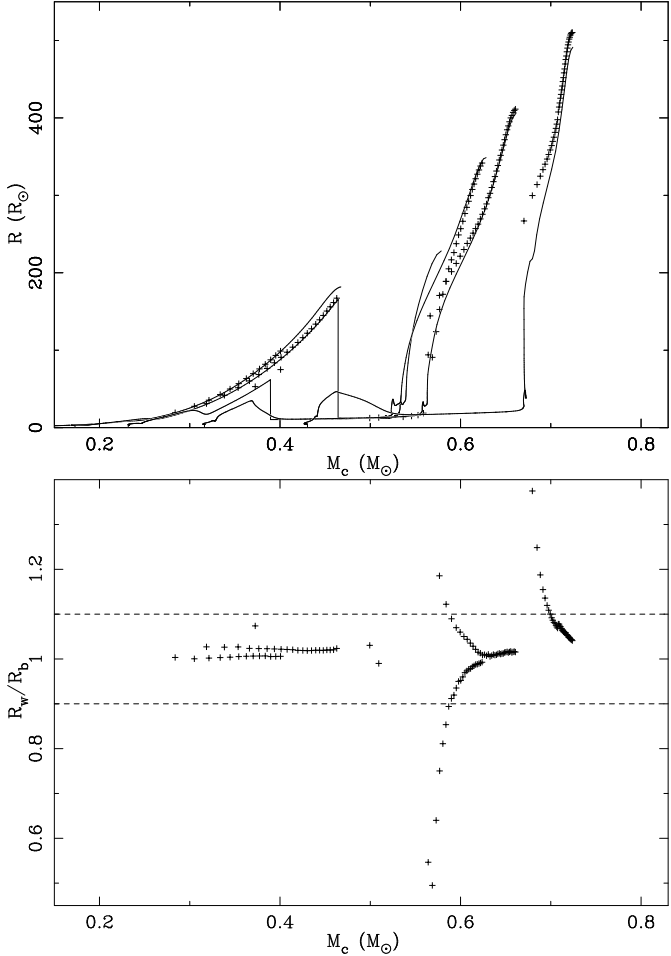

\resizebox{\hsize}{!}{\rotatebox{-90}{\includegraphics{5066f07a.eps}}} 
\resizebox{\hsize}{!}{\rotatebox{-90}{\includegraphics{5066f07b.eps}}} 
\caption{
 Comparison of a selection of grid models with $C_\mathrm{rml} = 1.0$ with initial masses of 
 1.3, 1.6, 2.0, 2.5 and 3.2\,$M_\odot$ to the basic models ($C_\mathrm{rml} = 0.0$).  
 {\it Upper panel} ({\bf a}): Comparison of the radius of the models with a stellar wind (solid lines) 
 and the radius of a basic model with the same mass and core mass (plusses).
 {\it Lower panel} ({\bf b}): The fraction of the radius of the wind model $R_\mathrm{w}$ over the radius of the
 basic model $R_\mathrm{b}$ with the same total and core mass.  Each data point corresponds to a point in the upper panel.
 Of the data points in the upper panel, 7 out of 143 (5\%) lie outside the plot boundaries in the lower panel. 
 The dashed horizontal lines show the region where agreement is better than 10\%, where 83\%\ of the data points
 lie. The 1.0\,$M_\odot$ model was left out because there are only a few basic models with lower mass, the higher-mass models
 were left out because they lose very little mass (see Fig.\,\ref{fig:wind_mi-mf}).
 \label{fig:wind_comp}  }
\end{figure}

The question is whether the properties of the model with reduced mass due to the wind are the same 
as those for a conservative model of that mass.  
In order to answer this question, we have compared the models from the `wind grid' to the basic, conservative 
models.  As the wind reduces the total mass of a model star, it usually reaches masses that are equal to that of
several models in the conservative grid.  As this happens, we interpolate linearly within the mass-losing
model to find the exact moment where its mass equals the mass of the conservative model.  We then use the 
helium-core mass of the interpolated mass-losing model to find the moment where the conservative model
has the same core mass and we calculate its radius and envelope binding energy, again by linear interpolation.  
This way we can compare the two models at the moment in evolution where they have the same total mass and
the same core mass.  This comparison is done in Fig.\,\ref{fig:wind_comp}. 
 Figure\,\ref{fig:wind_comp}a directly compares the radii of the two sets of models, in Fig.\,\ref{fig:wind_comp}b 
 the ratio of the two radii is shown.

Of the data points in Fig.\,\ref{fig:wind_comp}b 83\%\ lie between 0.9 and 1.1 and 61\%\ between 0.95 
and 1.05.  For the wind models with $C_\mathrm{rml} = 0.2$ these numbers are 99\%\ and 97\%, and for the models 
with $C_\mathrm{rml} = 0.5$ they are 94\%\ and 85\%\ respectively.  As can be expected, the models that have a
lower -- and perhaps a more realistic -- mass-loss rate compare better to the conservative models.  We see in 
Fig.\,\ref{fig:wind_comp}a that many of the points that lie farther from unity need only a small shift in core 
mass to give a perfect match.  This shift is certainly less than $0.05\,M_\odot$, which is what we will 
adopt for the uncertainty of the white-dwarf masses in Sect.\,\ref{sec:mtp2}.  We conclude here that there is 
sufficient agreement between a model that reaches a certain total mass because it suffers from mass loss and a 
conservative model of the same mass.  The agreement is particularly good for stars high up on the FGB or AGB,
where the density contrast between core and envelope is very large.


\section{Second mass-transfer phase}
\label{sec:mtp2}

For the formation of two white dwarfs in a close binary system, two phases of mass
transfer must happen.  We will call the binary system before the first mass
transfer the {\it initial binary}, with masses and orbital period $M_\mathrm{1i}$,
$M_\mathrm{2i}$ and $P_\mathrm{i}$.  If one considers mass loss due to stellar wind
before the first mass-transfer episode, these parameters are not necessarily equal
to the ZAMS parameters, especially for large `initial' periods.
The binary between the two mass-transfer phases
is referred to as the {\it intermediate binary} with $M_\mathrm{1m}$, $M_\mathrm{2m}$ 
and $P_\mathrm{m}$.  After the two mass-transfer episodes, we obtain the
{\it final binary} with parameters $M_\mathrm{1f}$, $M_\mathrm{2f}$ and $P_\mathrm{f}$, 
that should correspond to the values that are now observed and listed in 
Table\,\ref{tab:obswds}.  The subscripts `1' and `2' are used for the initial primary 
and secondary as defined in Sect.\,\ref{sec:observed}.

In the first mass transfer, the primary star fills its
Roche lobe and loses mass, that may or may not be accreted by the secondary.
This leads to the formation of the intermediate binary, that consists of the first 
white dwarf and a secondary of unknown mass.
In the second mass-transfer phase, the secondary fills its Roche lobe and loses
its envelope.  The second mass transfer results in the observed double white
dwarf binaries that are listed in Table\,\ref{tab:obswds} and must account for 
significant orbital shrinkage.  This is because the youngest white dwarf
must have been the core of its progenitor, the secondary in the intermediate binary.
Stars with cores between 0.3 and 0.7\,$M_\odot$ usually have radii of several tens
to several hundreds of solar radii, and the orbital separation of the binaries they
reside in must be even larger than that.  The orbital separation of the observed 
systems is typically only in the order of a few solar radii (Table\,\ref{tab:obswds}).  
Giant stars with large radii have deep convective envelopes and when such a star fills 
its Roche lobe, the ensuing mass transfer will be unstable and occur on a very short,
dynamical timescale, especially if the donor is much more massive than its companion.  
It is thought that the envelope of such a star can engulf its companion and 
this event is referred to as a {\it common envelope}.  The companion and the core 
of the donor orbit inside the common envelope and drag forces will release energy 
from the orbit, causing the orbit to shrink and the two degenerate stars to spiral in.
The freed orbital energy will heat the envelope and eventually expel it. This
way, the hypothesis of the common envelope with spiral-in can phenomenologically explain 
the formation of close double-white-dwarf binaries.

\subsection{The treatment of a spiral-in}
\label{sec:ce_treatment}

In order to estimate the orbital separation
of the post-common envelope system quantitatively, it is often assumed that the 
orbital energy of the system is decreased by an amount that is equal to the binding 
energy of the envelope of the donor star \citep{1984ApJ...277..355W}:
\begin{equation}
  U_\mathrm{b,e} ~=~
  - \alpha_\mathrm{ce}\, \left[ \frac{G M_\mathrm{1f} M_\mathrm{2f}}{2 a_\mathrm{f}}\, -\, 
  \frac{G M_\mathrm{1m}  M_\mathrm{2m}}{2 a_\mathrm{m}} \right].
  \label{eq:forwardce}
\end{equation}
The parameter $\alpha_\mathrm{ce}$ is the {\it common-envelope parameter}
that expresses the efficiency by which the orbital energy is deposited in the
envelope.  Intuitively one would expect that $\alpha_\mathrm{ce} \approx 1$.  
However, part of the liberated orbital energy might be radiated away from the envelope during the process, 
without contributing to its expulsion, thereby lowering $\alpha_\mathrm{ce}$.  
Conversely, if the common-envelope phase would last long enough that the donor star 
can produce a significant amount of energy by nuclear fusion, or if energy is released 
by accretion on to the white dwarf, this energy will 
support the expulsion and thus increase $\alpha_\mathrm{ce}$.  

In the forward calculation of a spiral-in the final orbital separation $a_\mathrm{f}$ 
depends strongly on the parameter $\alpha_\mathrm{ce}$, which must therefore be known.
In this section we will try to establish the binary systems that were the possible
progenitors of the observed double white dwarfs and we will therefore perform {\it backward}
calculations.  The advantage of this is that we start as close as possible to the observations
thus introducing as little uncertainty as possible.  The problem with this strategy is that we 
do not know the mass of the secondary progenitor beforehand.  We will have to consider this
mass as a free parameter and assume a range of possible values for it.
The grid of single-star models of Sect.\,\ref{sec:gbmodels} provides us 
with the total mass, core mass, radius and envelope binding energy at every moment 
of evolution, for a range of total masses between 0.8 and 10\,$M_\odot$.  
It is then not necessary to know the common-envelope parameter, and we 
can in fact use the binding energy to calculate the value for $\alpha_\mathrm{ce}$ 
that is needed to shrink the orbit of a model to the period of the observed double white dwarf. 
The accuracy with which $\alpha_\mathrm{ce}$ can be calculated 
scales directly with the accuracy of the binding energy. As
discussed below Eq.\,\ref{eq:be}, $U_\mathrm{b,e}$ and as a consequence 
$\alpha_\mathrm{ce}$ do not depend strongly on the exact way in
which the core is defined, for the low-mass stars which determine
the outcome of our calculations (see Table\,\ref{tab:solutions} below).
We make two assumptions about the evolution of the two stars 
during the common envelope to perform these backward calculations:
\begin{enumerate}
\item the core mass of the donor does not change,
\item the mass of the companion does not change.
\end{enumerate}
The first assumption will be valid if the timescale on which the common envelope
takes place is much shorter than the nuclear-evolution timescale of the giant donor.  
This is certainly true, since the mass transfer occurs on the dynamical timescale of
the donor.  The second assumption is supported firstly by the fact that the companion 
is a white dwarf, a degenerate object that has a low Eddington accretion limit and is furthermore 
difficult to hit directly by a mass stream from the donor.  The white dwarf could accrete
matter in the Bondi-Hoyle fashion \citep{1944MNRAS.104..273B}. This would not change the mass 
of the white dwarf significantly but could release appreciable amounts of energy.
Secondly, a common envelope 
is established very shortly after the beginning of the mass transfer, so that the mass 
stream disappears and the white dwarf is orbiting inside the fast-expanding envelope 
rather than accreting mass from the donor.  In the terminology used here, the second
assumption can be written as $M_\mathrm{1m} = M_\mathrm{1f}$.

From the two assumptions above it follows that the mass of the second white dwarf, 
the one that is formed in the spiral-in, is equal to the helium-core mass of the donor 
at the moment it fills its Roche lobe.  There is therefore a unique moment in the evolution 
of a given model star at which it could cause a common envelope with spiral-in
and produce a white dwarf of the proper mass.  Recall from Fig.\,\ref{fig:basic_mc-r}b
that although the second dredge-up reduces notably the helium-core mass of the more 
massive models in the grid, there is no overlap in core mass in the phases where 
the star could fill its Roche lobe on the first giant branch (solid lines) or asymptotic 
giant branch (dashed lines).  The moment where the model star could produce a white dwarf
of the desired mass in a common envelope with spiral-in is therefore defined by two conditions:
\begin{enumerate}
\item the helium-core mass of the model reaches the mass of the white dwarf,
\item the model star has its largest radius so far in its evolution.
\end{enumerate}
The second restriction is necessary because stars can shrink appreciably 
during their evolution, as noted in Sect.\,\ref{sec:basic_models}.
If the core of a model star 
obtains the desired mass at a point in the evolution where the star is smaller than it
has been at some point in the past, then the star cannot fill its Roche lobe at the
right moment to produce a white dwarf of the proper mass and therefore this star cannot be
the progenitor of the white dwarf.  This way, each model star has at most one moment
in its evolution where it could fill its Roche lobe and produce the observed double
white dwarf.  If such a moment does not exist, the model star is rejected as a possible
progenitor of the second white dwarf.

If the model star could be the progenitor of the youngest white dwarf in the 
observed system, the computer model gives us the radius of the donor star, that must be
equal to the Roche-lobe radius.  Under the assumption that the mass
of the first white dwarf does not change in the common envelope, the mass ratio of the two
stars $q_\mathrm{2m} \equiv M_\mathrm{2m}/M_\mathrm{1m}$ and the Roche-lobe radius of the 
secondary star $R_\mathrm{Rl2m}$ give us the orbital separation before the spiral-in 
$a_\mathrm{m}$, where we use the fit by \citet{1983ApJ...268..368E}
\begin{equation}
  R_\mathrm{Rl2m} ~=~ a_\mathrm{m}\, \frac{0.49\,q_\mathrm{2m}^{2/3}}{0.6\,q_\mathrm{2m}^{2/3} + \ln\left( 1 + q_\mathrm{2m}^{1/3} \right)}, ~~~ 0 < q_\mathrm{2m} < \infty.
  \label{eq:eggleton}
\end{equation}
Kepler's law finally provides us with the orbital period $P_\mathrm{m}$ of the intermediate system.  
The stellar model also gives the binding energy of the envelope
of the donor $U_\mathrm{b,e}$ at the onset of the common envelope and we can use
Eq.\,\ref{eq:forwardce} to determine the common-envelope parameter $\alpha_\mathrm{ce}$.
We will use $\alpha_\mathrm{ce}$ to judge the validity of the model star to be the progenitor of the
second white dwarf.  There are several reasons why a numerical solution can be rejected.
Firstly, the proposed donor could be a massive star
with a relatively small radius.  Then $a_\mathrm{m}$ will be small and it might happen
that $a_\mathrm{m}\,<\,a_\mathrm{f}\,\frac{M_\mathrm{2m}}{M_\mathrm{2f}}$, so that
$\alpha_\mathrm{ce}\,<\,0$.  This means that energy is needed to change the orbit
from $a_\mathrm{m}$ to $a_\mathrm{f}$, or even that $a_\mathrm{m}\,<\,a_\mathrm{f}$ and 
a spiral-in (if it can be called that) to the desired orbit will not
lead to expulsion of the common envelope.  Secondly, as mentioned above, $\alpha_\mathrm{ce}$
is expected to be close, though not necessarily equal, to unity.  However if the 
parameter is either much smaller or much larger than 1, we will consider the spiral-in 
to be `physically unbelievable'.  We arbitrarily chose the boundaries between
which $\alpha_\mathrm{ce}$ must lie for a believable spiral-in to be a factor of ten either way:
$0.1\,\leq\,\alpha_\mathrm{ce}\,\leq\,10$.  We think that the actual value
for $\alpha_\mathrm{ce}$ should be more constrained than that because common-envelope 
evolution is thought to last only a short time so that there is little time to generate or 
radiate large amounts of energy, but keep the range as
broad as it is to be certain that all possible progenitor systems are considered in
our sample.

\subsection{Results of the spiral-in calculations}
\label{sec:ce_results}

\begin{figure}
\resizebox{\hsize}{!}{\rotatebox{-90}{\includegraphics{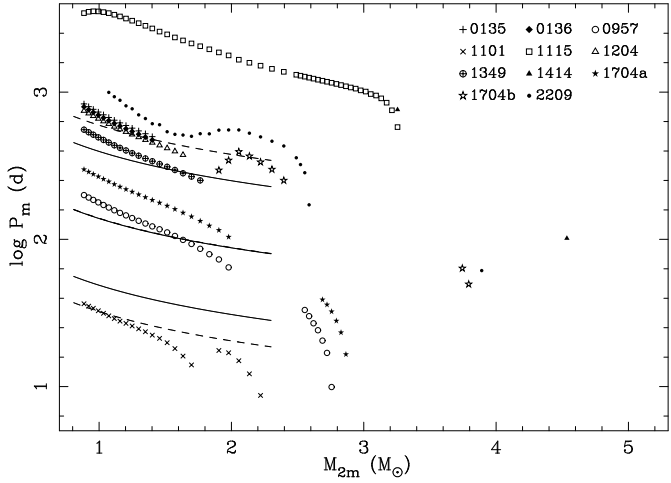}}} 
\caption{
 Results of the spiral-in calculations, each individual symbol is a solution  
 of the calculations and thus represents one pre-CE binary.   
 The figure shows the
 logarithm of the orbital period of the intermediate
 binary $P_\mathrm{m}$ as a function of the secondary mass $M_\mathrm{2m}$.  
 Different symbols represent different observed systems, as explained in the legend.  System
 1704a is the system listed in Table\,\ref{tab:obswds}, 1704b is the same system, but with 
 the reverse mass ratio.
 For solutions with $M_\mathrm{2m} < 2.5\,M_\odot$, only every third solution is plotted
 for clarity. Around $M_\mathrm{2m}\!=\!1.2$ and $\log P_\mathrm{m}\!=\!2.8$ the symbols of 
 WD\,0135--052, WD\,0136+768 and WD\,1204+450 overlap due to the fact that they have similar
 white-dwarf masses.
 For comparison we show the lines
 of the solutions for (top to bottom) WD\,0136+768, WD\,0957--666 and WD\,1101+364 taken from 
 \citet{2000A&A...360.1011N}, as described in the text.
 \label{fig:ce_allwds}  }
\end{figure}

We will now apply the stellar models of Sect.\,\ref{sec:basic_models} as described in the
previous section to calculate potential progenitors to the observed double white dwarfs as
listed in Table\,\ref{tab:obswds}.  As input parameters we took the values $P_\mathrm{f} = P_\mathrm{orb}$
and $M_\mathrm{2f} = M_2$ from the table, and assumed that $M_\mathrm{1f}\,\equiv\,M_2/q_2$, 
where $q_2$ is the observed mass ratio listed in Table\,\ref{tab:obswds}.  We thus ignore
for the moment any uncertainty in the observed masses.
Figure\,\ref{fig:ce_allwds} shows the orbital period $P_\mathrm{m}$ as a function of the 
secondary mass $M_\mathrm{2m}$.  
Each symbol is a solution to the spiral-in calculations
and represents an intermediate binary system that consists of the first white dwarf of mass
$M_\mathrm{1m}\!=\!M_\mathrm{1f}$, a companion of mass $M_\mathrm{2m}$ and an orbital period $P_\mathrm{m}$.
The secondary of this system will fill its Roche lobe at the moment when its helium-core
mass is equal to the mass of the observed white dwarf $M_\mathrm{2f}$, and can thus produce
the observed double-white-dwarf system with a common-envelope parameter that lies between
0.1 and 10.  

The solutions for each system in Fig.\,\ref{fig:ce_allwds} seem to lie on curves that roughly run from long orbital periods
for low-mass donors to short periods for higher-mass secondaries.  This is to be expected, 
partially because higher-mass stars have smaller radii at a certain core mass than stars of lower mass
(see Fig.\,\ref{fig:basic_mc-r}) and thus fill their Roche lobes at shorter orbital periods,
but mainly because the orbital period of a Roche-lobe filling star falls off approximately with the 
square root of its mass.
The Figure also shows gaps between the solutions, for instance for WD\,0957--666 and WD\,1704+481a,
between progenitor masses $M_\mathrm{2m}$ of about 2 and $2.5\,M_\odot$.  These gaps arise because the 
low-mass donors on the left side of the gap ignite helium degenerately when the core mass is 
$0.47\,M_\odot$, after which the star shrinks, whereas for stars with masses close to $2\,M_\odot$
helium ignition is non-degenerate and occurs at lower core masses, reaching a minimum for 
stars with a mass of $2.05\,M_\odot$, where helium ignition occurs when the helium-core mass
amounts to $0.33\,M_\odot$ (see Fig.\,\ref{fig:basic_mc-r}).  Thus, for white dwarfs with masses 
between 0.33 and $0.47\,M_\odot$ there is a range of masses between about 1.5 and
$3\,M_\odot$ for which the progenitor has just ignited helium in the core, and thus shrunk, 
when it reaches the desired helium-core mass.  

The dip and gap in Fig.\,\ref{fig:ce_allwds} for WD\,1101+364 (with $M_\mathrm{2f}\approx0.29M_\odot$) 
around $M_\mathrm{2m}=1.8\,M_\odot$ can be attributed to the first dredge-up that occurs for low-mass 
stars ($M < 2.2\,M_\odot$) early on the first giant branch.  Stars with these low masses shrink slightly 
due to this dredge-up that occurs at core masses between about 0.2 and $0.33\,M_\odot$, the higher 
core masses for the more massive stars (see Fig.\,\ref{fig:basic_mc-r}a).  
Stars at the low-mass ($M_\mathrm{2m}$) side of the gap obtain the desired core mass just after the
dredge-up, are relatively small and fill their Roche lobes at short periods.  Stars with masses that 
lie in the gap reach that core mass while shrinking and cannot fill their Roche lobes for that reason.
Stars at the high-mass end of the gap fill their Roche lobes just before the dredge-up so that
this happens when they are relatively large and therefore this happens at longer orbital periods.

For comparison we display as solid lines in Fig.\,\ref{fig:ce_allwds} the results for the systems WD\,0136+768, WD\,0957--666 
and WD\,1101+364 (from top to bottom), as found by \citet{2000A&A...360.1011N} and shown in their Fig.\,1.  
The differences between their and our results stem in part from the fact that the 
values for the observed masses have been updated by observations since their paper was published.  To compensate for 
this we include dashed lines for the two systems for which this is the case.  The dashed lines
were calculated with their method but the values for the observed masses as listed in this paper.  By 
comparing the lines to the symbols for the same systems, we see that they lie in the same region of the 
plot and in the first order approach they give about the same results.  However, the slopes in the two
sets of results are clearly different.  This can be attributed to the fact that \citet{2000A&A...360.1011N}
used a power law to describe the radius of a star as a function of its core mass only. The change in 
orbital period with mass in their calculations is the result of changing the total mass in Kepler's law.
Furthermore, they assumed that all stars with masses between 0.8 and $2.3\,M_\odot$ have a solution, 
whereas we find limits and gaps, partially due to the fact that we take into account the fact that stars 
shrink and partially because in Fig.\,\ref{fig:ce_allwds} only solutions with a restricted $\alpha_\mathrm{ce}$
are allowed.  On the other hand, we allow stars more massive than $2.3\,M_\odot$ as possible progenitors.

\begin{figure}
\resizebox{\hsize}{!}{\rotatebox{-90}{\includegraphics{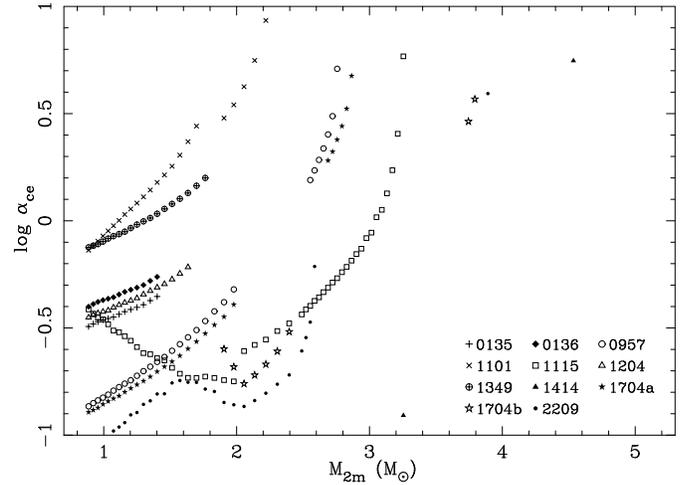}}} 
\caption{
 The logarithm of the common-envelope parameter $\alpha_\mathrm{ce}$ for the solutions of the 
 spiral-in calculations shown in Fig.\,\ref{fig:ce_allwds}.
 Different symbols represent different observed systems.  
 For $M_\mathrm{2m} < 2.5\,M_\odot$ every third solution is plotted only.
 \label{fig:ce_ace}  }
\end{figure}

\begin{figure}
\resizebox{\hsize}{!}{\rotatebox{-90}{\includegraphics{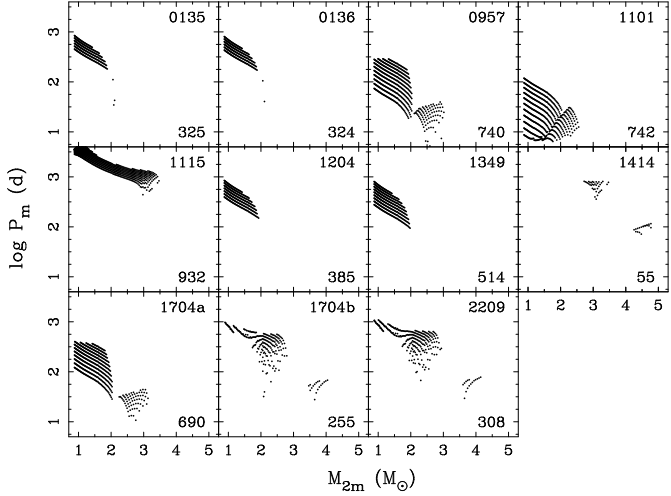}}} 
\caption{
 Results of the spiral-in calculations.  This figure is similar to Fig.\,\ref{fig:ce_allwds}
 and shows the orbital period of the pre-CE system as a function of the secondary mass.  The 
 solutions for each system are plotted in a separate panel, as labelled in the upper-right 
 corner. All solutions with acceptable $\alpha_\mathrm{ce}$ are plotted.  The number of 
 solutions for each system is shown in the lower-right corner. We assumed an
 uncertainty in $M_\mathrm{2f}$ of $0.05\,M_\odot$ and calculated $M_\mathrm{1f}$ using $q_2$.
 \label{fig:ce_allwds1}  }
\end{figure}

In Fig.\,\ref{fig:ce_ace}, we display the common-envelope parameter $\alpha_\mathrm{ce}$
for a selection of the solutions with $0.1\,\leq\,\alpha_\mathrm{ce}\,\leq\,10$ as a function 
of the unknown intermediate secondary mass $M_\mathrm{2m}$.  Each of the plot symbols has a
corresponding symbol in Fig.\,\ref{fig:ce_allwds}.  
To produce these two figures, we have so far implicitly
assumed that the masses of the two components are exact, so that there is at
most one acceptable solution for each progenitor mass.  This is of course unrealistic and it might
keep us from finding an acceptable solution.  At this stage we therefore introduce an uncertainty on the
values for $M_2$ in Table\,\ref{tab:obswds} and take $M_\mathrm{2f}$ = $M_2 - 0.05\,M_\odot, M_2 - 0.04\,M_\odot, 
\ldots, M_2 + 0.05\,M_\odot$.  Meanwhile we assume that the mass ratio and orbital period have negligible
observational error, because these errors are much smaller than those on the masses, and obtain 
the mass for the first white dwarf from $M_\mathrm{1f} = M_\mathrm{2f}/q_2$.
Thus we have 11 pairs of values for $M_\mathrm{1f}$ and $M_\mathrm{2f}$ for each observed system, which
we use as input for our spiral-in calculations.  The results are shown in Fig.\,\ref{fig:ce_allwds1}.

If we compare Fig.\,\ref{fig:ce_allwds} and Fig.\,\ref{fig:ce_allwds1}, we see that the wider range 
in input masses results in a wider range of solutions, similar to those
we found in Fig.\,\ref{fig:ce_allwds}, but extended in orbital period.  This can be understood 
qualitatively, since lowering the white-dwarf mass demands a lower helium-core mass in the progenitor
and thus a less evolved progenitor with a smaller radius at the onset of Roche-lobe overflow.  
Conversely, higher white-dwarf masses need more evolved progenitors that fill their Roche lobes at 
longer orbital periods.  
The introduction of this uncertainty clearly results in a larger and
more realistic set of solutions for the spiral-in calculations and therefore should be taken 
into account.

Each system in Fig.\,\ref{fig:ce_allwds1} is a possible progenitor of one of the ten observed double 
white dwarfs listed in Table\,\ref{tab:obswds}.  We now turn to the question whether and how these 
intermediate systems can be produced.


\section{First mass-transfer phase}
\label{sec:mtp1}

The solutions of the spiral-in calculations we found in the previous section are in our 
nomenclature {\it intermediate binaries}, that consist of one white dwarf and a non-degenerate
companion.  In this section we will look for an initial binary that consists of two 
zero-age main-sequence (ZAMS) stars of which the primary evolves, fills its Roche lobe, 
loses its hydrogen envelope, possibly transfers it to the secondary, so that one of the 
intermediate binaries of Fig.\,\ref{fig:ce_allwds1} is produced.  
The nature of this first mass transfer is a priori unknown.  In the following subsections
we will consider (1) stable and conservative mass transfer that will result in expansion of the
orbit in most cases, (2) a common envelope with spiral-in
based on energy balance (see Eq.\,\ref{eq:forwardce}) that usually gives rise to appreciable
orbital shrinkage and (3) envelope ejection due to dynamically unstable mass loss based on angular-momentum balance, as introduced 
by \citet{1967AcA....17....7P} and already used by \citet{2000A&A...360.1011N} for the same purpose, 
which can take place without much change in the orbital period.

\subsection{Conservative mass transfer}
\label{sec:cons_mt}

In this section we will find out which of the spiral-in solutions of Fig.\,\ref{fig:ce_allwds1}
may be produced by stable, conservative mass transfer.  We use the binary evolution code described in 
Sect.\,\ref{sec:code}.  For simplicity, we ignore stellar wind and the effect of stellar spin on the structure
of the star.  Because we assume conservative evolution, the total mass of the binary is constant, so that 
$M_\mathrm{1i} + M_\mathrm{2i} ~=~ M_\mathrm{1m} + M_\mathrm{2m}$, where the last two quantities 
are known.  Also, we ignore angular momentum exchange between spin and orbit by tidal forces, 
so that the orbital angular momentum is conserved.  This implies that
\begin{equation}
  \frac{P_\mathrm{m}}{P_\mathrm{i}} = \left( \frac{M_\mathrm{1i} \, M_\mathrm{2i}}{M_\mathrm{1m} \, M_\mathrm{2m}} \right)^3.
\label{eq:cons_mt}
\end{equation}

Because of the large number of possible intermediate systems we will first remove all such systems 
for which it can a priori be shown that they cannot be produced by conservative mass transfer.  
These systems have orbital periods that are either too short or too long to be formed this
way.  We can find a lower limit to the intermediate period as a function of secondary mass 
$M_\mathrm{2m}$ using the fact that the total mass of the initial system must be equal to the 
sum of the mass of the observed white dwarf $M_1$ and $M_\mathrm{2m}$.  We distributed this mass equally 
over two ZAMS stars and set the Roche-lobe radii equal to the two ZAMS radii.  By substituting the 
initial and desired masses in Eq.\,\ref{eq:cons_mt} we find a lower limit to the period of the 
intermediate binary, which we will call $P_\mathrm{min}$.

For each mass of the intermediate binary, an upper limit to the
intermediate period $P_\mathrm{m}$ can be estimated in two steps, as
follows.  First, we follow \citet{2000A&A...360.1011N} in noting that a
maximum period after mass transfer is reached for an initial mass
ratio $q_\mathrm{2i,opt}=0.66$.\footnote{ \citet{2000A&A...360.1011N} find
$q_\mathrm{2i,opt}=0.62$, because the equation they use for the Roche lobe is 
different from the Eq.\,\ref{eq:eggleton} that we use.  They also
use a different condition for the stability of mass transfer from an evolving star.
Our calculations show that if we would use Eq.\,\ref{eq:mce} instead, the upper limit
to $P_\mathrm{m}$ would drop, so that we can be confident that the limit we find is
indeed an upper limit.}  
To reach the highest value for $P_\mathrm{m}$ the initial mass ratio must be 
optimal as defined above, and in addition the initial period must be the
longest period $P_\mathrm{i,opt}$ for which the mass transfer is stable.
We use the conditions by \citet{2000MNRAS.315..543H} 
who define this point as the moment where the mass of the convective envelope $M_\mathrm{CE}$ exceeds a 
certain fraction of the total mass of the hydrogen envelope $M_\mathrm{E}$ for the first time:
\begin{equation}
\begin{array}{lll}
M_\mathrm{CE} &= {\textstyle \frac{2}{5}}\,M_\mathrm{E}, ~~~ & M_\mathrm{1i,opt} \leq 1.995\,M_\odot, \\
M_\mathrm{CE} &= {\textstyle \frac{1}{3}}\,M_\mathrm{E}, ~~~ & M_\mathrm{1i,opt} > 1.995\,M_\odot, 
\end{array}
\label{eq:mce}
\end{equation}
for $Z = 0.02$.  We then interpolate in our grid of stellar models of Sect.\,\ref{sec:gbmodels} 
to find the radius of a star with the desired mass at the base of the giant branch ($R_\mathrm{BGB}$).  
By assuming that this radius is equal to the Roche-lobe radius and using Eq.\,\ref{eq:eggleton}, we find
the optimum initial period $P_\mathrm{i,opt}$.  For a given binary mass, a unique initial binary system
is thus defined by $M_\mathrm{1i,opt}$, $M_\mathrm{2i,opt}$ and $P_\mathrm{i,opt}$.
We then use Eq.\,\ref{eq:cons_mt} to find for this initial system the maximum intermediate period, which 
we will call $P_\mathrm{max}$.  All intermediate systems, that result from our spiral-in calculations and 
have longer orbital periods than $P_\mathrm{max}$, can not be formed by conservative mass transfer.

\begin{figure}
\resizebox{\hsize}{!}{\rotatebox{-90}{\includegraphics{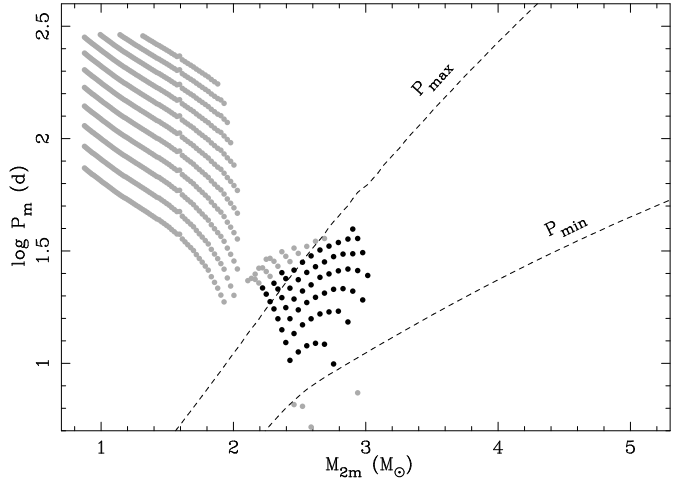}}} 
\caption{
 Results of the spiral-in calculations for WD\,0957--666 with period limits for a conservative first
 mass transfer.  This figure contains the same data as the third panel in Fig.\,\ref{fig:ce_allwds1} 
 (symbols) plus the period limits $P_\mathrm{min}$ and $P_\mathrm{max}$ (dashed lines).  The solutions 
 that lie between these limits are shown in black, the others in grey.  See the main text for details.
 \label{fig:ce_wd0957}  }
\end{figure}

The lower and upper limits for the orbital period between which a conservative solution must lie
for WD\,0957--666 are shown in Fig.\,\ref{fig:ce_wd0957} together with the intermediate systems found from 
the spiral-in calculations.  
Black dots represent solutions that 
lie between the limits and could match the outcome of a conservative model,
grey dots lie outside these limits and cannot be created by conservative mass transfer.  
There is a slight difference between the dashed lines and the division between filled and open
symbols in the Figure, because the spiral-in solutions are shown with the uncertainty in the
masses described in the previous section, whereas the period limits are only shown for the measured 
$M_2$ and $q_2$ (see Table\,\ref{tab:obswds}) for clarity.

\begin{figure}
\resizebox{\hsize}{!}{\rotatebox{-90}{\includegraphics{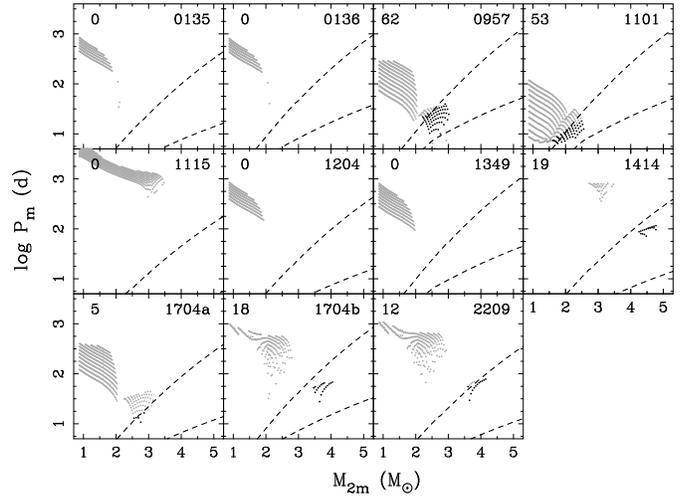}}} 
\caption{
 Results of the spiral-in calculations with period limits for conservative 
 mass transfer as in Fig.\,\ref{fig:ce_wd0957}, but for all systems.  The number 
 in the upper left corner of each panel is the number of systems that lie between 
 the period limits. 
 \label{fig:ce_allwds2}  }
\end{figure}

\begin{figure}
\resizebox{\hsize}{!}{\rotatebox{-90}{\includegraphics{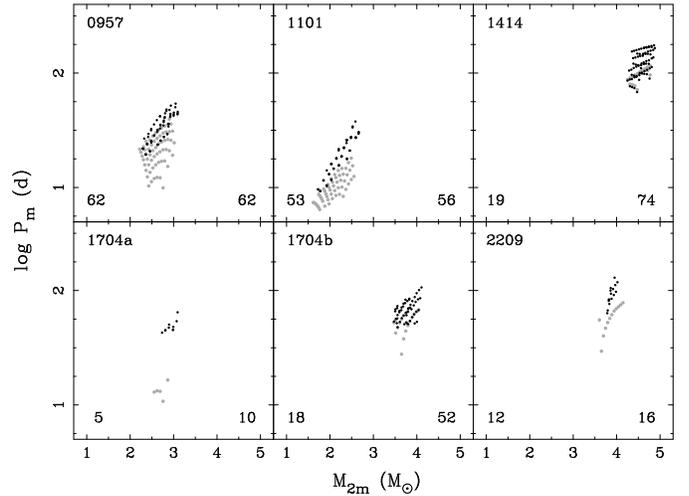}}} 
\caption{
 Results of the spiral-in calculations (grey symbols), obtained as in Fig.\,\ref{fig:ce_wd0957},
 and the solutions of calculations of conservative evolution (black symbols). Only the six systems shown have 
 spiral-in solutions within the period limits (see Fig.\,\ref{fig:ce_allwds2}). The numbers in the lower left and lower right
 corners are the numbers of plotted spiral-in solutions and conservative solutions respectively.
 \label{fig:ce_stable}  }
\end{figure}

After selecting the spiral-in solutions that lie between these period limits for all eleven 
systems, we find that such solutions exist for only six of the observed binaries, as shown in
Fig.\,\ref{fig:ce_allwds2}.
We tried to model these intermediate systems with the binary evolution code described in Sect.\,\ref{sec:code}.  
Because of the large number of allowed spiral-in solutions for WD\,0957--666 and WD\,1101+364, we decided to 
model about half of the solutions for these two systems and all of the solutions for the other four.
Because we assume that during this part of the evolution mass and orbital angular momentum are conserved, 
the only free parameter is the initial mass ratio $q_\mathrm{1i} \equiv M_\mathrm{1i}/M_\mathrm{2i}$.  
For each of the spiral-in solutions we selected, we chose
five different values for $q_\mathrm{1i}$, evenly spread in the logarithm: 1.1, 1.3, 1.7, 2.0 and 2.5.
The total number of conservative models that we calculated is 570, of which 270 resulted
in a double white dwarf. The majority of the rest either experienced dynamical mass transfer
or evolved into a contact system.  A few models were discarded because of numerical problems.
The results of the calculations for the conservative first mass transfer are compared to the 
solutions of the spiral-in calculations in Fig.\,\ref{fig:ce_stable}.

\begin{figure}
\resizebox{\hsize}{!}{\rotatebox{-90}{\includegraphics{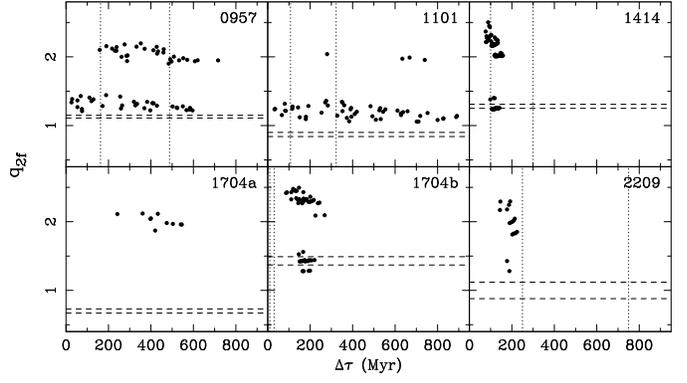}}} 
\caption{
 The mass ratio of model double white dwarfs formed by a conservative first mass 
 transfer and a common envelope with spiral-in, against the age difference between the two components.  
 The dashed horizontal lines show the observed range of possible mass ratios as shown in 
 Table\,\ref{tab:obswds}.  The dotted vertical lines are the estimated cooling-age differences
 $\pm 50\%$ (see Table\,\ref{tab:obswds}).
 \label{fig:ce_stableq2}  }
\end{figure}

The systems that result from our conservative models generally have longer orbital periods 
than the intermediate systems that we are looking for.  This means that stable mass transfer in the models 
continues beyond the point where the desired masses and orbital period are reached.  The result is that 
$M_\mathrm{1m}$ is too small and that $M_\mathrm{2m}$ and $P_\mathrm{m}$ are too large.  The reason that mass
transfer continues is that the donor star is not yet sufficiently evolved:  the helium core is 
still small and there is sufficient envelope mass to keep the Roche lobe filled. 
White dwarfs of higher mass would result from larger values of $q_\mathrm{1i}$.  This way, the initial
primary is more massive and the initial period is longer, so that the star fills its Roche lobe at a slightly 
later stage in evolution.  Both effects increase the mass of the resulting white dwarf.  
However, if one chooses the initial mass ratio too high, the system evolves into a contact binary or, for even higher $q_\mathrm{1i}$, mass 
transfer becomes dynamically unstable. In both cases the required intermediate system will not be 
produced.  These effects put an upper limit to the initial mass ratio for which mass transfer is still stable, 
and thus an upper limit to the white-dwarf mass that can be produced with stable mass transfer for a given 
system mass.
Our calculations show that conservative models with an initial mass ratio of 2.5 produce no double white
dwarfs. Apparently this value of $q_\mathrm{1i}$ is beyond the upper limit.
The solutions in Fig.\,\ref{fig:ce_stableq2} with a final mass ratio close to or in agreement
with the observations come predominantly from the models with initial mass ratios 
of 1.7 and 2.0.  

Because small deviations in the masses and orbital period of the intermediate systems can 
still lead to acceptable double white dwarfs, we monitor the evolution of these systems to the point
where the secondary fills its Roche lobe and determine the mass of the second white dwarf $M_\mathrm{2f}$
from the helium-core mass of the secondary at that point.  Because the secondary in the intermediate binary is slightly too massive 
in most cases, it is smaller at a given core mass (see Fig.\,\ref{fig:basic_mc-r}) so that the mass
of the second white dwarf becomes larger than desired.  Combined with an undermassive first white dwarf
this results in a too large mass ratio $q_\mathrm{2f}$.  This is shown in Fig.\,\ref{fig:ce_stableq2}, where 
the values for $q_\mathrm{2f}$ for our conservative models are compared to the observations.  
The Figure also shows the difference in age of the system between the moment where the second white dwarf
was formed and the moment when the first white dwarf was formed ($\Delta\tau$).  This difference should be
similar to the observed difference in cooling age between the two components of the binary
(see Table\,\ref{tab:obswds}).  The vertical dotted lines show this observed cooling-age difference with 
an uncertainty of 50\%.

Figure\,\ref{fig:ce_stableq2} shows that of the six systems presented, only two have a mass ratio within
the observed range, although values for the other systems may be close.
We see that the mass ratios of the solutions for most of the systems are divided in two groups and the difference in 
mass ratio can amount to a factor of 2 between them.  The division arises because in most models 
the common envelope is supposed to occur on the short giant branch of stars that are more massive than
$2\,M_\odot$.  If the secondary is slightly smaller and the orbital period slightly longer than it should
be, the star can ignite helium in its core and start shrinking before it has expanded sufficiently to fill its Roche lobe.  
When this star expands again after core helium exhaustion, it has a much more massive helium core and produces 
a much more massive white dwarf than desired (see Fig.\,\ref{fig:basic_mc-r}).  Thus, a small offset in 
the parameters of the model after the first mass-transfer phase can result in large differences after 
the spiral-in.  Of the 270 stable models shown in Fig.\,\ref{fig:ce_stableq2}, 126 (47\%) are 
in the group with lower mass ratios ($q_\mathrm{2f}\mathrel{\hbox{\rlap{\lower.55ex \hbox {$\sim$}} \kern-.3em \raise.4ex \hbox{$<$}}} 1.7$).

The modelled mass ratios for the systems WD\,0957--666 and WD\,1101+364 are close to the observed values, and we find that this 
is especially true for the models on the low-mass end of the range in observed white-dwarf masses we used.  
This can be understood, because the maximum mass of a white dwarf that can be created with conservative mass
transfer is set by the total mass in the system.  The system mass is determined by the spiral-in
calculations in Sect.\,\ref{sec:ce_results}, where we find that the total mass that is available to create these two 
systems lies between about $2$ and $3.5\,M_\odot$. This system mass is simply insufficient to create white dwarfs
with the observed masses.  If we would extend the uncertainty in the observed masses to allow lower 
white-dwarf masses, it seems likely that we could explain these two double white dwarfs with a conservative 
mass-transfer phase followed by a common envelope with spiral-in.  The same could possibly be achieved with stable, non-conservative 
mass transfer.  
If the mass were transferred to the accretor and subsequently partially lost from the system in an isotropic wind,
this would stabilise the mass transfer.  The mass transfer would then still be stable for slightly longer initial periods, 
so that higher initial primary masses are allowed.  Both effects result in higher white-dwarf masses.

All 126 stable solutions in the lower group of mass ratios ($q_\mathrm{2f}\mathrel{\hbox{\rlap{\lower.55ex \hbox {$\sim$}} \kern-.3em \raise.4ex \hbox{$<$}}} 1.7$) have 
$\alpha_\mathrm{ce}>1$ and 83 (66\%) have $\alpha_\mathrm{ce}<5$.  If we become more demanding and 
insist that $\alpha_\mathrm{ce}$ should be less than 2, we are left with 14 solutions, all for WD\,0957--666.
These solutions all have $\alpha_\mathrm{ce}>1.6$.  If we additionally demand that the age difference of
these models be less than 50\%\ from the observed cooling-age difference, only 6 solutions are left 
with age differences roughly between 190 and 410\,Myr, $\alpha_\mathrm{ce}>1.8$ and 
$1.32\leq q_\mathrm{2f}\leq1.44$.

We conclude that although the evolutionary channel of conservative mass transfer followed
by a spiral-in can explain some of the observed systems, evolution along this channel cannot
produce all observed double white dwarfs.   We must therefore reject this formation channel as the single 
mechanism to create the white-dwarf binaries.  The reason that it fails to explain some 
of the observed white dwarfs is that the observed masses for the first white dwarfs in these systems 
are too high to be explained by conservative mass transfer in a binary with the total mass that is set
by the spiral-in calculations.  Allowing for mass loss from the system during mass
transfer could result in better matches.  However it is clear from Fig.\,\ref{fig:ce_allwds2}
that this will certainly not work for at least 5 of the 10 observed systems because their orbital periods are too large.
We will need to consider other prescriptions in addition to stable mass transfer to produce the observed 
white-dwarf primaries for these systems.

\subsection{Unstable mass transfer}
\label{sec:unst_mt}

In this section we try to explain the formation of the first white dwarf in the intermediate systems
shown in Fig.\,\ref{fig:ce_allwds1} by unstable mass transfer.  Mass transfer occurs on the dynamical 
timescale if the donor is evolved and has a deep convective envelope.  There are two prescriptions that 
predict the change in orbital period in such an event.  The first is a classical common envelope with a spiral-in, based on 
energy conservation as we have used in Sect.\,\ref{sec:mtp2}.  The second prescription was introduced in this context by 
\citet{2000A&A...360.1011N} and further explored by \citet{2005MNRAS.356..753N}, and uses angular-momentum 
balance to calculate the change in orbital period.  Where the first prescription results in a strong 
orbital shrinkage (spiral-in) for all systems, in the second prescription this is not necessarily the case, 
so that the orbital period may or may not change drastically using the same efficiency parameter, 
while the envelope of the donor star is lost.

In both scenarios we are looking for an initial binary of which the components have masses $M_\mathrm{1i}$ 
and $M_\mathrm{2i}$.  The primary will evolve fastest, fill its Roche lobe and eject its envelope due to
dynamically unstable mass loss, so that its core becomes exposed and 
forms a white dwarf with mass $M_\mathrm{1m}$.  We assume that the mass of the
secondary star does not change during this process, so that $M_\mathrm{2i} = M_\mathrm{2m}$.  We use the 
model stars from Sect.\,\ref{sec:basic_models} as the possible progenitors for the first white dwarf. The 
orbital period before the envelope ejection is again determined by setting the radius of the model star equal 
to the Roche-lobe radius and applying Eq.\,\ref{eq:eggleton}, where the subscripts `m' must be replaced by `i'.

Because we demand that $M_\mathrm{1i} > M_\mathrm{2i}$, the original secondary can be any but the most massive
star from our grid and the total number of possible binaries in our grid is $\sum_{n=1}^{198} n = 19701$
for each system we want to model.  The total number of systems that we try to model is 121: the 11 
observed systems (the 10 from Table\,\ref{tab:obswds} plus the system WD\,1704+481b) times 11 different 
assumptions for the masses of the observed stars (between $\pm 0.05\,M_\odot$
from the observed value).  We have thus tried slightly less than 2.4 million initial binaries to find 
acceptable progenitors to these systems.  All these possible progenitor systems have been filtered by the 
following criteria, in addition to the ones already mentioned in Sect.\,\ref{sec:mtp2}:
\begin{enumerate}
  \item the radius of the star is larger than the radius at the base of the giant branch $R>R_\mathrm{BGB}$,
        which point is defined by Eq.\,\ref{eq:mce}, 
  \item the mass ratio is larger than the critical mass ratio for dynamical mass transfer $q > q_\mathrm{crit}$ 
        as defined by Eq.\,57 of \citet{2002MNRAS.329..897H}.  Together with the previous criterion, this 
	ensures that the mass transfer can be considered to proceed on the dynamical timescale,
  \item the time since the ZAMS after which the first white dwarf is created $\tau_1$ is less than the same
        for the second white dwarf ($\tau_2$) and, additionally, $\tau_2 <$ 13\,Gyr.
\end{enumerate}

After we filter the approximately 2.4\,million possible progenitor systems with the criteria above, about 204,000
systems are left in the sample (8.5\%) for which two subsequent envelope-ejection scenarios could result in the desired 
masses, provided that we can somehow explain the change in orbital period that is needed to obtain 
the observed periods.  For each of the two prescriptions for dynamical mass loss we will see whether this sample
contains physically acceptable solutions in the sections that follow.

\subsubsection{Classical common envelope with spiral-in}
\label{sec:ace1}

The treatment of a classical common envelope with spiral-in based on energy conservation has been described in detail in 
Sect.\,\ref{sec:mtp2} and therefore need not be reiterated here.  In the calculations described above,
Eq.\,\ref{eq:forwardce} provides us with the parameter $\alpha_\mathrm{ce1}$ for the first spiral-in.  
In order to use Eq.\,\ref{eq:forwardce} the subscripts `m' must be replaced by `i' and the subscripts `f' by `m'.  
The values of the common-envelope parameter for the first spiral-in must be physically acceptable and we
demand that $0.1\!\leq\!\alpha_\mathrm{ce1}\!\leq\!10$.
When we apply this criterion to the results of our calculations, only 25 possible progenitors out of the 
204,000 binaries in our sample survive. All 25 survivors are solutions for WD\,0135--052 and have 
$\alpha_\mathrm{ce1} \mathrel{\hbox{\rlap{\lower.55ex \hbox {$\sim$}} \kern-.3em \raise.4ex \hbox{$>$}}} 2.5$.  

We find that of the systems that pass the criterion in the second spiral-in and have
$0.1\!\leq\!\alpha_\mathrm{ce2}\!\leq\!10$, most (99\%) need a negative $\alpha_\mathrm{ce1}$ in order to satisfy 
Eq.\,\ref{eq:forwardce}, so that we reject them.  We can clearly conclude 
that the scenario of two subsequent classical common envelopes with spiral-in can be rejected as the formation 
scenario for any of the observed double white dwarfs.  This confirms the conclusions of \citet{2000A&A...360.1011N} 
and \citet{2005MNRAS.356..753N}, based on the value of the product $\alpha_\mathrm{ce}\,\lambda_\mathrm{env}$,
where $\lambda_\mathrm{env}$ is the envelope-structure parameter defined in Eq.\,\ref{eq:lambda}.

\subsubsection{Envelope ejection with angular-momentum balance}
\label{sec:gce1}

The idea to determine the change in orbital period in a common envelope from balance of angular momentum
originates from \citet{1967AcA....17....7P}.  In \citet{2000A&A...360.1011N} and \citet{2005MNRAS.356..753N}
the prescription was used to model observed double white dwarfs.  The principle is similar to that of 
a classical common envelope, here with an efficiency parameter that we will call $\gamma$ in the 
general case.  In this section we will use three slightly different prescriptions for mass loss with 
angular-momentum balance requiring three different definitions of $\gamma$.  For all three prescriptions 
the mass loss of the donor is dynamically unstable and its envelope is ejected from the system.  Because not 
all of these prescriptions necessarily involve an envelope that engulfs both stars, we shall refer to them 
as envelope ejection or dynamical mass loss rather than common-envelope evolution.  The first prescription is that 
defined by \citet{2000A&A...360.1011N}, where a common envelope is established first, after which the mass is
lost from its surface.  The mass thus carries the average angular momentum of the system and we will call the
parameter for this prescription $\gamma_\mathrm{s}$.  In the second prescription the mass is first transferred and 
then re-emitted with the specific orbital angular momentum of the accretor.  We will designate $\gamma_\mathrm{a}$ 
for this prescription.  In the third prescription the mass is lost directly from the donor in an isotropic wind and 
the corresponding parameter is $\gamma_\mathrm{d}$.  We will call the companion to the donor star `accretor', 
even if no matter is actually accreted.

The prescription of dynamical mass loss with the average specific angular momentum of the initial system
for the first mass-transfer phase, using this and earlier subscript conventions, is:
\begin{equation}
\frac{J_\mathrm{i} - J_\mathrm{m}}{J_\mathrm{i}} ~=~ \gamma_\mathrm{s1}\, \frac{M_\mathrm{1i} - M_\mathrm{1m}}{M_\mathrm{1i} + M_\mathrm{2i}},
\label{eq:gce}
\end{equation}
where $J$ is the total orbital angular momentum \citep{2000A&A...360.1011N}. Our demands for a physically acceptable
solution to explain the observed binaries is now $0.1\!\leq\!\gamma_\mathrm{s1}\!\leq\!10$ for the first envelope
ejection and $0.1\!\leq\!\alpha_\mathrm{ce2}\!\leq\!10$ for the second.  From the set of about 204,000 solutions
we found above, almost 150,000 (72\%) meet these demands and nearly 134,000 solutions (66\%) have values for
$\gamma_\mathrm{s1}$ between 0.5 and 2, in which all observed systems are represented.

\begin{figure*}
\resizebox{0.5\hsize}{!}{\rotatebox{-90}{\includegraphics{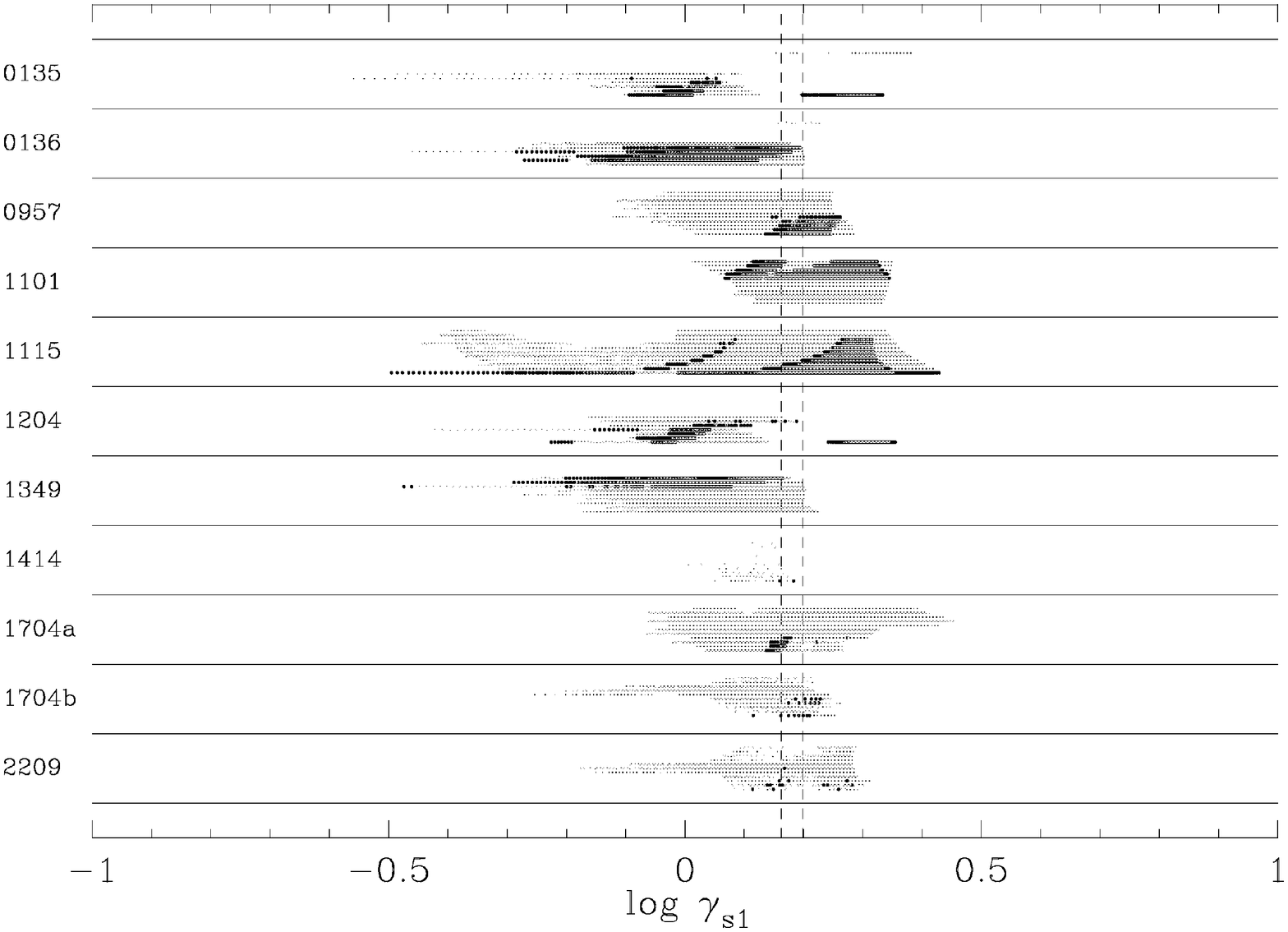}}} 
\resizebox{0.5\hsize}{!}{\rotatebox{-90}{\includegraphics{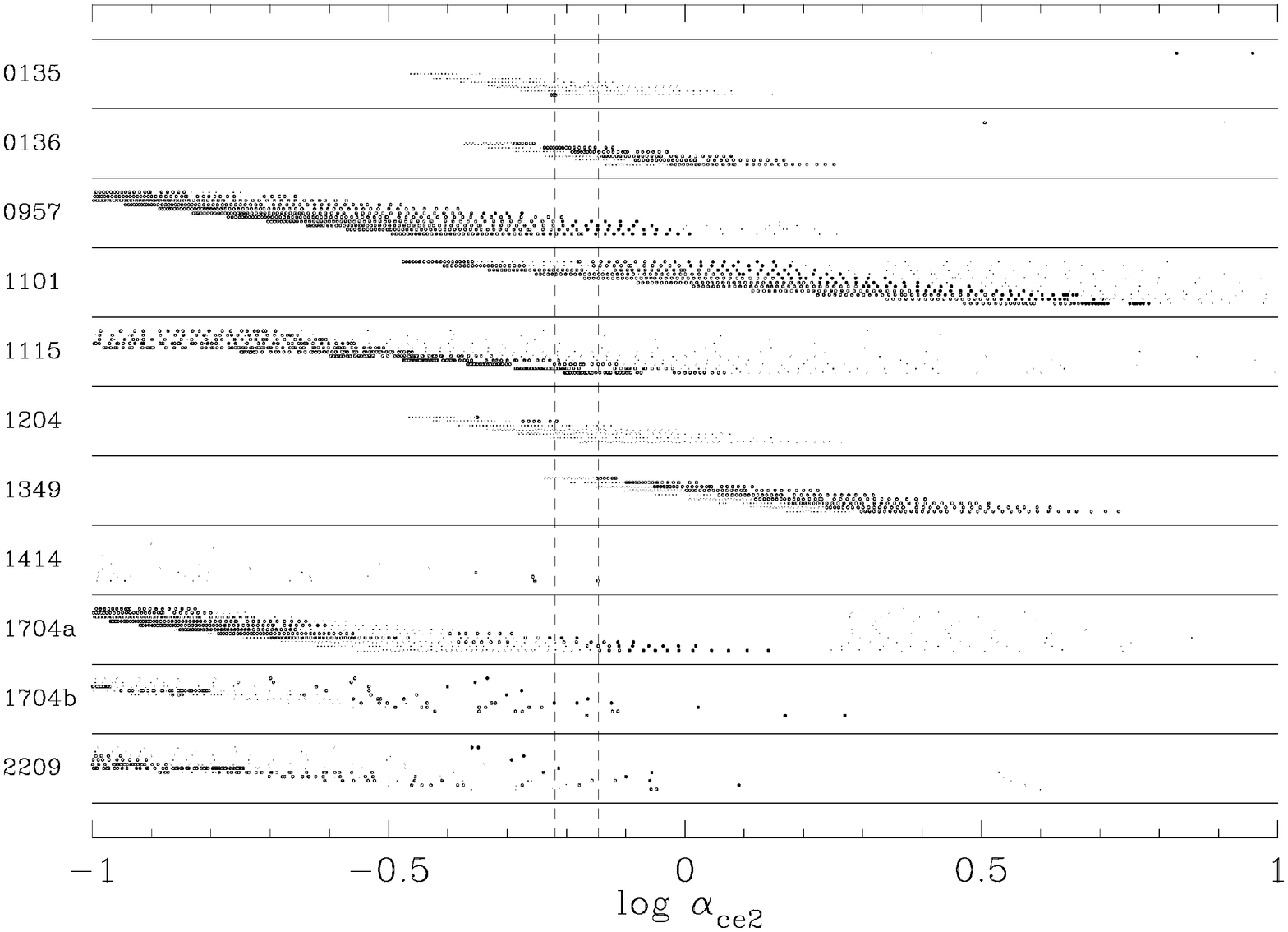}}} 
\caption{
 The distribution of the logarithm of the envelope-ejection parameters for solutions of the double dynamical mass-loss 
 scenario. Each dot represents a system that evolves through an episode of dynamical mass loss with 
 $\gamma_\mathrm{s1}$ and then a common envelope with spiral-in with $\alpha_\mathrm{ce2}$ to form 
 one of the observed white dwarfs. Only solutions for which $0.1\!\leq\!\gamma_\mathrm{s1}\!\leq\!10$ 
 {\it and} $0.1\!\leq\!\alpha_\mathrm{ce2}\!\leq\!10$ are shown.  Hence, each dot in one panel has a 
 counterpart in the other panel.  The dashed vertical lines show $\gamma_\mathrm{s1}\!=\!1.45, 1.58$ 
 in the left panel and $\alpha_\mathrm{ce2}\!=\!0.61, 0.72$ in the right panel.
 Black dots in one panel indicate solutions that lie between the dashed lines in the other panel;
 black dots between black lines therefore lie in the preferred range in both panels.
 The vertical position of each dot within the line for its system shows the deviation from the observed 
 secondary mass $M_2$. From the lower to the upper dots in a line these masses are: 
 $M_\mathrm{2f} = M_2-0.05, M_2-0.04, \ldots, M_2+0.05\,M_\odot$.  The total number of solutions in 
 the Figure is nearly 150,000. The Figure is made after \citet{2005MNRAS.356..753N}.
 \label{fig:cece_ga}  }
\end{figure*}

We tried to constrain the ranges for $\gamma_\mathrm{s1}$ and $\alpha_\mathrm{ce2}$ as much as possible, 
thereby keeping at least one solution for each observed system.  We can write these ranges as 
$(\gamma_0-\frac{\Delta\gamma}{2} , \gamma_0+\frac{\Delta\gamma}{2})$ and $(\alpha_0-\frac{\Delta\alpha}{2} , \alpha_0+\frac{\Delta\alpha}{2})$,
where $\gamma_0$ and $\alpha_0$ are the central values and $\Delta\gamma$ and $\Delta\alpha$ 
are the widths of each range.  We independently varied $\gamma_0$ and $\alpha_0$
and for each pair we took the smallest values of $\Delta\gamma$ and $\Delta\alpha$ for which
there is at least one solution for each observed system that lies within both ranges.
The set of smallest ranges thus obtained is considered to be the best 
range for $\gamma_\mathrm{s1}$ and $\alpha_\mathrm{ce2}$ that can explain all systems.
Because it is harder to trifle with the angular-momentum budget than with that of energy, we kept the relative 
width of the range for $\gamma_\mathrm{s1}$ twice as small as that for $\alpha_\mathrm{ce2}$ 
($2\frac{\Delta\gamma}{\gamma_0} = \frac{\Delta\alpha}{\alpha_0}$).  Our 
calculations show that changing this factor merely redistributes the widths over the two ranges without
affecting the central values much and thus precisely which factor we use seems to be unimportant 
for the result.  We find that the set of narrowest
ranges that contain a solution for each system is $1.45\!\leq\!\gamma_\mathrm{s1}\!\leq\!1.58$ and 
$0.61\!\leq\!\alpha_\mathrm{ce2}\!\leq\!0.72$.  These results are plotted in Fig.\,\ref{fig:cece_ga}.

We can alternatively treat the second envelope ejection with the angular-momentum prescription as well, where 
we need to introduce a factor $\gamma_\mathrm{s2}$ by replacing all subscripts `m' by `f' and all subscripts 
`i' by `m' in Eq.\,\ref{eq:gce}.  Again we search for the narrowest ranges of $\gamma_\mathrm{s1}$ and 
$\gamma_\mathrm{s2}$ that contain at least one solution per observed system.  We now force the relative widths 
of the two ranges to be equal.  The best solution is then
$1.16\!\leq\!\gamma_\mathrm{s1}\!\leq\!1.22$ and $1.62\!\leq\!\gamma_\mathrm{s2}\!\leq\!1.69$.

In both prescriptions above ($\gamma_\mathrm{s1}\alpha_\mathrm{ce2}$ and $\gamma_\mathrm{s1}\gamma_\mathrm{s2}$) we 
find that the values for $\gamma$ lie significantly above unity.  This is in accordance with the findings 
of \citet{2000A&A...360.1011N} and \citet{2005MNRAS.356..753N}, 
but slightly discomforting because there is no 
physical reason as to why $\gamma$ should have this, or indeed any other, particular value.
We will therefore rewrite Eq.\,\ref{eq:gce} for the case where the mass is lost with the specific orbital angular momentum of 
one of the stars in the binary, so that we can expect that $\gamma \approx 1$.
In order to do this we use the equations derived by \citet{1997A&A...327..620S} in their Section 2.1.  They define 
the fractions of mass lost by the donor: $\alpha_\mathrm{w}$ the fraction that is lost in an isotropic wind from the
donor, $\beta_\mathrm{w}$ the fraction that is transferred and then lost in a wind by the companion and $\epsilon_\mathrm{w}$ the fraction
that is accreted by the companion, where we introduced the subscript `w' to avoid confusion with $\alpha_\mathrm{ce}$.  
We assume that no matter is accreted, so that $\alpha_\mathrm{w}+\beta_\mathrm{w}=1$ and $\epsilon_\mathrm{w}=0$ and 
ignore the finite size of the star by putting $A_\mathrm{w}=1$.  Their Eq.\,24 then gives (replacing their notation by ours):
\begin{equation}
\frac{J_\mathrm{m}}{J_\mathrm{i}} = \left(\frac{q_\mathrm{m}}{q_\mathrm{i}}\right)^{\alpha_\mathrm{w}} \frac{1+q_\mathrm{i}}{1+q_\mathrm{m}} 
 \, \exp\left[ \beta_\mathrm{w} \left( q_\mathrm{m} - q_\mathrm{i} \right) \right],
\label{eq:soberman1}
\end{equation}
where we will consider the cases where $\alpha_\mathrm{w}\!=\!0$; $\beta_\mathrm{w}\!=\!1$, describing isotropic 
re-emission by the accretor, and $\alpha_\mathrm{w}\!=\!1$; $\beta_\mathrm{w}\!=\!0$ for an isotropic wind from 
the donor.  Their $q$ is defined as $m_\mathrm{donor}/m_\mathrm{accretor}$.  We can now replace Eq.\,\ref{eq:gce} 
with similar expressions for these two cases:
\begin{equation}
\frac{J_\mathrm{i} - J_\mathrm{m}}{J_\mathrm{i}} \,=\, \gamma_\mathrm{a1} \left( 1 - \frac{M_\mathrm{tot,i}}{M_\mathrm{tot,m}} \, 
  \exp\left[ \frac{M_\mathrm{1m} - M_\mathrm{1i}}{M_\mathrm{2}} \right] \right)   \\ ~~ (\alpha_\mathrm{w}\!=0),
\label{eq:gcea}
\end{equation}
\begin{equation}
\frac{J_\mathrm{i} - J_\mathrm{m}}{J_\mathrm{i}} \,=\, \gamma_\mathrm{d1}\, \frac{M_\mathrm{1i} - M_\mathrm{1m}}{M_\mathrm{1m} + M_\mathrm{2m}}\, \frac{M_\mathrm{2i}}{M_\mathrm{1i}} ~~~ (\alpha_\mathrm{w}=1),
\label{eq:gceb}
\end{equation}
where we introduced the parameters $\gamma_\mathrm{a1}$ and $\gamma_\mathrm{d1}$ to describe deviations from
the idealised case of purely isotropic mass loss.  We treat $\gamma_\mathrm{a1}$ and $\gamma_\mathrm{d1}$ as free
parameters, but expect that $\gamma \approx 1$ in the picture that mass is preferentially lost from either donor
or accretor.
We used $M_\mathrm{tot} = M_1 + M_2$ to denote the total mass of the system.
The results of the analysis described above, but now for the modified definitions of $\gamma$, for the $\gamma\alpha$ 
and $\gamma\gamma$ scenarios, each with $\alpha_\mathrm{w}=0$, $\beta_\mathrm{w}=1$ (re-emission from the accretor, 
$\gamma_\mathrm{a}\!\approx\!1$) and $\alpha_\mathrm{w}=1$, $\beta_\mathrm{w}=0$ (wind from the donor, 
$\gamma_\mathrm{d}\!\approx\!1$) are shown in Table\,\ref{tab:cece_results} and compared to the previous results.

\begin{table}
\centering
\begin{tabular}{lllllll}
\hline \hline                                                                                                    
Prescription					& $\gamma_1$  		 & $\gamma_{0,1}$   &$\gamma_2/\alpha_\mathrm{ce2}$ 	  & $\gamma_{0,2}$/$\alpha_{0,2}$ \\
\hline
$\gamma_\mathrm{s1}\alpha_\mathrm{ce2}$		& 1.45--1.58             & 1.52             & $\alpha\!:$ 0.61--0.72              & $\alpha\!:$ 0.66 \\
$\gamma_\mathrm{s1}\gamma_\mathrm{s2}$		& 1.16--1.22             & 1.19             & $\gamma\!:$ 1.62--1.69              & $\gamma\!:$ 1.65 \\
$\gamma_\mathrm{a1}\alpha_\mathrm{ce2}$		& 2.04--2.26             & 2.15             & $\alpha\!:$ 0.54--0.67              & $\alpha\!:$ 0.59 \\
$\gamma_\mathrm{a1}\gamma_\mathrm{a2}$  	& 1.34--1.36             & 1.35             & $\gamma\!:$ 1.39--1.42              & $\gamma\!:$ 1.40 \\
$\gamma_\mathrm{d1}\alpha_\mathrm{ce2}$  	& 0.92--1.08             & 1.00             & $\alpha\!:$ 0.47--0.64              & $\alpha\!:$ 0.56 \\
$\gamma_\mathrm{d1}\gamma_\mathrm{d2}$  	& 0.91--1.07             & 0.99             & $\gamma\!:$ 2.55--3.02              & $\gamma\!:$ 2.78 \\
\hline
\end{tabular}
\caption{Narrowest ranges for $\gamma$ and $\alpha_\mathrm{ce}$ that 
contain at least one solution to the envelope-ejection scenario per observed system and their central values. 
The six different prescriptions are explained in the main text.  
\label{tab:cece_results}  }
\end{table}

We see that the values for $\gamma$ change drastically, as may be expected.  The fact that
the values for $\alpha_\mathrm{ce}$ change slightly has to do with the fact that we now select different solutions 
to the calculations than before.  The prescription $\gamma_\mathrm{a1}\gamma_\mathrm{a2}$ has very narrow ranges.
Numerically, the prescription $\gamma_\mathrm{d1}\alpha_\mathrm{ce2}$ seems very attractive:
$\gamma_\mathrm{d1} \approx 1.0$ and $\alpha_\mathrm{ce2} \approx 0.6$.  Although the value for 
$\alpha_\mathrm{ce2}$ is lower than unity, it may not be unrealistic that 40\,\%\ of the freed orbital 
energy is emitted by radiation.  This is the scenario where the mass is lost in an isotropic wind by the 
donor in the first dynamical mass-loss episode and the second mass loss is a canonical common envelope with spiral-in.
We also notice that in order to get a sufficiently strong spiral-in from an envelope ejection by a strong 
donor wind (the $\gamma_\mathrm{d1}\gamma_\mathrm{d2}$ prescription), the mass has to be lost with almost 
three times the specific angular momentum of the donor.

\subsection{Formation by multiple prescriptions}
\label{sec:multiple_prescriptions}

So far, we assumed that all ten observed double white dwarfs were formed by one and the same prescription.  
Although some prescriptions are clearly better in explaining the formation of all the observed systems than others, 
none of them is completely satisfactory, mainly because the parameters $\gamma$ or $\alpha_\mathrm{ce}$   
are far from the desired values.  Furthermore, there is no reason why the ten systems should all 
have evolved according to the same prescription in nature if there are several options available.  We therefore
slightly change our strategy here by assuming that different envelope-ejection prescriptions, described in 
Sect.\,\ref{sec:gce1}, can play a role in the formation of each of the observed systems.

For the dynamical mass loss, we now demand that $\gamma$ and $\alpha_\mathrm{ce}$ are close to unity.
Because angular momentum should be better conserved than energy, we accept solutions with 
$0.90\!<\!\gamma_\mathrm{a,d}\!<\!1.10$ and $0.80\!<\!\alpha_\mathrm{ce}\!<\!1.20$.  For the prescription 
by Eq.\,\ref{eq:gce}, \citet{2005MNRAS.356..753N} show that all systems can be explained with 
$1.50\!<\!\gamma_\mathrm{s}\!<\!1.75$, which we adapt to $1.46\!<\!\gamma_\mathrm{s}\!<\!1.79$ to give it the same 
relative width.  The total number of solutions between these ranges is about 42500, divided over 6 prescriptions 
and 11 systems (on average 3.5\%\ of the solutions selected in Sect.\,\ref{sec:unst_mt}).  For each observed system 
and each prescription, we look whether there is at least one solution with a envelope-ejection parameter within 
these ranges.  The results are shown as the first symbol in each entry of Table\,\ref{tab:comp_prescriptions}.
\begin{table*}
\centering
\begin{tabular}{lllllllll}
\hline \hline                                                                                                    
System	& 1: $\gamma_\mathrm{s}\alpha_\mathrm{ce}$ 	& 2: $\gamma_\mathrm{s}\gamma_\mathrm{s}$	& 3: $\gamma_\mathrm{a}\alpha_\mathrm{ce}$	& 4: $\gamma_\mathrm{a}\gamma_\mathrm{a}$  & 5: $\gamma_\mathrm{d}\alpha_\mathrm{ce}$ 		& 6: $\gamma_\mathrm{d}\gamma_\mathrm{a}$		& Opt. res.		& Best prescr.  	\\	
\hline
 0135     & $-/-$          & $+/$$\sim$     & $+/$$\sim$     & $-/-$          & $+/$$\sim$     & $+/$$\sim$     & $+/$$\sim$	  & 2,3,5,6 \\
 0136     & $+/+$          & $+/+$          & $+/$$\sim$     & $+/$$\sim$     & $+/+$          & $+/+$          & $+/+$      	  & 1,2,5,6 \\
 0957     & $+/+$          & $+/+$          & $-/-$          & $+/-$          & $+/+$          & $+/+$          & $+/+$      	  & 1,2,5,6 \\
 1101     & $+/$$\sim$     & $+/-$          & $+/-$          & $-/-$          & $+/$$\sim$     & $+/$$\sim$     & $+/$$\sim$	  & 1,5,6 \\
 1115     & $+/$$\sim$     & $+/+$          & $+/$$\sim$     & $+/$$\sim$     & $+/+$          & $+/+$          & $+/+$      	  & 2,5,6 \\
\multicolumn{9}{c}{}\\						        	      
 1204     & $-/-$          & $+/-$          & $+/-$          & $+/-$          & $+/-$          & $+/+$          & $+/+$      	  & 6 \\
 1349     & $+/+$          & $+/+$          & $+/+$          & $+/+$          & $+/+$          & $+/+$          & $+/+$      	  & 1,2,3,4,5,6 \\
 1414     & $-/-$          & $+/+$          & $-/-$          & $+/+$          & $-/-$          & $+/+$          & $+/+$      	  & 2,4,6 \\
 1704a    & $+/-$          & $+/-$          & $-/-$          & $-/-$          & $-/-$          & $-/-$          & $+/-$      	  & 1,2 \\
 1704b    & $+/-$          & $+/-$          & $-/-$          & $+/-$          & $+/-$          & $+/-$          & $+/-$      	  & 1,2,4,5,6 \\
 2209     & $+/+$          & $+/+$          & $-/-$          & $-/-$          & $+/$$\sim$     & $+/+$          & $+/+$      	  & 1,2,6 \\
\hline
\end{tabular}
\caption{Comparison of the different prescriptions used to reconstruct the observed double white dwarfs. 
The symbols $+$, $\sim$ and $-$ mean that the model solutions are in good, moderate or bad agreement with the observations.
The first of the two symbols in each column is based on the mass ratio only and the second includes the age difference.
The method for obtaining the first symbol in each entry is described in Sect.\,\ref{sec:multiple_prescriptions}, that for the 
second symbol in Sect.\,\ref{sec:age_difference}.
The symbols in the headers of the columns labelled 1--6 are explained in the main text.
The columns for $\gamma_\mathrm{a}\gamma_\mathrm{d}$ and $\gamma_\mathrm{d}\gamma_\mathrm{d}$ were left out because they do 
not contain any solutions.  The last two columns show the optimum result and the prescriptions that give this result (1--6).
For the different prescriptions we demanded that $1.46\!<\!\gamma_\mathrm{s}\!<\!1.79$, $0.90\!<\!\gamma_\mathrm{a}\!<\!1.10$, 
$0.90\!<\!\gamma_\mathrm{d}\!<\!1.10$ and $0.80\!<\!\alpha_\mathrm{ce}\!<\!1.20$.
\label{tab:comp_prescriptions}  }
\end{table*}
The plus signs show which prescription can explain the mass ratio of an observed double white dwarf with envelope ejection 
parameters within the chosen ranges of $\gamma$ and $\alpha_\mathrm{ce}$.  The table confirms that the prescription 
$\gamma_\mathrm{s}\gamma_\mathrm{s}$ can explain all observed systems in this sense.  The prescription 
$\gamma_\mathrm{d}\gamma_\mathrm{a}$ can do the same if we are satisfied with a solution for WD\,1704+481 
a {\it or} b.  We did not include the columns for 
$\gamma_\mathrm{a}\gamma_\mathrm{d}$ and $\gamma_\mathrm{d}\gamma_\mathrm{d}$, because they cannot explain any of the
observed systems.  This can readily be understood, since in these two scenarios the last envelope is lost by a wind
from the donor. This process will usually widen the orbit, rather than cause the spiral-in
that is needed to explain the observed binaries.  We have to expand the $\gamma$-ranges to 0.25--1.75 in order to get the 
first solution for just a single system with one of these two prescriptions and only by allowing values for $\gamma$
up to 2.8, we can find a solution for each binary.

\subsection{Constraining the age difference}
\label{sec:age_difference}

The large number of solutions found in the previous section allows us to increase the number of 
selection criteria that we use to qualify a solution as physically acceptable.  We now include the
age difference of the components in our model systems and demand that it is comparable to the observed cooling-age 
difference for that system.  The age difference in our models is the difference in age at which each of the
components fills its Roche lobe and causes dynamical mass loss.  In addition, we reject the solutions for 
which the initial mass ratio $q_\mathrm{1i} < 1.03$.  In these binaries, both components have probably evolved
beyond the main sequence and it is uncertain what would happen to the secondary at the envelope ejection of 
the primary.

Table\,\ref{tab:cece_results2} lists the number of model solutions for each prescription and each system.
\begin{table*}
\centering
\begin{tabular}{llllllll}
\hline \hline                                                                                                    
System	& Obs. $\Delta\tau$ &	\multicolumn{6}{l}{Number of solutions and model age differences (Myr)} \\
	& (Myr) & 1: $\gamma_\mathrm{s}\alpha_\mathrm{ce}$ 	& 2: $\gamma_\mathrm{s}\gamma_\mathrm{s}$	& 3: $\gamma_\mathrm{a}\alpha_\mathrm{ce}$	& 4: $\gamma_\mathrm{a}\gamma_\mathrm{a}$  & 5: $\gamma_\mathrm{d}\alpha_\mathrm{ce}$ 		& 6: $\gamma_\mathrm{d}\gamma_\mathrm{a}$		\\	
\hline
 0135  &   175--525 &  {\it      0}         $\!\!$ &  {\it      5},  636--1138$\!\!$ &  {\it     44}, 1454--3598$\!\!$ &  {\it      0}         $\!\!$ &  {\it    113}, 1198--3608$\!\!$ &  {\it     31},   53--1495$\!\!$  \\
 0136  &   225--675 &  {\it    280},  170--775$\!\!$ &  {\it    497},  170--971$\!\!$ &  {\it    201}, 1356--7398$\!\!$ &  {\it    116}, 1356--4069$\!\!$ &  {\it   1373},  170--5403$\!\!$ &  {\it   1903},  161--3838$\!\!$  \\
 0957  &   163--488 &  {\it    184},  264--1079$\!\!$ &  {\it   5413},  139--5971$\!\!$ &  {\it      0}         $\!\!$ &  {\it      5}, 2946--3659$\!\!$ &  {\it    274},  126--1079$\!\!$ &  {\it   7552},   90--6226$\!\!$  \\
 1101  &   108--323 &  {\it   1412},  868--11331$\!\!$ &  {\it   4919}, 1253--11614$\!\!$ &  {\it      9}, 9999--12255$\!\!$ &  {\it      0}         $\!\!$ &  {\it     13},  990--1987$\!\!$ &  {\it     74},  363--2636$\!\!$  \\
 1115  &    80--240 &  {\it    635},  244--1141$\!\!$ &  {\it   3073},  205--1075$\!\!$ &  {\it     29},  648--1129$\!\!$ &  {\it      4},  523--596$\!\!$ &  {\it     83},  240--1344$\!\!$ &  {\it    180},  229--592$\!\!$  \\
\multicolumn{8}{c}{}\\		
 1204  &    40--120 &  {\it      0}         $\!\!$ &  {\it      5},  808--1082$\!\!$ &  {\it     76}, 1541--5364$\!\!$ &  {\it      4}, 1302--1505$\!\!$ &  {\it    152}, 1216--5629$\!\!$ &  {\it     42},   74--1495$\!\!$  \\
 1349  &     ? &    {\it    161},  215--760$\!\!$ &  {\it    653},  170--840$\!\!$ &  {\it    851}, 1690--11521$\!\!$ &  {\it    274}, 1356--6374$\!\!$ &  {\it   1892},  215--6505$\!\!$ &  {\it   4283},  157--6168$\!\!$  \\
 1414  &   100--300 &  {\it      0}         $\!\!$ &  {\it     14},   83--146$\!\!$ &  {\it      0}         $\!\!$ &  {\it     49},  290--920$\!\!$ &  {\it      0}         $\!\!$ &  {\it      6},  167--207$\!\!$  \\
 1704a &   -30-- -10 &  {\it     27},  858--1294$\!\!$ &  {\it    770},   52--2046$\!\!$ &  {\it      0}         $\!\!$ &  {\it      0}         $\!\!$ &  {\it      0}         $\!\!$ &  {\it      0}         $\!\!$  \\
 1704b &    10--30 &   {\it      6},  465--553$\!\!$ &  {\it    163},  307--933$\!\!$ &  {\it      0}         $\!\!$ &  {\it      9}, 1060--1313$\!\!$ &  {\it      6},  465--553$\!\!$ &  {\it    311},  205--1498$\!\!$  \\
 2209  &   250--750 &  {\it      6},  552--967$\!\!$ &  {\it   1049},   61--967$\!\!$ &  {\it      0}         $\!\!$ &  {\it      0}         $\!\!$ &  {\it      2},  933--975$\!\!$ &  {\it    161},   61--1060$\!\!$  \\
\hline
\end{tabular}
\caption{Results for the various evolution scenarios for double white dwarfs with 
two unstable mass-transfer episodes.  The range of observed $\Delta\tau$ is the observed cooling-age difference 
$\pm\,50\%$.  Columns labelled 1 through 6 give the number of model solutions for each scenario followed by the range in age 
difference of these solutions in Megayears.  The columns with $\gamma_\mathrm{a}\gamma_\mathrm{d}$ and 
$\gamma_\mathrm{d}\gamma_\mathrm{d}$ were left out, because they do not contain any solutions.  In addition to
a value for $\gamma$ or $\alpha$ in the desired ranges (see Table\,\ref{tab:comp_prescriptions}), we demand 
that $q_\mathrm{1i} > 1.03$.
\label{tab:cece_results2}  }
\end{table*} 
The columns labelled 1--6 represent the same prescriptions as those columns in Table\,\ref{tab:comp_prescriptions}.  The first number in each 
of these columns is the number of solutions that is found within the same ranges for $\gamma$ and 
$\alpha_\mathrm{ce}$ as we used in Table\,\ref{tab:comp_prescriptions}.  This means that a minus sign in that table corresponds
to a zero in Table\,\ref{tab:cece_results2}.  Behind the entries with a positive number of solutions
the range of age difference that these solutions span is shown.  Again, the columns for $\gamma_\mathrm{a}\gamma_\mathrm{d}$ 
and $\gamma_\mathrm{d}\gamma_\mathrm{d}$ are not displayed, because they do not contain any solutions for any system. 
All prescriptions that are listed in Table\,\ref{tab:cece_results2} provide a solution 
for more than one observed system, and each observed system has at least one prescription that provides it with 
a solution.  The number of solutions per combination of prescription and observed system ranges from zero to several 
hundreds and the age differences of the accepted models lie between 36\,Myr and more than 12\,Gyr.

We will use Table\,\ref{tab:cece_results2} to compare the age differences of the models to the observed values
and use this comparison to judge the `quality' of the model solutions.  
We will assume that if the age difference in the model lies within 50\%\ of the measured cooling-age difference 
(the range in the second column of Table\,\ref{tab:cece_results2}) that this is a good agreement which we will 
assign the symbol `$+$'.  If the difference is larger than that, but smaller than a factor of five we will call 
it `close' and assign a `$\sim$'.  Cases where the nearest solution has an age difference that is more than a 
factor of five from the observed value is considered `bad' and assigned the symbol `$-$'.  If we do this for 
all cases, we obtain the second symbol for each entry in Table\,\ref{tab:comp_prescriptions}, which we can use to 
directly compare the quality of the solutions for each prescription and each observed system.  

If we compare the first and second symbol of each entry in Table\,\ref{tab:comp_prescriptions}, we see that imposing
the extra demand of a model age difference ($\Delta\tau$) that is comparable to the observations rejects a significant 
number of solutions.  We find that from the 42500 solutions that we found for the six prescriptions, based 
on final mass ratio only, 39400 are left if we demand that $q_\mathrm{1i} > 1.03$, 31100 if we additionally demand 
that the model value of $\Delta\tau$ lies within a factor of 5 from the observed value, and 26800 if we demand that 
$\Delta\tau$ given by the model lies less that 50\%\ from observed.  This means that the plusses in some entries
in the Table now become tildes or minuses, because no model solution can be found with an acceptable age difference
for the given combination of observed system and envelope-ejection prescription.  This is especially the case for
WD\,0135--052, WD\,1101+364 and WD\,1704+481 and to a lesser extent WD\,1204+450.  

\begin{figure*}
\resizebox{0.5\hsize}{!}{\rotatebox{-90}{\includegraphics{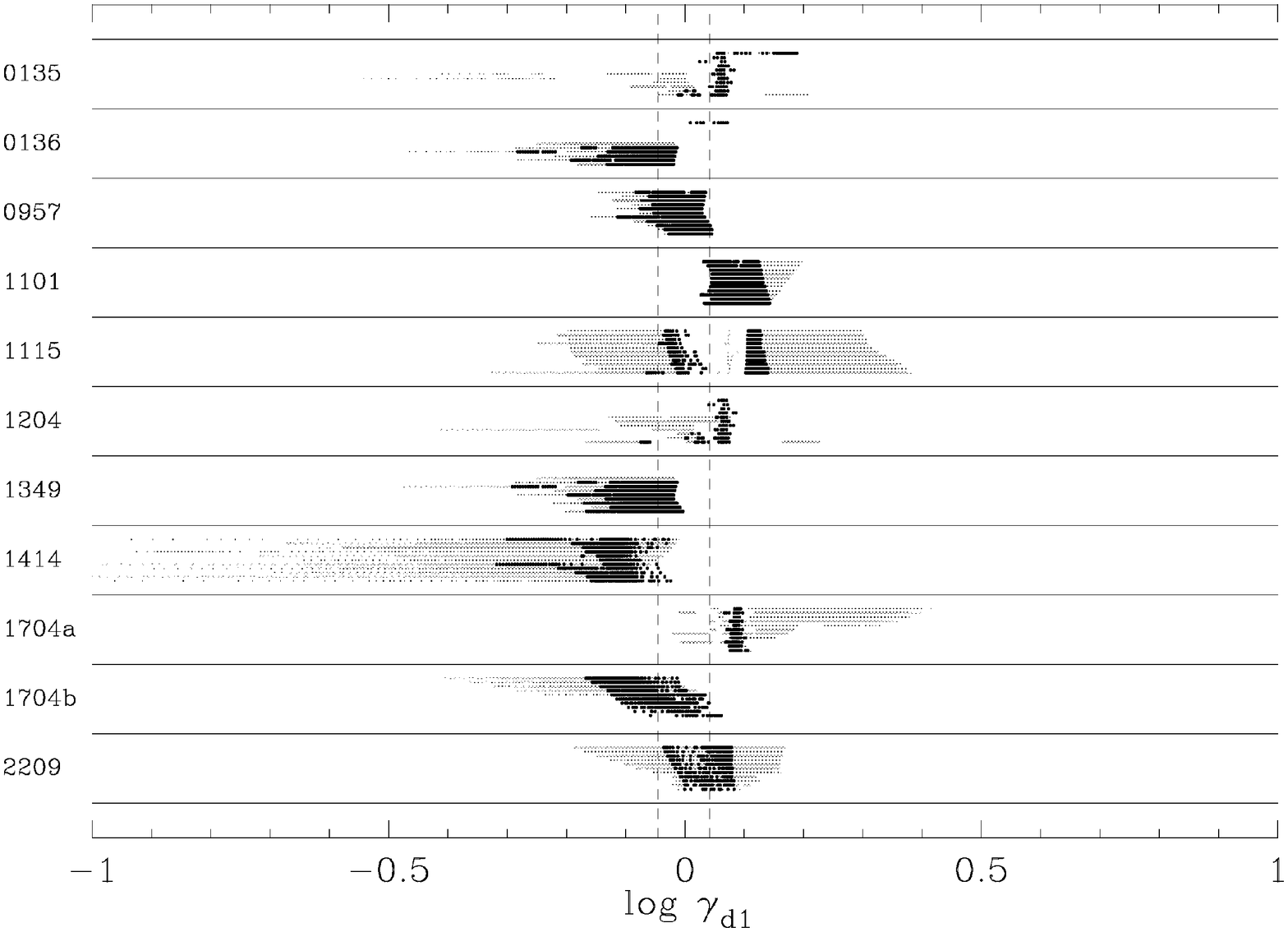}}} 
\resizebox{0.5\hsize}{!}{\rotatebox{-90}{\includegraphics{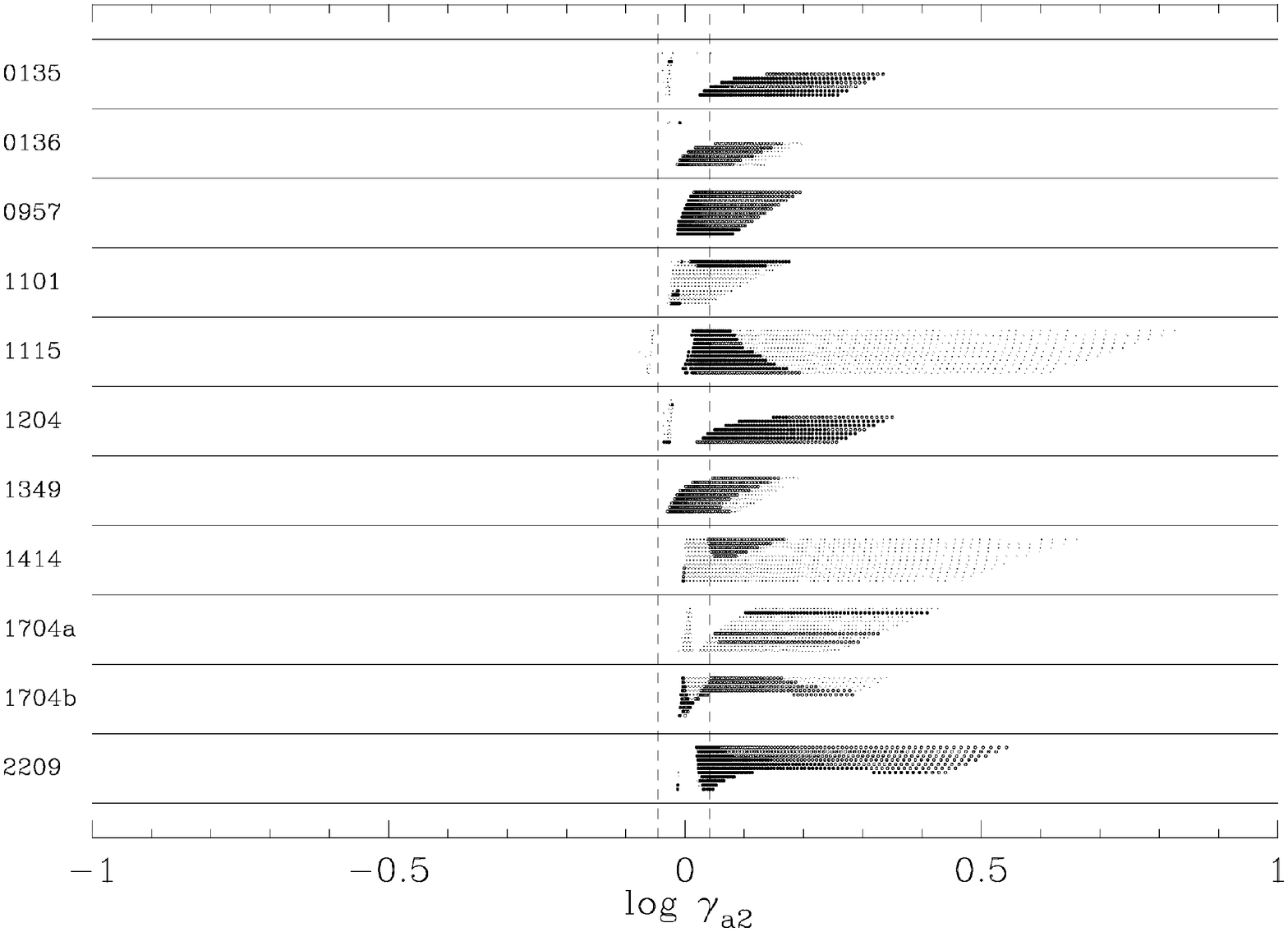}}} 
\caption{
 Solutions for the $\gamma_\mathrm{d}\gamma_\mathrm{a}$ scenario for each system.  The representation is the same as
 in Fig.\,\ref{fig:cece_ga}.  The vertical dashed lines are at $\gamma = 0.9$ and $\gamma = 1.1$ in both panels.  
 Black dots are solutions that lie between these lines in the other panel. Black dots that lie between the lines thus 
 have $0.9 \leq \gamma_\mathrm{d1} \leq 1.1$ and $0.9 \leq \gamma_\mathrm{a2} \leq 1.1$.  The total number of solutions 
 exceeds 200,000.
 \label{fig:cece_gdga}  }
\end{figure*}

The prescription 
$\gamma_\mathrm{d}\gamma_\mathrm{a}$ now gives slightly better results than $\gamma_\mathrm{s}\gamma_\mathrm{s}$,
and we find that $\gamma_\mathrm{d}\gamma_\mathrm{a}$ 
gives the same or a better result than $\gamma_\mathrm{s}\gamma_\mathrm{s}$ for each system except WD\,1704+481a,
for which the prescription $\gamma_\mathrm{d}\gamma_\mathrm{a}$ does not provide a solution with the correct mass ratio.
The envelope-ejection prescription $\gamma_\mathrm{d}\gamma_\mathrm{a}$ can explain all systems except 
WD\,1704+481 fairly well, as we shall see in the next section that the two tildes for WD\,0135--052 and especially 
WD\,1101+364 are very close to the demands for a plus.  Figure\,\ref{fig:cece_gdga} shows the solutions for the
$\gamma_\mathrm{d}\gamma_\mathrm{a}$ scenario for each system schematically.

\subsection{Description of selected solutions}
The goal of this research is, of course, to find out whether we can somehow explain the formation of the observed 
double white dwarfs.  If this is the case, we hope to learn firstly which prescriptions govern this formation and 
secondly what the progenitor systems are that evolve to the observed white-dwarf binaries.  
Although we do not find one prescription that can explain all observed systems in a perfectly satisfying way, 
the prescription $\gamma_\mathrm{d}\gamma_\mathrm{a}$ comes very close.  In this section we present 
the evolution of some of the best solutions among our calculations.  We list the five main
parameters that describe the evolutionary scenario of a solution (two initial masses, the initial period and the two 
envelope-ejection parameters) and two minor ones (the actual final masses that the models give) in a table.
Because we present solutions for six different formation prescriptions and eleven observed systems that may have more 
than one `best' solution this table is quite large.  This is particularly the case because we want to remove the 
arbitrarily chosen boundaries that we have used so far to qualify a solution.  We therefore list at least one 
solution per prescription per system, independent of how far its parameters lie from the preferred values.  
We chose to publish the complete table in electronic form only and give an excerpt of it in this paper in 
Table\,\ref{tab:solutions}.  In this way, the reader may verify how particular models do or do not work.

We manually picked the `best' solutions for a given combination of formation prescription and observed system, in 
the sense that the solution has a $\gamma$ close to unity (or, in case of $\gamma_\mathrm{s}$, close to 1.63), 
an $\alpha_\mathrm{ce}$ close to the range of 0.5--1.0 and an age difference that is close to the observed value.
In the cases where there are different solutions that each excel in a different one of these three properties, 
we may present more than one solution.  If there are several solutions that are similar on these grounds, we 
prefer those with lower initial masses and with $q_\mathrm{1i} > 1.03$.  We then leave it to the reader to
judge whether these solutions are acceptable.
The values for $q_\mathrm{2f}$ and $P_\mathrm{f}$ are identical to the value listed in Table\,\ref{tab:obswds} 
and we do not list $q_\mathrm{2f}$ in Table\,\ref{tab:solutions}.  The intermediate masses are also left out of the table, 
because no matter is accreted during the dynamical mass loss and thus $M_\mathrm{1m} = M_\mathrm{1f}$ and 
$M_\mathrm{2m} = M_\mathrm{2i}$ in our models.  The numbering of the solutions in the excerpted table is the 
same as in the complete version.
  
 \begin{table*}
 \centering
 \begin{tabular}{rlrrrrrrrrrrrrrrrr}
 \hline \hline
 Nr. & WD & $\!\!$Presc. &  $\gamma_\mathrm{1}$ & $\!\!\gamma_\mathrm{2}$,  & \multicolumn{2}{c}{$\Delta\tau$ (Myr)} & $\!\!\!\!\Delta(\Delta\tau)$ & $\tau_2$  & $M_\mathrm{1i}$ & $M_\mathrm{2i}$ & $q_\mathrm{1i}$ & $P_\mathrm{i}$  &  $q_\mathrm{2m}$ & $P_\mathrm{m}$  &  $M_\mathrm{1f}$ & $M_\mathrm{2f}$  & $P_\mathrm{f}$  \\
 & & & & $\!\!\!\alpha_\mathrm{ce2}$  & Obs & Mdl & \% & Gyr   & $M_\odot$ & $M_\odot$ &  & d   &   & d  &  $M_\odot$ & $M_\odot$ & d   \\
 \hline
   4 &$\!\!\!$  0135  &$\!\!\!\!$ $\gamma_\mathrm{s}\gamma_\mathrm{s}$  &  1.75 &$\!\!\!$  1.57 &  350 &$\!\!\!\!\!$  636 &    82. &  1.36 &     2.59 &$\!\!\!$  2.06 &$\!\!\!$  1.26 &$\!\!\!$   150.8 &     3.56 &$\!\!\!$   110.9 &     0.58 &$\!\!\!$  0.52 &$\!\!\!$ 1.56 \\
  10 &$\!\!\!$  0135  &$\!\!\!\!$ $\gamma_\mathrm{d}\gamma_\mathrm{a}$  &  1.11 &$\!\!\!$  0.94 &  350 &$\!\!\!\!\!$  118 &    66. &  0.41 &     3.30 &$\!\!\!$  2.90 &$\!\!\!$  1.14 &$\!\!\!$   36.28 &     6.21 &$\!\!\!$   41.10 &     0.47 &$\!\!\!$  0.42 &$\!\!\!$ 1.56 \\
  12 &$\!\!\!$  0136  &$\!\!\!\!$ $\gamma_\mathrm{s}\alpha_\mathrm{ce}$ &  1.48 &$\!\!\!$  1.01 &  450 &$\!\!\!\!\!$  449 &    0.2 &  3.73 &     1.44 &$\!\!\!$  1.38 &$\!\!\!$  1.04 &$\!\!\!$   72.27 &     4.16 &$\!\!\!$   265.9 &     0.33 &$\!\!\!$  0.42 &$\!\!\!$ 1.41 \\
  21 &$\!\!\!$  0136  &$\!\!\!\!$ $\gamma_\mathrm{d}\gamma_\mathrm{a}$  &  0.96 &$\!\!\!$  1.05 &  450 &$\!\!\!\!\!$  450 &    0.0 &  2.39 &     1.70 &$\!\!\!$  1.59 &$\!\!\!$  1.07 &$\!\!\!$   106.1 &     4.36 &$\!\!\!$   371.4 &     0.37 &$\!\!\!$  0.46 &$\!\!\!$ 1.41 \\
  24 &$\!\!\!$  0957  &$\!\!\!\!$ $\gamma_\mathrm{s}\alpha_\mathrm{ce}$ &  1.62 &$\!\!\!$  1.00 &  325 &$\!\!\!\!\!$  427 &    31. &  1.16 &     2.34 &$\!\!\!$  2.00 &$\!\!\!$  1.17 &$\!\!\!$   8.110 &     6.66 &$\!\!\!$   28.52 &     0.30 &$\!\!\!$  0.34 &$\!\!\!$ 0.06 \\
  32 &$\!\!\!$  0957  &$\!\!\!\!$ $\gamma_\mathrm{d}\gamma_\mathrm{a}$  &  1.00 &$\!\!\!$  1.01 &  325 &$\!\!\!\!\!$  317 &    2.5 &  1.52 &     1.98 &$\!\!\!$  1.83 &$\!\!\!$  1.08 &$\!\!\!$   26.17 &     5.61 &$\!\!\!$   79.26 &     0.33 &$\!\!\!$  0.37 &$\!\!\!$ 0.06 \\
  44 &$\!\!\!$  1101  &$\!\!\!\!$ $\gamma_\mathrm{d}\gamma_\mathrm{a}$  &  1.10 &$\!\!\!$  0.98 &  215 &$\!\!\!\!\!$  322 &    50. &  0.74 &     2.87 &$\!\!\!$  2.34 &$\!\!\!$  1.23 &$\!\!\!$   22.02 &     5.97 &$\!\!\!$   28.23 &     0.39 &$\!\!\!$  0.34 &$\!\!\!$ 0.14 \\
  46 &$\!\!\!$  1115  &$\!\!\!\!$ $\gamma_\mathrm{s}\alpha_\mathrm{ce}$ &  1.79 &$\!\!\!$  1.00 &  160 &$\!\!\!\!\!$  239 &    49. &  0.50 &     3.70 &$\!\!\!$  2.94 &$\!\!\!$  1.26 &$\!\!\!$   1693. &     3.58 &$\!\!\!$   980.4 &     0.82 &$\!\!\!$  0.69 &$\!\!\!$30.09 \\
  57 &$\!\!\!$  1115  &$\!\!\!\!$ $\gamma_\mathrm{d}\gamma_\mathrm{a}$  &  0.97 &$\!\!\!$  1.04 &  160 &$\!\!\!\!\!$  240 &    50. &  0.32 &     5.42 &$\!\!\!$  3.42 &$\!\!\!$  1.58 &$\!\!\!$   201.2 &     3.84 &$\!\!\!$   1012. &     0.89 &$\!\!\!$  0.75 &$\!\!\!$30.09 \\
 \multicolumn{18}{l}{} \\
  60 &$\!\!\!$  1204  &$\!\!\!\!$ $\gamma_\mathrm{s}\gamma_\mathrm{s}$  &  1.89 &$\!\!\!$  1.25 &   80 &$\!\!\!\!\!$   80 &    0.5 &  0.31 &     3.60 &$\!\!\!$  3.21 &$\!\!\!$  1.12 &$\!\!\!$   69.10 &     6.07 &$\!\!\!$   39.45 &     0.53 &$\!\!\!$  0.46 &$\!\!\!$ 1.60 \\
  68 &$\!\!\!$  1204  &$\!\!\!\!$ $\gamma_\mathrm{d}\gamma_\mathrm{a}$  &  1.09 &$\!\!\!$  0.92 &   80 &$\!\!\!\!\!$  100 &    25. &  0.38 &     3.34 &$\!\!\!$  2.98 &$\!\!\!$  1.12 &$\!\!\!$   15.47 &     6.32 &$\!\!\!$   19.99 &     0.47 &$\!\!\!$  0.41 &$\!\!\!$ 1.60 \\
  70 &$\!\!\!$  1349  &$\!\!\!\!$ $\gamma_\mathrm{s}\alpha_\mathrm{ce}$ &  1.45 &$\!\!\!$  1.01 &    0 &$\!\!\!\!\!$  461 &    0.0 &  4.52 &     1.35 &$\!\!\!$  1.32 &$\!\!\!$  1.03 &$\!\!\!$   105.9 &     3.77 &$\!\!\!$   364.5 &     0.35 &$\!\!\!$  0.44 &$\!\!\!$ 2.21 \\
  77 &$\!\!\!$  1349  &$\!\!\!\!$ $\gamma_\mathrm{d}\gamma_\mathrm{a}$  &  0.95 &$\!\!\!$  0.98 &    0 &$\!\!\!\!\!$  101 &    0.0 &  1.58 &     1.86 &$\!\!\!$  1.81 &$\!\!\!$  1.03 &$\!\!\!$   63.44 &     5.19 &$\!\!\!$   241.2 &     0.35 &$\!\!\!$  0.44 &$\!\!\!$ 2.21 \\
  80 &$\!\!\!$  1414  &$\!\!\!\!$ $\gamma_\mathrm{s}\gamma_\mathrm{s}$  &  1.49 &$\!\!\!$  1.76 &  200 &$\!\!\!\!\!$  112 &    44. &  0.87 &     2.55 &$\!\!\!$  2.43 &$\!\!\!$  1.05 &$\!\!\!$   525.4 &     4.09 &$\!\!\!$   1783. &     0.59 &$\!\!\!$  0.76 &$\!\!\!$ 0.52 \\
  85 &$\!\!\!$  1414  &$\!\!\!\!$ $\gamma_\mathrm{d}\gamma_\mathrm{a}$  &  0.95 &$\!\!\!$  0.99 &  200 &$\!\!\!\!\!$  188 &    5.9 &  0.43 &     3.51 &$\!\!\!$  3.09 &$\!\!\!$  1.14 &$\!\!\!$   70.81 &     5.99 &$\!\!\!$   358.3 &     0.52 &$\!\!\!$  0.66 &$\!\!\!$ 0.52 \\
  89 &$\!\!\!$  1704a &$\!\!\!\!$ $\gamma_\mathrm{s}\alpha_\mathrm{ce}$ &  2.05 &$\!\!\!$  0.43 &  -20 &$\!\!\!\!\!$    7 &   135. &  1.36 &     2.03 &$\!\!\!$  1.90 &$\!\!\!$  1.07 &$\!\!\!$   252.8 &     3.51 &$\!\!\!$   96.02 &     0.54 &$\!\!\!$  0.38 &$\!\!\!$ 0.14 \\
  98 &$\!\!\!$  1704a &$\!\!\!\!$ $\gamma_\mathrm{d}\gamma_\mathrm{a}$  &  1.11 &$\!\!\!$  1.13 &  -20 &$\!\!\!\!\!$   52 &   360. &  1.41 &     2.06 &$\!\!\!$  1.88 &$\!\!\!$  1.09 &$\!\!\!$   40.37 &     3.66 &$\!\!\!$   65.66 &     0.51 &$\!\!\!$  0.36 &$\!\!\!$ 0.14 \\
 101 &$\!\!\!$  1704b &$\!\!\!\!$ $\gamma_\mathrm{s}\alpha_\mathrm{ce}$ &  1.74 &$\!\!\!$  0.76 &   20 &$\!\!\!\!\!$  285 &  1326. &  0.75 &     2.76 &$\!\!\!$  2.55 &$\!\!\!$  1.08 &$\!\!\!$   49.12 &     6.40 &$\!\!\!$   107.3 &     0.40 &$\!\!\!$  0.57 &$\!\!\!$ 0.14 \\
 107 &$\!\!\!$  1704b &$\!\!\!\!$ $\gamma_\mathrm{d}\alpha_\mathrm{ce}$ &  1.03 &$\!\!\!$  0.15 &   20 &$\!\!\!\!\!$  182 &   810. &  2.23 &     1.68 &$\!\!\!$  1.65 &$\!\!\!$  1.01 &$\!\!\!$   212.1 &     4.08 &$\!\!\!$   478.6 &     0.41 &$\!\!\!$  0.58 &$\!\!\!$ 0.14 \\
 113 &$\!\!\!$  2209  &$\!\!\!\!$ $\gamma_\mathrm{s}\gamma_\mathrm{s}$  &  1.62 &$\!\!\!$  1.69 &  500 &$\!\!\!\!\!$  477 &    4.5 &  1.37 &     2.40 &$\!\!\!$  2.00 &$\!\!\!$  1.20 &$\!\!\!$   165.4 &     3.64 &$\!\!\!$   258.7 &     0.55 &$\!\!\!$  0.55 &$\!\!\!$ 0.28 \\
 122 &$\!\!\!$  2209  &$\!\!\!\!$ $\gamma_\mathrm{d}\gamma_\mathrm{a}$  &  1.04 &$\!\!\!$  1.05 &  500 &$\!\!\!\!\!$  340 &    32. &  0.50 &     4.15 &$\!\!\!$  2.94 &$\!\!\!$  1.41 &$\!\!\!$   98.45 &     4.67 &$\!\!\!$   294.3 &     0.63 &$\!\!\!$  0.63 &$\!\!\!$ 0.28 \\
 \hline
 \end{tabular}
 \caption{Selected model solutions for the double envelope-ejection scenario.  This table is an
 excerpt of the total list of 122 entries.  The first nine columns show the number of the entry, the double white dwarf that
 the model is a solution for, the prescription used, the two envelope-ejection parameters, the age difference of the two
 components as observed and in the model ($\Delta\tau$) in Myr, the relative difference between the observed and model age difference, defined as
 $\Delta(\Delta\tau) \equiv \left|\frac{\Delta\tau_\mathrm{mod} - \Delta\tau_\mathrm{obs}}{\Delta\tau_\mathrm{obs}}\right|$
 in \%, the time of the formation of the double white dwarf since the ZAMS ($\tau_2$) in Gyr.
 The last nine columns list binary parameters: the initial (ZAMS) masses, mass ratio and orbital period, 
 the intermediate mass ratio and period and the final masses and period.
 \label{tab:solutions}  }
 \end{table*}

The complete table contains 122 solutions.  The initial binaries have primary masses 
between 1.27\,$M_\odot$ and 5.42\,$M_\odot$.
Of the 122 solutions, 49\%\ have an initial primary mass less than 2.5\,$M_\odot$ and 93\%\ of the primaries are 
less massive than 4\,$M_\odot$.  Thus, the models suggest that the double white dwarfs are formed by low-mass 
stars, as may be required to explain the observed numbers of these binaries.  Of the initial systems, 89\%\ 
have orbital periods between 10 and 1000\,days.  All proposed solutions undergo a first envelope ejection described
by angular-momentum balance of some sort, which allows the orbital period to increase during such a 
mass-transfer phase.  In 80\%\ of the selected solutions this is the case, and for 61\%\ of the solutions
the intermediate orbital period is twice or more as long as the initial period.
Of the 122 solutions, 57\%\ have initial mass ratios $q_\mathrm{1i}<1.15$
while for 17\%\ $q_\mathrm{1i}>1.5$.  We included only 3 solutions with $q_\mathrm{1i}<1.03$.  
It could be that these initial systems evolve into a double common envelope, where the two 
white dwarfs are formed simultaneously and the second white dwarf is undermassive.  On the other hand, 
because the orbital period increases in most of the first envelope ejections, the outcome of such a common envelope 
is uncertain.  We should treat these solutions with some scepticism.

We now briefly discuss the individual solutions that are listed in the excerpted Table\,\ref{tab:solutions}.
For WD\,0135--052 we find solutions with acceptable $\gamma$'s or with acceptable $\Delta\tau$, but not with both.
In solution\,4 the age difference differs 82\%\ from observed while the values for $\gamma_\mathrm{s}$ lie less than 
10\%\ from 1.63.  Solution\,10 ($\gamma_\mathrm{d}\gamma_\mathrm{a}$) has
a value for $\gamma_\mathrm{d}$ just outside the 10\%\ region (10.8\%) and a value for $\Delta\tau$ that lies just
outside the 50\%\ region (66\%).  These two solutions therefore have properties that lie not too far from the 
preferred values.  Solutions\,12 ($\gamma_\mathrm{s}\alpha_\mathrm{ce}$) and 21 ($\gamma_\mathrm{d}\gamma_\mathrm{a}$) 
for WD\,0136+768 have envelope-ejection parameters within 10\%\ and 5\%\ respectively from the preferred values, while 
the model ages hardly differ at all from the observed values.  Both solutions are also a solution for 
$\gamma_\mathrm{s}\gamma_\mathrm{s}$ (in the complete table). In solutions\,24 and 32 for WD\,0957--666 we find 
envelope-ejection parameters with less than 1\%\ from the preferred values and the age differences well within 
50\%\ from observed. For WD\,1101+364, solution is very close to the requirements for a plus sign in 
Table\,\ref{tab:comp_prescriptions}: $\Delta\tau$ lies just less than 50\%\ from observed, however 
$\gamma_\mathrm{d}=1.103$, so that it falls just outside the 10\%\ range from the preferred value.
Solutions\,46 and 57 for PG\,1115+116 have envelope-ejection parameters 10\%\ and 5\%\ respectively from the 
desired values and acceptable values for $\Delta\tau$. The initial masses are high for these solutions, in accordance with the 
fact that these stars are required to form white dwarfs with masses as high as 0.8\,$M_\odot$.  

WD\,1204+450 has few solutions that have an age difference as small as observed.  Solutions that do have the
observed age difference, as solution\,60, often have values for $\gamma$ or $\alpha$ that lie far from desired.
The prescription $\gamma_\mathrm{d}\gamma_\mathrm{a}$ is the only one that provides solutions that are qualified
as `good' in Table\,\ref{tab:comp_prescriptions}. Solution\,68 has values for $\gamma$ within 10\%\ from unity and
an age difference 25\%\ from the observed value.
For WD\,1349+144 the cooling ages are not known, although the similar Balmer spectra of the two 
components \citep{2003whdw.conf...43K} seem to suggest that $\Delta\tau$ is small.  
Solutions\,70 and 77 give solutions with envelope-ejection parameters slightly more that 10\%\ and 5\%\ from the
preferred values respectively and have age differences of 461 and 101\,Myr.  The complete table shows that
in addition to this solutions with values for $\Delta\tau$ such as 129, 215, 384, 975 and 1776\,Myr can be
found, so that they span a large range within which the actual age difference is likely to lie.
The prescriptions $\gamma_\mathrm{s}\gamma_\mathrm{s}$ and $\gamma_\mathrm{d}\gamma_\mathrm{a}$ give good solutions
for the system HE\,1414--0848.  The solutions\,80 and 85 have values for $\gamma$ that lie within 10\%\ and 5\%\
respectively from the desired values, while their age differences are 44\%\ and 6\%\ from observed.

Since the observed age difference of WD\,1704+481a is $-20$\,Myr, we introduced a system with the reversed mass
ratio (WD\,1704+481b) and hence an age difference of $+20$\,Myr.  Interestingly enough, the solutions with closest
age difference for WD\,1704+481b have $\Delta\tau\!\mathrel{\hbox{\rlap{\lower.55ex \hbox {$\sim$}} \kern-.3em \raise.4ex \hbox{$>$}}}\!180$\,Myr, a factor of nine or more than observed, as is the
case for solution\,107, though it has $\alpha_\mathrm{ce}=0.15$.  Solution\,109 has nice values for $\gamma$, but
an age difference of 205\,Myr.  However, for WD\,1704+481a we find solutions with reasonable envelope-ejection 
parameters (13\%\ from the preferred value) and an age difference of around 50\,Myr, like solution\,98, and with 
parameters that are more off, but with an age difference of only 7\,Myr as in solution\,89.  The system WD\,1704+481a 
seems therefore better explained by our models than the system with the reverse mass ratio.  Because the observed 
cooling-age difference is only in the order of a few per cent of the total age of the system (see Table\,\ref{tab:obswds}), 
a change of 10\%\ in the determined cooling age of one of the two components is sufficient to alter the age difference 
from $-20$\,Myr to $+50$\,Myr, which would bring it in the range that can be explained by our models.
For HE\,2209--1444, we present solutions\,113 and 122, that have envelope-ejection parameters less than 5\%\ from the desired 
values and age differences that lie 5\%\ and 32\%\ respectively from the observed cooling-age difference.

Summarising, we find that we can explain the mass ratios and orbital periods of the ten observed systems well with
the scenarios $\gamma_\mathrm{s}\gamma_\mathrm{s}$ and $\gamma_\mathrm{d}\gamma_\mathrm{a}$.  If we also take into 
account the observed age difference between the components, the scenario $\gamma_\mathrm{d}\gamma_\mathrm{a}$ still
works well.  However, the allowed ranges for $\gamma$ and age difference must be extended a little to include 
WD\,0135--052 ($\gamma_\mathrm{d1}\!=\!1.11$, $\Delta\!(\Delta\tau)\!=\!66\%$) and WD\,1101+364 
($\gamma_\mathrm{d1}\!=\!1.11$), while the small age difference observed for WD\,1704+481 cannot be explained.


\section{Discussion}
\label{sec:discussion}

\subsection{Comparison to other work}

In this paper we investigate the formation scenarios for double white dwarfs first put forward by 
\citet{2000A&A...360.1011N}.  Their paper is based on three double white dwarfs and we expanded this 
to the set of ten that has been observed so far.  Rather than using analytical 
approximations, we used a stellar evolution code to do most of the calculations.  One of the advantages 
of this is that we calculate the binding energy of the donor star at the onset of a common envelope, so 
that we can directly calculate the common-envelope parameter $\alpha_\mathrm{ce}$ without the need of 
the envelope-structure parameter $\lambda_\mathrm{env}$, that turns out to be far from constant during 
the evolution of a star (see Fig.\,\ref{fig:basic_mc-lam}).  This allows us to demand physically 
acceptable values for $\alpha_\mathrm{ce}$.

The use of an evolution code instead of analytical expressions obviously gives more accurate values
for instance for the core-mass\,--\,radius relation.  Our main conclusions are nevertheless the same as those of 
\citet{2000A&A...360.1011N}, even though they are based on a larger sample of observed binaries: stable,
conservative mass 
transfer followed by a common envelope with spiral-in based on energy balance cannot explain the formation of the 
observed systems, and neither can the $\alpha_\mathrm{ce}\alpha_\mathrm{ce}$ scenario of two such spiral-ins.  
We therefore arrive at the same conclusion, that a third mass-transfer prescription is needed
to explain the first mass-transfer phase of these systems and we use their envelope-ejection
prescription, based on angular-momentum balance (Eq.\,\ref{eq:gce}).

\citet{2005MNRAS.356..753N} use more advanced fits to stellar models, but still need the envelope-structure 
parameter $\lambda_\mathrm{env}$ so that it is difficult to interpret the values they find for the product
$\alpha_\mathrm{ce}\,\lambda_\mathrm{env}$.  They use the same ten observed double-lined white dwarfs as we 
do, next to a number of single-lined systems.  They also conclude that a $\gamma$-envelope ejection is needed 
for the first mass transfer and find, like \citet{2000A&A...360.1011N}, that all observed systems can be
explained by $1.50<\gamma_\mathrm{s}<1.75$, for both mass-transfer phases.  Alternatively the second 
mass-transfer episode can be reconstructed with $0<\alpha_\mathrm{ce}\,\lambda_\mathrm{env}<4$.  However, \citet{2005MNRAS.356..753N}
do not discuss the coupling of the two solution sets for the two phases, {\it e.g.} it is not described how 
many of the solutions with $1.50<\gamma_\mathrm{s1}<1.75$ have $\gamma_\mathrm{s2}$ in the same range.  
We introduced slightly different definitions for the $\gamma$-algorithm in Eqs.\,\ref{eq:gcea} and \ref{eq:gceb}, 
so that we can demand that $\gamma$ is in the order of unity.  We find indeed that we can explain the 
observed masses and periods with $\gamma_\mathrm{a},\gamma_\mathrm{d}\sim 1.0$.  

We add to the treatment by \citet{2000A&A...360.1011N} and \citet{2005MNRAS.356..753N} in demanding that, 
in addition to the masses and orbital period, the age difference of our models must be comparable to the 
observed value. It turns out that this puts a strong constraint on the selection of model solutions for all 
three definitions of $\gamma$.  However, we can still explain most systems, although we need mass 
loss described by both $\gamma_\mathrm{a}$ and $\gamma_\mathrm{d}$ to do so.

The description for dynamical mass loss with the specific angular momentum of
the donor star (Eq.\ref{eq:gceb}) is similar to the scenario of a tidally-enhanced stellar wind
\citep{1988MNRAS.231..823T,1988ApJ...334..357T}.  In this scenario the mass loss from a (sub)giant 
due to stellar wind increases up to a factor of 150 with respect to Reimers' empirical law 
\citep{1975MSRSL...8..369R} when the star is close to filling its Roche lobe.  \citet{1988MNRAS.231..823T}
postulate the enhanced wind to explain for instance observed pre-Algol systems such as Z\,Her.  In this 
binary the more evolved star is 10\%\ less massive than its main-sequence companion, while only filling 
about half of its Roche lobe.

\citet{1998MNRAS.296.1019H} uses this tidally-enhanced stellar wind in his research on the formation of
double degenerates and concludes, among others, that his models that include this enhanced stellar wind 
give a better explanation of the observed double-degenerate binaries than models that do not include it.  
The enhanced mass loss makes subsequent mass transfer due to Roche-lobe overflow dynamically more stable. 
Envelope ejection due to dynamical mass loss is then more often prevented and binaries evolve to longer orbital periods before
the second mass transfer, which is then more likely to produce a CO white dwarf.  Thus, the enhanced-wind
scenario increases the ratio of CO-helium double white dwarfs to helium-helium binaries.

Envelope ejection described by Eq.\ref{eq:gceb} is essentially the same as the limiting case in which most or all 
of the envelope is lost due to an enhanced wind. \citet{1988MNRAS.231..823T} show that the tidally-enhanced 
wind can indeed prevent Roche-lobe overflow altogether because the envelope is completely lost by 
the wind and the core becomes exposed.  Without an enhanced wind, this happens for binaries with an initial
mass ratio of 2 or less only if they have initial periods of more than 1000\,days.  When the tidally-enhanced wind is 
included, core exposure without Roche-lobe overflow occurs for these binaries with initial periods as short 
as 10--30\,days.

\subsection{Alternative formation scenario for massive white dwarfs}
\begin{figure}
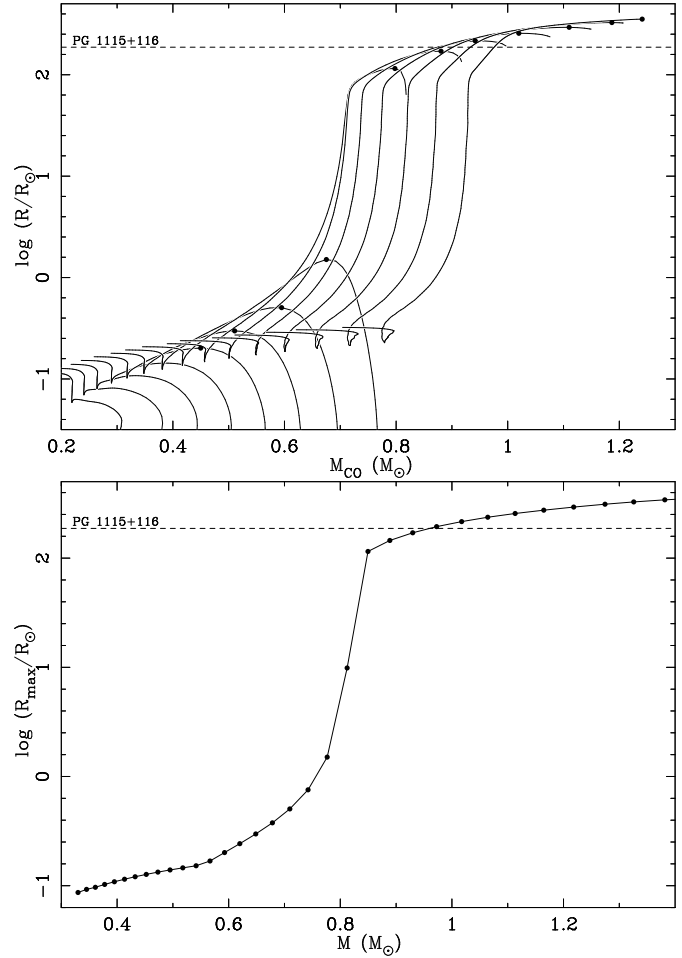

\resizebox{\hsize}{!}{\rotatebox{-90}{\includegraphics{5066f17a.eps}}} 
\resizebox{\hsize}{!}{\rotatebox{-90}{\includegraphics{5066f17b.eps}}} 
\caption{
 {\it Upper panel} ({\bf a}): The radius of a helium star as a function of
 its CO-core mass, for a selection of 15 models with total masses between $0.41$ and $1.43\,M_\odot$.
 The dots show where the maximum radii are obtained and are used for the lower panel.
 The dashed line is the Roche-lobe radius for the intermediate primary of PG\,1115+116 according
 to our solution\,57.
 {\it Lower panel} ({\bf b}): The maximum radius of a low-mass helium star as a function of
 its total mass, for a selection of 33 models with masses between $0.33$ and $1.4\,M_\odot$.
 The dots are the data points, the solid line connects them to guide the eye.
 \label{fig:hestars}  }
\end{figure}

In the present research we have assumed that after envelope ejection occurs, the core of the Roche-lobe 
filling giant becomes a helium or CO white dwarf with no further evolution other than cooling.  However,
helium cores that are more massive than $0.33\,M_\odot$ are not degenerate and those more massive 
than about $0.5\,M_\odot$ will burn most of the helium in their cores and produce a CO core.  If exposed, 
they are in effect helium stars.  For helium stars less massive than about $0.75\,M_\odot$ the radius hardly changes 
during the helium (shell) burning, but stars more massive than that experience a giant phase.  This is
shown in Fig.\,\ref{fig:hestars}a, where the radius of a selection of helium-star models is plotted as a 
function of the CO-core mass.
For the more massive models in the Figure, the stars expand from the order of a few tenths of a solar radius to a 
few hundred solar radii.  Thus, helium stars with a core mass $M_\mathrm{CO}\mathrel{\hbox{\rlap{\lower.55ex \hbox {$\sim$}} \kern-.3em \raise.4ex \hbox{$>$}}} 0.7\,M_\odot$ may 
and those with $M_\mathrm{CO}\mathrel{\hbox{\rlap{\lower.55ex \hbox {$\sim$}} \kern-.3em \raise.4ex \hbox{$>$}}} 0.8\,M_\odot$ must become giants and could fill their Roche lobes 
as a consequence.  The black dots in Fig.\,\ref{fig:hestars}a indicate the maximum radius for a certain 
helium-star model and if we plot the maximum radii of these and other models as a function of the total 
mass of the star, we obtain Fig.\,\ref{fig:hestars}b.  This Figure shows that a helium star more massive 
than about $0.83\,M_\odot$ must evolve through a giant phase (see {\it e.g.} Paczy{\'n}ski 1971; Habets 
1986\nocite{1971AcA....21....1P,1986A&A...167...61H}). 

There are two double white dwarfs in the observed sample with $M_2 > 0.6\,M_\odot$, PG\,1115+116 (both
components) and HE\,1414--0848 (the secondary).  The evolutionary scenarios in Table\,\ref{tab:solutions}
suggest that all these stars emerge from the envelope ejection with a CO core, except solution\,57 for
PG\,1115+116, where the 5.42\,$M_\odot$ primary progenitor produces a 0.89\,$M_\odot$ helium core before
helium ignites.  The Roche-lobe radius of the 0.89\,$M_\odot$ helium star in the intermediate binary is
187\,$R_\odot$ according to this solution and shown as the dashed line in Fig.\,\ref{fig:hestars}.

The Figure shows that the mass and Roche-lobe radius of this star are in 
the proper range to fit the helium-giant scenario. 
We show a simple numerical example to illustrate this scenario.  The dot in Fig.\,\ref{fig:hestars}a at 
$M_\mathrm{CO}$=0.88\,$M_\odot$ and $R=171\,R_\odot$, just below the dashed line, is the point where 
the model of 0.93$\,M_\odot$ from our grid of helium-star models reaches its largest radius.  The star 
thus has an envelope mass of only 0.05\,$M_\odot$ and with a mass ratio of almost 4, mass transfer 
would be stable (Eq.\,57 of \citet{2002MNRAS.329..897H}).  If we assume that this star would be the 
primary of solution\,57 in Table\,\ref{tab:solutions} and that 0.04\,$M_\odot$ would be transferred
conservatively, the orbital period after the mass transfer would be 1115\,d, so that the period would 
not change drastically and the ensuing second envelope ejection would be similar to the one found in 
solution\,57. If the mass were lost in a wind, which could be triggered by the fact that the star expands,
but for which the Roche lobe need not be filled, the orbital period would change less than 2\%\ to 1031\,d.
It seems that a complete, detailed model could be found to explain this system along these lines.

Both components in HE\,1414--0848 are DA white dwarfs \citep{2002A&A...386..957N}, as is the secondary
of PG\,1115+116.  The hydrogen in the spectra of these stars suggests that the surface layer that
formed after the envelope was ejected is still present.  
However, the primary in PG\,1115+116 is a DB white dwarf.  As \citet{2002MNRAS.334..833M} point 
out, the giant phase of a helium star could be the explanation for this and the scenario sketched above 
might indeed describe the formation of this system.


\section{Conclusions}
\label{sec:conclusions}

We investigated several formation scenarios for the observed ten double white dwarfs listed in 
Table\,\ref{tab:obswds} and present the best models in Table\,\ref{tab:solutions}.
We draw five main conclusions:

\begin{itemize}
\item The scenario where the first mass-transfer phase is stable and conservative, followed
by a common envelope with spiral-in based on energy conservation (see Eq.\,\ref{eq:forwardce})
cannot explain the observed masses and periods of all double white dwarfs.

\item The scenario with envelope ejection based on angular-momentum balance followed by
ejection of the second envelope with either energy or angular-momentum balance can explain the observed
masses and orbital periods very well.

\item Including the age difference as a quality criterion for model solutions produces strong
restrictions to the selection of solutions and makes it much more difficult to find acceptable
solutions.

\item By taking into account the possibilities that mass is lost either from the donor or from 
the accretor, we show that the formation of the close double white dwarfs can be explained
if the mass carries the specific angular momentum of one of the two binary members.

\item In particular, the scenario in which the envelope of the primary is lost with the
specific orbital angular momentum of the donor star, followed by envelope loss of the secondary
with the specific orbital angular momentum of its companion star ($\gamma_\mathrm{d}\gamma_\mathrm{a}$)
can explain the observed masses, orbital periods and age differences well.

\end{itemize}

\begin{acknowledgements}
We thank P.P. Eggleton for making his binary evolution code available to us.
We would like to thank the anonymous referee for pointing out an error in 
Eq.\,\ref{eq:soberman1} in the first submitted version of this article.
\end{acknowledgements}


\bibliographystyle{aa}
\bibliography{5066}

 \setcounter{table}{5}
 \begin{table*}
 \centering
 \begin{tabular}{rlrrrrrrrrrrrrrrrr}
 \hline \hline
 Nr. & WD & $\!\!$Presc. &  $\gamma_\mathrm{1}$ & $\!\!\gamma_\mathrm{2}$,  & \multicolumn{2}{c}{$\Delta\tau$ (Myr)} & $\!\!\!\!\Delta(\Delta\tau)$ & $\tau_2$  & $M_\mathrm{1i}$ & $M_\mathrm{2i}$ & $q_\mathrm{1i}$ & $P_\mathrm{i}$  &  $q_\mathrm{2m}$ & $P_\mathrm{m}$  &  $M_\mathrm{1f}$ & $M_\mathrm{2f}$  & $P_\mathrm{f}$  \\
 & & & & $\!\!\!\alpha_\mathrm{ce2}$  & Obs & Mdl & \% & Gyr   & $M_\odot$ & $M_\odot$ &  & d   &   & d  &  $M_\odot$ & $M_\odot$ & d   \\
 \hline
 \multicolumn{18}{l}{} \\
   1 &$\!\!\!$  0135  &$\!\!\!\!$ $\gamma_\mathrm{s}\alpha_\mathrm{ce}$ &  2.02 &$\!\!\!$  0.87 &  350 &$\!\!\!\!\!$  377 &    7.7 &  3.14 &     1.51 &$\!\!\!$  1.46 &$\!\!\!$  1.04 &$\!\!\!$   504.5 &     3.12 &$\!\!\!$   264.7 &     0.47 &$\!\!\!$  0.42 &$\!\!\!$ 1.56 \\
   2 &$\!\!\!$  0135  &$\!\!\!\!$ $\gamma_\mathrm{s}\alpha_\mathrm{ce}$ &  0.81 &$\!\!\!$  0.61 &  350 &$\!\!\!\!\!$  899 &   157. &  2.17 &     2.11 &$\!\!\!$  1.63 &$\!\!\!$  1.29 &$\!\!\!$   33.22 &     3.20 &$\!\!\!$   372.5 &     0.51 &$\!\!\!$  0.46 &$\!\!\!$ 1.56 \\
   3 &$\!\!\!$  0135  &$\!\!\!\!$ $\gamma_\mathrm{s}\gamma_\mathrm{s}$  &  2.02 &$\!\!\!$  1.73 &  350 &$\!\!\!\!\!$  377 &    7.7 &  3.14 &     1.51 &$\!\!\!$  1.46 &$\!\!\!$  1.04 &$\!\!\!$   504.5 &     3.12 &$\!\!\!$   264.7 &     0.47 &$\!\!\!$  0.42 &$\!\!\!$ 1.56 \\
   4 &$\!\!\!$  0135  &$\!\!\!\!$ $\gamma_\mathrm{s}\gamma_\mathrm{s}$  &  1.75 &$\!\!\!$  1.57 &  350 &$\!\!\!\!\!$  636 &    82. &  1.36 &     2.59 &$\!\!\!$  2.06 &$\!\!\!$  1.26 &$\!\!\!$   150.8 &     3.56 &$\!\!\!$   110.9 &     0.58 &$\!\!\!$  0.52 &$\!\!\!$ 1.56 \\
   5 &$\!\!\!$  0135  &$\!\!\!\!$ $\gamma_\mathrm{a}\alpha_\mathrm{ce}$ &  1.01 &$\!\!\!$  0.61 &  350 &$\!\!\!\!\!$  899 &   157. &  2.17 &     2.11 &$\!\!\!$  1.63 &$\!\!\!$  1.29 &$\!\!\!$   33.22 &     3.20 &$\!\!\!$   372.5 &     0.51 &$\!\!\!$  0.46 &$\!\!\!$ 1.56 \\
   6 &$\!\!\!$  0135  &$\!\!\!\!$ $\gamma_\mathrm{a}\alpha_\mathrm{ce}$ &  1.01 &$\!\!\!$  0.80 &  350 &$\!\!\!\!\!$ 1825 &   421. &  2.09 &     3.42 &$\!\!\!$  1.65 &$\!\!\!$  2.07 &$\!\!\!$   32.08 &     3.38 &$\!\!\!$   291.9 &     0.49 &$\!\!\!$  0.44 &$\!\!\!$ 1.56 \\
   7 &$\!\!\!$  0135  &$\!\!\!\!$ $\gamma_\mathrm{a}\gamma_\mathrm{a}$  &  1.01 &$\!\!\!$  1.21 &  350 &$\!\!\!\!\!$  899 &   157. &  2.17 &     2.11 &$\!\!\!$  1.63 &$\!\!\!$  1.29 &$\!\!\!$   33.22 &     3.20 &$\!\!\!$   372.5 &     0.51 &$\!\!\!$  0.46 &$\!\!\!$ 1.56 \\
   8 &$\!\!\!$  0135  &$\!\!\!\!$ $\gamma_\mathrm{d}\alpha_\mathrm{ce}$ &  1.36 &$\!\!\!$  0.89 &  350 &$\!\!\!\!\!$  261 &    25. &  3.02 &     1.51 &$\!\!\!$  1.48 &$\!\!\!$  1.03 &$\!\!\!$   505.1 &     3.16 &$\!\!\!$   260.8 &     0.47 &$\!\!\!$  0.42 &$\!\!\!$ 1.56 \\
   9 &$\!\!\!$  0135  &$\!\!\!\!$ $\gamma_\mathrm{d}\gamma_\mathrm{a}$  &  1.36 &$\!\!\!$  1.22 &  350 &$\!\!\!\!\!$  377 &    7.7 &  3.14 &     1.51 &$\!\!\!$  1.46 &$\!\!\!$  1.04 &$\!\!\!$   504.5 &     3.12 &$\!\!\!$   264.7 &     0.47 &$\!\!\!$  0.42 &$\!\!\!$ 1.56 \\
  10 &$\!\!\!$  0135  &$\!\!\!\!$ $\gamma_\mathrm{d}\gamma_\mathrm{a}$  &  1.11 &$\!\!\!$  0.94 &  350 &$\!\!\!\!\!$  118 &    66. &  0.41 &     3.30 &$\!\!\!$  2.90 &$\!\!\!$  1.14 &$\!\!\!$   36.28 &     6.21 &$\!\!\!$   41.10 &     0.47 &$\!\!\!$  0.42 &$\!\!\!$ 1.56 \\
  11 &$\!\!\!$  0135  &$\!\!\!\!$ $\gamma_\mathrm{d}\gamma_\mathrm{a}$  &  1.06 &$\!\!\!$  0.95 &  350 &$\!\!\!\!\!$   61 &    82. &  0.26 &     3.79 &$\!\!\!$  3.42 &$\!\!\!$  1.11 &$\!\!\!$   26.62 &     6.16 &$\!\!\!$   51.56 &     0.56 &$\!\!\!$  0.50 &$\!\!\!$ 1.56 \\
 \multicolumn{18}{l}{} \\
  12 &$\!\!\!$  0136  &$\!\!\!\!$ $\gamma_\mathrm{s}\alpha_\mathrm{ce}$ &  1.48 &$\!\!\!$  1.01 &  450 &$\!\!\!\!\!$  449 &    0.2 &  3.73 &     1.44 &$\!\!\!$  1.38 &$\!\!\!$  1.04 &$\!\!\!$   72.27 &     4.16 &$\!\!\!$   265.9 &     0.33 &$\!\!\!$  0.42 &$\!\!\!$ 1.41 \\
  13 &$\!\!\!$  0136  &$\!\!\!\!$ $\gamma_\mathrm{s}\gamma_\mathrm{s}$  &  1.51 &$\!\!\!$  1.62 &  450 &$\!\!\!\!\!$  450 &    0.0 &  2.39 &     1.70 &$\!\!\!$  1.59 &$\!\!\!$  1.07 &$\!\!\!$   106.1 &     4.36 &$\!\!\!$   371.4 &     0.37 &$\!\!\!$  0.46 &$\!\!\!$ 1.41 \\
  14 &$\!\!\!$  0136  &$\!\!\!\!$ $\gamma_\mathrm{s}\gamma_\mathrm{s}$  &  1.48 &$\!\!\!$  1.66 &  450 &$\!\!\!\!\!$  449 &    0.2 &  3.73 &     1.44 &$\!\!\!$  1.38 &$\!\!\!$  1.04 &$\!\!\!$   72.27 &     4.16 &$\!\!\!$   265.9 &     0.33 &$\!\!\!$  0.42 &$\!\!\!$ 1.41 \\
  15 &$\!\!\!$  0136  &$\!\!\!\!$ $\gamma_\mathrm{a}\alpha_\mathrm{ce}$ &  1.58 &$\!\!\!$  0.93 &  450 &$\!\!\!\!\!$ 1212 &   169. &  1.85 &     2.46 &$\!\!\!$  1.72 &$\!\!\!$  1.43 &$\!\!\!$   39.79 &     4.82 &$\!\!\!$   294.3 &     0.36 &$\!\!\!$  0.45 &$\!\!\!$ 1.41 \\
  16 &$\!\!\!$  0136  &$\!\!\!\!$ $\gamma_\mathrm{a}\alpha_\mathrm{ce}$ &  1.01 &$\!\!\!$  0.74 &  450 &$\!\!\!\!\!$ 1776 &   295. &  2.26 &     2.72 &$\!\!\!$  1.61 &$\!\!\!$  1.69 &$\!\!\!$   14.85 &     4.42 &$\!\!\!$   360.4 &     0.37 &$\!\!\!$  0.46 &$\!\!\!$ 1.41 \\
  17 &$\!\!\!$  0136  &$\!\!\!\!$ $\gamma_\mathrm{a}\gamma_\mathrm{a}$  &  1.59 &$\!\!\!$  0.97 &  450 &$\!\!\!\!\!$  891 &    98. &  1.49 &     2.52 &$\!\!\!$  1.86 &$\!\!\!$  1.36 &$\!\!\!$   13.72 &     5.58 &$\!\!\!$   175.9 &     0.33 &$\!\!\!$  0.42 &$\!\!\!$ 1.41 \\
  18 &$\!\!\!$  0136  &$\!\!\!\!$ $\gamma_\mathrm{a}\gamma_\mathrm{a}$  &  1.01 &$\!\!\!$  1.03 &  450 &$\!\!\!\!\!$ 1995 &   343. &  2.55 &     2.59 &$\!\!\!$  1.55 &$\!\!\!$  1.67 &$\!\!\!$   9.960 &     4.55 &$\!\!\!$   265.1 &     0.34 &$\!\!\!$  0.43 &$\!\!\!$ 1.41 \\
  19 &$\!\!\!$  0136  &$\!\!\!\!$ $\gamma_\mathrm{d}\alpha_\mathrm{ce}$ &  0.93 &$\!\!\!$  0.99 &  450 &$\!\!\!\!\!$  317 &    30. &  3.90 &     1.40 &$\!\!\!$  1.37 &$\!\!\!$  1.03 &$\!\!\!$   74.07 &     4.11 &$\!\!\!$   269.8 &     0.33 &$\!\!\!$  0.42 &$\!\!\!$ 1.41 \\
  20 &$\!\!\!$  0136  &$\!\!\!\!$ $\gamma_\mathrm{d}\alpha_\mathrm{ce}$ &  0.95 &$\!\!\!$  1.00 &  450 &$\!\!\!\!\!$  299 &    34. &  2.68 &     1.59 &$\!\!\!$  1.53 &$\!\!\!$  1.04 &$\!\!\!$   75.88 &     4.50 &$\!\!\!$   269.5 &     0.34 &$\!\!\!$  0.43 &$\!\!\!$ 1.41 \\
  21 &$\!\!\!$  0136  &$\!\!\!\!$ $\gamma_\mathrm{d}\gamma_\mathrm{a}$  &  0.96 &$\!\!\!$  1.05 &  450 &$\!\!\!\!\!$  450 &    0.0 &  2.39 &     1.70 &$\!\!\!$  1.59 &$\!\!\!$  1.07 &$\!\!\!$   106.1 &     4.36 &$\!\!\!$   371.4 &     0.37 &$\!\!\!$  0.46 &$\!\!\!$ 1.41 \\
  22 &$\!\!\!$  0136  &$\!\!\!\!$ $\gamma_\mathrm{d}\gamma_\mathrm{a}$  &  0.96 &$\!\!\!$  1.01 &  450 &$\!\!\!\!\!$  279 &    38. &  1.85 &     1.81 &$\!\!\!$  1.72 &$\!\!\!$  1.05 &$\!\!\!$   80.54 &     4.82 &$\!\!\!$   294.3 &     0.36 &$\!\!\!$  0.45 &$\!\!\!$ 1.41 \\
 \multicolumn{18}{l}{} \\
  23 &$\!\!\!$  0957  &$\!\!\!\!$ $\gamma_\mathrm{s}\alpha_\mathrm{ce}$ &  1.74 &$\!\!\!$  1.00 &  325 &$\!\!\!\!\!$  341 &    4.8 &  1.16 &     2.25 &$\!\!\!$  2.00 &$\!\!\!$  1.12 &$\!\!\!$   15.22 &     6.66 &$\!\!\!$   28.52 &     0.30 &$\!\!\!$  0.34 &$\!\!\!$ 0.06 \\
  24 &$\!\!\!$  0957  &$\!\!\!\!$ $\gamma_\mathrm{s}\alpha_\mathrm{ce}$ &  1.62 &$\!\!\!$  1.00 &  325 &$\!\!\!\!\!$  427 &    31. &  1.16 &     2.34 &$\!\!\!$  2.00 &$\!\!\!$  1.17 &$\!\!\!$   8.110 &     6.66 &$\!\!\!$   28.52 &     0.30 &$\!\!\!$  0.34 &$\!\!\!$ 0.06 \\
  25 &$\!\!\!$  0957  &$\!\!\!\!$ $\gamma_\mathrm{s}\gamma_\mathrm{s}$  &  1.70 &$\!\!\!$  1.63 &  325 &$\!\!\!\!\!$  318 &    2.2 &  2.26 &     1.70 &$\!\!\!$  1.61 &$\!\!\!$  1.05 &$\!\!\!$   119.8 &     4.34 &$\!\!\!$   224.1 &     0.37 &$\!\!\!$  0.42 &$\!\!\!$ 0.06 \\
  26 &$\!\!\!$  0957  &$\!\!\!\!$ $\gamma_\mathrm{a}\alpha_\mathrm{ce}$ &  1.64 &$\!\!\!$  0.44 &  325 &$\!\!\!\!\!$  939 &   189. &  1.58 &     2.46 &$\!\!\!$  1.81 &$\!\!\!$  1.36 &$\!\!\!$   5.820 &     5.67 &$\!\!\!$   67.28 &     0.32 &$\!\!\!$  0.36 &$\!\!\!$ 0.06 \\
  27 &$\!\!\!$  0957  &$\!\!\!\!$ $\gamma_\mathrm{a}\gamma_\mathrm{a}$  &  1.89 &$\!\!\!$  1.00 &  325 &$\!\!\!\!\!$  644 &    98. &  1.22 &     2.55 &$\!\!\!$  1.98 &$\!\!\!$  1.29 &$\!\!\!$   11.16 &     5.89 &$\!\!\!$   79.27 &     0.34 &$\!\!\!$  0.38 &$\!\!\!$ 0.06 \\
  28 &$\!\!\!$  0957  &$\!\!\!\!$ $\gamma_\mathrm{a}\gamma_\mathrm{a}$  &  1.15 &$\!\!\!$  1.09 &  325 &$\!\!\!\!\!$ 1921 &   491. &  2.39 &     2.76 &$\!\!\!$  1.59 &$\!\!\!$  1.73 &$\!\!\!$   17.00 &     4.28 &$\!\!\!$   233.0 &     0.37 &$\!\!\!$  0.42 &$\!\!\!$ 0.06 \\
  29 &$\!\!\!$  0957  &$\!\!\!\!$ $\gamma_\mathrm{d}\alpha_\mathrm{ce}$ &  1.02 &$\!\!\!$  0.71 &  325 &$\!\!\!\!\!$  334 &    2.8 &  2.25 &     1.70 &$\!\!\!$  1.61 &$\!\!\!$  1.05 &$\!\!\!$   13.89 &     5.70 &$\!\!\!$   35.74 &     0.28 &$\!\!\!$  0.32 &$\!\!\!$ 0.06 \\
  30 &$\!\!\!$  0957  &$\!\!\!\!$ $\gamma_\mathrm{d}\alpha_\mathrm{ce}$ &  1.06 &$\!\!\!$  1.00 &  325 &$\!\!\!\!\!$  341 &    4.8 &  1.16 &     2.25 &$\!\!\!$  2.00 &$\!\!\!$  1.12 &$\!\!\!$   15.22 &     6.66 &$\!\!\!$   28.52 &     0.30 &$\!\!\!$  0.34 &$\!\!\!$ 0.06 \\
  31 &$\!\!\!$  0957  &$\!\!\!\!$ $\gamma_\mathrm{d}\alpha_\mathrm{ce}$ &  1.00 &$\!\!\!$  1.00 &  325 &$\!\!\!\!\!$  427 &    31. &  1.16 &     2.34 &$\!\!\!$  2.00 &$\!\!\!$  1.17 &$\!\!\!$   8.110 &     6.66 &$\!\!\!$   28.52 &     0.30 &$\!\!\!$  0.34 &$\!\!\!$ 0.06 \\
  32 &$\!\!\!$  0957  &$\!\!\!\!$ $\gamma_\mathrm{d}\gamma_\mathrm{a}$  &  1.00 &$\!\!\!$  1.01 &  325 &$\!\!\!\!\!$  317 &    2.5 &  1.52 &     1.98 &$\!\!\!$  1.83 &$\!\!\!$  1.08 &$\!\!\!$   26.17 &     5.61 &$\!\!\!$   79.26 &     0.33 &$\!\!\!$  0.37 &$\!\!\!$ 0.06 \\
  33 &$\!\!\!$  0957  &$\!\!\!\!$ $\gamma_\mathrm{d}\gamma_\mathrm{a}$  &  1.02 &$\!\!\!$  1.00 &  325 &$\!\!\!\!\!$  334 &    2.8 &  2.25 &     1.70 &$\!\!\!$  1.61 &$\!\!\!$  1.05 &$\!\!\!$   13.89 &     5.70 &$\!\!\!$   35.74 &     0.28 &$\!\!\!$  0.32 &$\!\!\!$ 0.06 \\
 \multicolumn{18}{l}{} \\
  34 &$\!\!\!$  1101  &$\!\!\!\!$ $\gamma_\mathrm{s}\alpha_\mathrm{ce}$ &  1.95 &$\!\!\!$  0.89 &  215 &$\!\!\!\!\!$  487 &   126. &  1.71 &     1.98 &$\!\!\!$  1.76 &$\!\!\!$  1.12 &$\!\!\!$   122.8 &     4.51 &$\!\!\!$   48.24 &     0.39 &$\!\!\!$  0.34 &$\!\!\!$ 0.14 \\
  35 &$\!\!\!$  1101  &$\!\!\!\!$ $\gamma_\mathrm{s}\alpha_\mathrm{ce}$ &  2.08 &$\!\!\!$  1.00 &  215 &$\!\!\!\!\!$  208 &    3.3 &  2.38 &     1.63 &$\!\!\!$  1.59 &$\!\!\!$  1.03 &$\!\!\!$   118.7 &     4.33 &$\!\!\!$   39.18 &     0.37 &$\!\!\!$  0.32 &$\!\!\!$ 0.14 \\
  36 &$\!\!\!$  1101  &$\!\!\!\!$ $\gamma_\mathrm{s}\gamma_\mathrm{s}$  &  1.72 &$\!\!\!$  1.31 &  215 &$\!\!\!\!\!$  322 &    50. &  0.74 &     2.87 &$\!\!\!$  2.34 &$\!\!\!$  1.23 &$\!\!\!$   22.02 &     5.97 &$\!\!\!$   28.23 &     0.39 &$\!\!\!$  0.34 &$\!\!\!$ 0.14 \\
  37 &$\!\!\!$  1101  &$\!\!\!\!$ $\gamma_\mathrm{s}\gamma_\mathrm{s}$  &  2.02 &$\!\!\!$  1.47 &  215 &$\!\!\!\!\!$  319 &    48. &  1.85 &     1.83 &$\!\!\!$  1.72 &$\!\!\!$  1.07 &$\!\!\!$   145.1 &     4.40 &$\!\!\!$   51.33 &     0.39 &$\!\!\!$  0.34 &$\!\!\!$ 0.14 \\
  38 &$\!\!\!$  1101  &$\!\!\!\!$ $\gamma_\mathrm{a}\alpha_\mathrm{ce}$ &  1.63 &$\!\!\!$  0.89 &  215 &$\!\!\!\!\!$ 1287 &   498. &  1.71 &     2.87 &$\!\!\!$  1.76 &$\!\!\!$  1.62 &$\!\!\!$   21.31 &     4.51 &$\!\!\!$   48.24 &     0.39 &$\!\!\!$  0.34 &$\!\!\!$ 0.14 \\
  39 &$\!\!\!$  1101  &$\!\!\!\!$ $\gamma_\mathrm{a}\gamma_\mathrm{a}$  &  2.46 &$\!\!\!$  0.98 &  215 &$\!\!\!\!\!$  295 &    37. &  0.71 &     2.87 &$\!\!\!$  2.37 &$\!\!\!$  1.21 &$\!\!\!$   22.05 &     6.05 &$\!\!\!$   26.43 &     0.39 &$\!\!\!$  0.34 &$\!\!\!$ 0.14 \\
  40 &$\!\!\!$  1101  &$\!\!\!\!$ $\gamma_\mathrm{a}\gamma_\mathrm{a}$  &  1.63 &$\!\!\!$  1.04 &  215 &$\!\!\!\!\!$ 1287 &   498. &  1.71 &     2.87 &$\!\!\!$  1.76 &$\!\!\!$  1.62 &$\!\!\!$   21.31 &     4.51 &$\!\!\!$   48.24 &     0.39 &$\!\!\!$  0.34 &$\!\!\!$ 0.14 \\
  41 &$\!\!\!$  1101  &$\!\!\!\!$ $\gamma_\mathrm{d}\alpha_\mathrm{ce}$ &  1.27 &$\!\!\!$  1.01 &  215 &$\!\!\!\!\!$  215 &    0.0 &  1.52 &     1.93 &$\!\!\!$  1.83 &$\!\!\!$  1.05 &$\!\!\!$   131.8 &     4.69 &$\!\!\!$   43.20 &     0.39 &$\!\!\!$  0.34 &$\!\!\!$ 0.14 \\
  42 &$\!\!\!$  1101  &$\!\!\!\!$ $\gamma_\mathrm{d}\alpha_\mathrm{ce}$ &  1.10 &$\!\!\!$  1.06 &  215 &$\!\!\!\!\!$ 1059 &   392. &  1.48 &     2.87 &$\!\!\!$  1.86 &$\!\!\!$  1.54 &$\!\!\!$   21.45 &     4.75 &$\!\!\!$   41.47 &     0.39 &$\!\!\!$  0.34 &$\!\!\!$ 0.14 \\
  43 &$\!\!\!$  1101  &$\!\!\!\!$ $\gamma_\mathrm{d}\gamma_\mathrm{a}$  &  1.11 &$\!\!\!$  0.98 &  215 &$\!\!\!\!\!$  268 &    25. &  0.69 &     2.87 &$\!\!\!$  2.40 &$\!\!\!$  1.20 &$\!\!\!$   22.08 &     6.13 &$\!\!\!$   24.87 &     0.39 &$\!\!\!$  0.34 &$\!\!\!$ 0.14 \\
  44 &$\!\!\!$  1101  &$\!\!\!\!$ $\gamma_\mathrm{d}\gamma_\mathrm{a}$  &  1.10 &$\!\!\!$  0.98 &  215 &$\!\!\!\!\!$  322 &    50. &  0.74 &     2.87 &$\!\!\!$  2.34 &$\!\!\!$  1.23 &$\!\!\!$   22.02 &     5.97 &$\!\!\!$   28.23 &     0.39 &$\!\!\!$  0.34 &$\!\!\!$ 0.14 \\
  45 &$\!\!\!$  1101  &$\!\!\!\!$ $\gamma_\mathrm{d}\gamma_\mathrm{a}$  &  1.10 &$\!\!\!$  0.95 &  215 &$\!\!\!\!\!$  363 &    69. &  1.10 &     2.34 &$\!\!\!$  2.03 &$\!\!\!$  1.15 &$\!\!\!$   5.150 &     6.79 &$\!\!\!$   5.880 &     0.30 &$\!\!\!$  0.26 &$\!\!\!$ 0.14 \\
 \end{tabular}
 \caption{Selected model solutions for the double envelope-ejection scenario.  This table is the total list of which an excerpt
 can be found as Table\,\ref{tab:solutions} in the article.  
 The first nine columns show the number of the entry, the double white dwarf that
 the model is a solution for, the prescription used, the two envelope-ejection parameters, the age difference of the two
 components as observed and in the model ($\Delta\tau$) in Myr, the relative difference between the observed and model age difference, defined as
 $\Delta(\Delta\tau) \equiv \left|\frac{\Delta\tau_\mathrm{mod} - \Delta\tau_\mathrm{obs}}{\Delta\tau_\mathrm{obs}}\right|$
 in \%, the time of the formation of the double white dwarf since the ZAMS ($\tau_2$) in Gyr.
 The last nine columns list binary parameters: the initial (ZAMS) masses, mass ratio and orbital period, 
 the intermediate mass ratio and period and the final masses and period. 
 {\it (The table is continued on the next page)}
 \label{tab:table_app1}  }
 \end{table*}
 
 \setcounter{table}{5}
 \begin{table*}
 \centering
 \begin{tabular}{rlrrrrrrrrrrrrrrrr}
 \hline \hline
 Nr. & WD & $\!\!$Presc. &  $\gamma_\mathrm{1}$ & $\!\!\gamma_\mathrm{2}$,  & \multicolumn{2}{c}{$\Delta\tau$ (Myr)} & $\!\!\!\!\Delta(\Delta\tau)$ & $\tau_2$  & $M_\mathrm{1i}$ & $M_\mathrm{2i}$ & $q_\mathrm{1i}$ & $P_\mathrm{i}$  &  $q_\mathrm{2m}$ & $P_\mathrm{m}$  &  $M_\mathrm{1f}$ & $M_\mathrm{2f}$  & $P_\mathrm{f}$  \\
 & & & & $\!\!\!\alpha_\mathrm{ce2}$  & Obs & Mdl & \% & Gyr   & $M_\odot$ & $M_\odot$ &  & d   &   & d  &  $M_\odot$ & $M_\odot$ & d   \\
 \hline
 \multicolumn{18}{l}{} \\
  46 &$\!\!\!$  1115  &$\!\!\!\!$ $\gamma_\mathrm{s}\alpha_\mathrm{ce}$ &  1.79 &$\!\!\!$  1.00 &  160 &$\!\!\!\!\!$  239 &    49. &  0.50 &     3.70 &$\!\!\!$  2.94 &$\!\!\!$  1.26 &$\!\!\!$   1693. &     3.58 &$\!\!\!$   980.4 &     0.82 &$\!\!\!$  0.69 &$\!\!\!$30.09 \\
  47 &$\!\!\!$  1115  &$\!\!\!\!$ $\gamma_\mathrm{s}\alpha_\mathrm{ce}$ &  1.95 &$\!\!\!$  1.00 &  160 &$\!\!\!\!\!$  203 &    27. &  0.73 &     2.90 &$\!\!\!$  2.59 &$\!\!\!$  1.12 &$\!\!\!$   2088. &     3.24 &$\!\!\!$   1017. &     0.80 &$\!\!\!$  0.67 &$\!\!\!$30.09 \\
  48 &$\!\!\!$  1115  &$\!\!\!\!$ $\gamma_\mathrm{s}\alpha_\mathrm{ce}$ &  1.90 &$\!\!\!$  1.00 &  160 &$\!\!\!\!\!$  165 &    2.9 &  0.50 &     3.38 &$\!\!\!$  2.94 &$\!\!\!$  1.15 &$\!\!\!$   1960. &     3.58 &$\!\!\!$   980.4 &     0.82 &$\!\!\!$  0.69 &$\!\!\!$30.09 \\
  49 &$\!\!\!$  1115  &$\!\!\!\!$ $\gamma_\mathrm{s}\gamma_\mathrm{s}$  &  1.80 &$\!\!\!$  1.53 &  160 &$\!\!\!\!\!$  190 &    19. &  0.42 &     3.89 &$\!\!\!$  3.13 &$\!\!\!$  1.24 &$\!\!\!$   2284. &     3.51 &$\!\!\!$   1314. &     0.89 &$\!\!\!$  0.75 &$\!\!\!$30.09 \\
  50 &$\!\!\!$  1115  &$\!\!\!\!$ $\gamma_\mathrm{s}\gamma_\mathrm{s}$  &  1.73 &$\!\!\!$  1.62 &  160 &$\!\!\!\!\!$  313 &    95. &  0.56 &     3.74 &$\!\!\!$  2.83 &$\!\!\!$  1.32 &$\!\!\!$   2340. &     3.17 &$\!\!\!$   1500. &     0.89 &$\!\!\!$  0.75 &$\!\!\!$30.09 \\
  51 &$\!\!\!$  1115  &$\!\!\!\!$ $\gamma_\mathrm{a}\alpha_\mathrm{ce}$ &  1.41 &$\!\!\!$  0.93 &  160 &$\!\!\!\!\!$  240 &    50. &  0.32 &     5.42 &$\!\!\!$  3.42 &$\!\!\!$  1.58 &$\!\!\!$   201.2 &     3.84 &$\!\!\!$   1012. &     0.89 &$\!\!\!$  0.75 &$\!\!\!$30.09 \\
  52 &$\!\!\!$  1115  &$\!\!\!\!$ $\gamma_\mathrm{a}\alpha_\mathrm{ce}$ &  1.00 &$\!\!\!$  0.98 &  160 &$\!\!\!\!\!$  831 &   419. &  0.94 &     4.89 &$\!\!\!$  2.37 &$\!\!\!$  2.07 &$\!\!\!$   153.1 &     3.01 &$\!\!\!$   1053. &     0.79 &$\!\!\!$  0.66 &$\!\!\!$30.09 \\
  53 &$\!\!\!$  1115  &$\!\!\!\!$ $\gamma_\mathrm{a}\gamma_\mathrm{a}$  &  1.49 &$\!\!\!$  1.03 &  160 &$\!\!\!\!\!$  229 &    43. &  0.31 &     5.42 &$\!\!\!$  3.47 &$\!\!\!$  1.56 &$\!\!\!$   201.5 &     3.88 &$\!\!\!$   832.5 &     0.89 &$\!\!\!$  0.75 &$\!\!\!$30.09 \\
  54 &$\!\!\!$  1115  &$\!\!\!\!$ $\gamma_\mathrm{a}\gamma_\mathrm{a}$  &  1.08 &$\!\!\!$  1.08 &  160 &$\!\!\!\!\!$  546 &   241. &  0.65 &     4.89 &$\!\!\!$  2.69 &$\!\!\!$  1.82 &$\!\!\!$   87.43 &     3.47 &$\!\!\!$   814.0 &     0.77 &$\!\!\!$  0.65 &$\!\!\!$30.09 \\
  55 &$\!\!\!$  1115  &$\!\!\!\!$ $\gamma_\mathrm{d}\alpha_\mathrm{ce}$ &  0.97 &$\!\!\!$  0.93 &  160 &$\!\!\!\!\!$  240 &    50. &  0.32 &     5.42 &$\!\!\!$  3.42 &$\!\!\!$  1.58 &$\!\!\!$   201.2 &     3.84 &$\!\!\!$   1012. &     0.89 &$\!\!\!$  0.75 &$\!\!\!$30.09 \\
  56 &$\!\!\!$  1115  &$\!\!\!\!$ $\gamma_\mathrm{d}\gamma_\mathrm{a}$  &  1.32 &$\!\!\!$  1.00 &  160 &$\!\!\!\!\!$  164 &    2.5 &  0.46 &     3.51 &$\!\!\!$  3.02 &$\!\!\!$  1.17 &$\!\!\!$   1378. &     3.90 &$\!\!\!$   427.6 &     0.77 &$\!\!\!$  0.65 &$\!\!\!$30.09 \\
  57 &$\!\!\!$  1115  &$\!\!\!\!$ $\gamma_\mathrm{d}\gamma_\mathrm{a}$  &  0.97 &$\!\!\!$  1.04 &  160 &$\!\!\!\!\!$  240 &    50. &  0.32 &     5.42 &$\!\!\!$  3.42 &$\!\!\!$  1.58 &$\!\!\!$   201.2 &     3.84 &$\!\!\!$   1012. &     0.89 &$\!\!\!$  0.75 &$\!\!\!$30.09 \\
 \multicolumn{18}{l}{} \\
  58 &$\!\!\!$  1204  &$\!\!\!\!$ $\gamma_\mathrm{s}\alpha_\mathrm{ce}$ &  2.15 &$\!\!\!$  0.83 &   80 &$\!\!\!\!\!$  136 &    70. &  5.37 &     1.27 &$\!\!\!$  1.25 &$\!\!\!$  1.01 &$\!\!\!$   630.3 &     2.65 &$\!\!\!$   274.3 &     0.47 &$\!\!\!$  0.41 &$\!\!\!$ 1.60 \\
  59 &$\!\!\!$  1204  &$\!\!\!\!$ $\gamma_\mathrm{s}\alpha_\mathrm{ce}$ &  1.05 &$\!\!\!$  0.96 &   80 &$\!\!\!\!\!$  225 &   181. &  1.58 &     2.06 &$\!\!\!$  1.81 &$\!\!\!$  1.14 &$\!\!\!$   32.89 &     3.58 &$\!\!\!$   256.2 &     0.51 &$\!\!\!$  0.44 &$\!\!\!$ 1.60 \\
  60 &$\!\!\!$  1204  &$\!\!\!\!$ $\gamma_\mathrm{s}\gamma_\mathrm{s}$  &  1.89 &$\!\!\!$  1.25 &   80 &$\!\!\!\!\!$   80 &    0.5 &  0.31 &     3.60 &$\!\!\!$  3.21 &$\!\!\!$  1.12 &$\!\!\!$   69.10 &     6.07 &$\!\!\!$   39.45 &     0.53 &$\!\!\!$  0.46 &$\!\!\!$ 1.60 \\
  61 &$\!\!\!$  1204  &$\!\!\!\!$ $\gamma_\mathrm{a}\alpha_\mathrm{ce}$ &  1.45 &$\!\!\!$  0.96 &   80 &$\!\!\!\!\!$  225 &   181. &  1.58 &     2.06 &$\!\!\!$  1.81 &$\!\!\!$  1.14 &$\!\!\!$   32.89 &     3.58 &$\!\!\!$   256.2 &     0.51 &$\!\!\!$  0.44 &$\!\!\!$ 1.60 \\
  62 &$\!\!\!$  1204  &$\!\!\!\!$ $\gamma_\mathrm{a}\alpha_\mathrm{ce}$ &  1.29 &$\!\!\!$  0.87 &   80 &$\!\!\!\!\!$  498 &   522. &  1.85 &     2.06 &$\!\!\!$  1.72 &$\!\!\!$  1.20 &$\!\!\!$   32.72 &     3.40 &$\!\!\!$   277.6 &     0.51 &$\!\!\!$  0.44 &$\!\!\!$ 1.60 \\
  63 &$\!\!\!$  1204  &$\!\!\!\!$ $\gamma_\mathrm{a}\gamma_\mathrm{a}$  &  2.42 &$\!\!\!$  0.94 &   80 &$\!\!\!\!\!$  143 &    78. &  0.42 &     3.34 &$\!\!\!$  2.87 &$\!\!\!$  1.17 &$\!\!\!$   15.40 &     6.08 &$\!\!\!$   37.29 &     0.47 &$\!\!\!$  0.41 &$\!\!\!$ 1.60 \\
  64 &$\!\!\!$  1204  &$\!\!\!\!$ $\gamma_\mathrm{a}\gamma_\mathrm{a}$  &  1.06 &$\!\!\!$  1.08 &   80 &$\!\!\!\!\!$ 1387 &  1634. &  1.66 &     3.34 &$\!\!\!$  1.79 &$\!\!\!$  1.87 &$\!\!\!$   14.56 &     3.79 &$\!\!\!$   172.9 &     0.47 &$\!\!\!$  0.41 &$\!\!\!$ 1.60 \\
  65 &$\!\!\!$  1204  &$\!\!\!\!$ $\gamma_\mathrm{d}\alpha_\mathrm{ce}$ &  0.72 &$\!\!\!$  0.96 &   80 &$\!\!\!\!\!$  225 &   181. &  1.58 &     2.06 &$\!\!\!$  1.81 &$\!\!\!$  1.14 &$\!\!\!$   32.89 &     3.58 &$\!\!\!$   256.2 &     0.51 &$\!\!\!$  0.44 &$\!\!\!$ 1.60 \\
  66 &$\!\!\!$  1204  &$\!\!\!\!$ $\gamma_\mathrm{d}\alpha_\mathrm{ce}$ &  1.47 &$\!\!\!$  0.87 &   80 &$\!\!\!\!\!$  138 &    72. &  4.66 &     1.32 &$\!\!\!$  1.30 &$\!\!\!$  1.01 &$\!\!\!$   606.6 &     2.76 &$\!\!\!$   263.4 &     0.47 &$\!\!\!$  0.41 &$\!\!\!$ 1.60 \\
  67 &$\!\!\!$  1204  &$\!\!\!\!$ $\gamma_\mathrm{d}\gamma_\mathrm{a}$  &  1.10 &$\!\!\!$  0.95 &   80 &$\!\!\!\!\!$   74 &    7.8 &  0.26 &     3.89 &$\!\!\!$  3.42 &$\!\!\!$  1.14 &$\!\!\!$   38.82 &     5.96 &$\!\!\!$   51.85 &     0.57 &$\!\!\!$  0.50 &$\!\!\!$ 1.60 \\
  68 &$\!\!\!$  1204  &$\!\!\!\!$ $\gamma_\mathrm{d}\gamma_\mathrm{a}$  &  1.09 &$\!\!\!$  0.92 &   80 &$\!\!\!\!\!$  100 &    25. &  0.38 &     3.34 &$\!\!\!$  2.98 &$\!\!\!$  1.12 &$\!\!\!$   15.47 &     6.32 &$\!\!\!$   19.99 &     0.47 &$\!\!\!$  0.41 &$\!\!\!$ 1.60 \\
 \multicolumn{18}{l}{} \\
  69 &$\!\!\!$  1349  &$\!\!\!\!$ $\gamma_\mathrm{s}\alpha_\mathrm{ce}$ &  1.56 &$\!\!\!$  1.05 &    0 &$\!\!\!\!\!$  129 &    0.0 &  2.17 &     1.68 &$\!\!\!$  1.63 &$\!\!\!$  1.03 &$\!\!\!$   108.7 &     4.48 &$\!\!\!$   352.9 &     0.37 &$\!\!\!$  0.46 &$\!\!\!$ 2.21 \\
  70 &$\!\!\!$  1349  &$\!\!\!\!$ $\gamma_\mathrm{s}\alpha_\mathrm{ce}$ &  1.45 &$\!\!\!$  1.01 &    0 &$\!\!\!\!\!$  461 &    0.0 &  4.52 &     1.35 &$\!\!\!$  1.32 &$\!\!\!$  1.03 &$\!\!\!$   105.9 &     3.77 &$\!\!\!$   364.5 &     0.35 &$\!\!\!$  0.44 &$\!\!\!$ 2.21 \\
  71 &$\!\!\!$  1349  &$\!\!\!\!$ $\gamma_\mathrm{s}\gamma_\mathrm{s}$  &  1.56 &$\!\!\!$  1.56 &    0 &$\!\!\!\!\!$  129 &    0.0 &  2.17 &     1.68 &$\!\!\!$  1.63 &$\!\!\!$  1.03 &$\!\!\!$   93.59 &     4.58 &$\!\!\!$   317.3 &     0.36 &$\!\!\!$  0.45 &$\!\!\!$ 2.21 \\
  72 &$\!\!\!$  1349  &$\!\!\!\!$ $\gamma_\mathrm{s}\gamma_\mathrm{s}$  &  1.52 &$\!\!\!$  1.63 &    0 &$\!\!\!\!\!$  384 &    0.0 &  2.55 &     1.63 &$\!\!\!$  1.55 &$\!\!\!$  1.05 &$\!\!\!$   112.7 &     4.25 &$\!\!\!$   379.4 &     0.37 &$\!\!\!$  0.46 &$\!\!\!$ 2.21 \\
  73 &$\!\!\!$  1349  &$\!\!\!\!$ $\gamma_\mathrm{a}\alpha_\mathrm{ce}$ &  1.01 &$\!\!\!$  1.02 &    0 &$\!\!\!\!\!$ 1776 &    0.0 &  2.26 &     2.72 &$\!\!\!$  1.61 &$\!\!\!$  1.69 &$\!\!\!$   14.85 &     4.42 &$\!\!\!$   360.4 &     0.37 &$\!\!\!$  0.46 &$\!\!\!$ 2.21 \\
  74 &$\!\!\!$  1349  &$\!\!\!\!$ $\gamma_\mathrm{a}\gamma_\mathrm{a}$  &  1.35 &$\!\!\!$  0.97 &    0 &$\!\!\!\!\!$  975 &    0.0 &  1.53 &     2.59 &$\!\!\!$  1.83 &$\!\!\!$  1.41 &$\!\!\!$   10.16 &     5.38 &$\!\!\!$   206.9 &     0.34 &$\!\!\!$  0.43 &$\!\!\!$ 2.21 \\
  75 &$\!\!\!$  1349  &$\!\!\!\!$ $\gamma_\mathrm{d}\alpha_\mathrm{ce}$ &  0.97 &$\!\!\!$  1.05 &    0 &$\!\!\!\!\!$  129 &    0.0 &  2.17 &     1.68 &$\!\!\!$  1.63 &$\!\!\!$  1.03 &$\!\!\!$   108.7 &     4.48 &$\!\!\!$   352.9 &     0.37 &$\!\!\!$  0.46 &$\!\!\!$ 2.21 \\
  76 &$\!\!\!$  1349  &$\!\!\!\!$ $\gamma_\mathrm{d}\alpha_\mathrm{ce}$ &  0.96 &$\!\!\!$  1.02 &    0 &$\!\!\!\!\!$  215 &    0.0 &  2.26 &     1.68 &$\!\!\!$  1.61 &$\!\!\!$  1.04 &$\!\!\!$   108.5 &     4.42 &$\!\!\!$   360.4 &     0.37 &$\!\!\!$  0.46 &$\!\!\!$ 2.21 \\
  77 &$\!\!\!$  1349  &$\!\!\!\!$ $\gamma_\mathrm{d}\gamma_\mathrm{a}$  &  0.95 &$\!\!\!$  0.98 &    0 &$\!\!\!\!\!$  101 &    0.0 &  1.58 &     1.86 &$\!\!\!$  1.81 &$\!\!\!$  1.03 &$\!\!\!$   63.44 &     5.19 &$\!\!\!$   241.2 &     0.35 &$\!\!\!$  0.44 &$\!\!\!$ 2.21 \\
  78 &$\!\!\!$  1349  &$\!\!\!\!$ $\gamma_\mathrm{d}\gamma_\mathrm{a}$  &  0.97 &$\!\!\!$  1.03 &    0 &$\!\!\!\!\!$  235 &    0.0 &  2.17 &     1.70 &$\!\!\!$  1.63 &$\!\!\!$  1.04 &$\!\!\!$   106.4 &     4.48 &$\!\!\!$   352.9 &     0.37 &$\!\!\!$  0.46 &$\!\!\!$ 2.21 \\
 \multicolumn{18}{l}{} \\
  79 &$\!\!\!$  1414  &$\!\!\!\!$ $\gamma_\mathrm{s}\alpha_\mathrm{ce}$ &  1.52 &$\!\!\!$  0.71 &  200 &$\!\!\!\!\!$  188 &    5.9 &  0.43 &     3.51 &$\!\!\!$  3.09 &$\!\!\!$  1.14 &$\!\!\!$   70.81 &     5.99 &$\!\!\!$   358.3 &     0.52 &$\!\!\!$  0.66 &$\!\!\!$ 0.52 \\
  80 &$\!\!\!$  1414  &$\!\!\!\!$ $\gamma_\mathrm{s}\gamma_\mathrm{s}$  &  1.49 &$\!\!\!$  1.76 &  200 &$\!\!\!\!\!$  112 &    44. &  0.87 &     2.55 &$\!\!\!$  2.43 &$\!\!\!$  1.05 &$\!\!\!$   525.4 &     4.09 &$\!\!\!$   1783. &     0.59 &$\!\!\!$  0.76 &$\!\!\!$ 0.52 \\
  81 &$\!\!\!$  1414  &$\!\!\!\!$ $\gamma_\mathrm{a}\alpha_\mathrm{ce}$ &  2.13 &$\!\!\!$  0.71 &  200 &$\!\!\!\!\!$  196 &    1.9 &  0.43 &     3.56 &$\!\!\!$  3.09 &$\!\!\!$  1.15 &$\!\!\!$   51.80 &     5.99 &$\!\!\!$   358.3 &     0.52 &$\!\!\!$  0.66 &$\!\!\!$ 0.52 \\
  82 &$\!\!\!$  1414  &$\!\!\!\!$ $\gamma_\mathrm{a}\gamma_\mathrm{a}$  &  1.10 &$\!\!\!$  1.04 &  200 &$\!\!\!\!\!$  290 &    45. &  0.47 &     3.99 &$\!\!\!$  3.02 &$\!\!\!$  1.32 &$\!\!\!$   41.73 &     5.08 &$\!\!\!$   1363. &     0.59 &$\!\!\!$  0.76 &$\!\!\!$ 0.52 \\
  83 &$\!\!\!$  1414  &$\!\!\!\!$ $\gamma_\mathrm{a}\gamma_\mathrm{a}$  &  1.04 &$\!\!\!$  1.04 &  200 &$\!\!\!\!\!$  325 &    63. &  0.50 &     3.99 &$\!\!\!$  2.94 &$\!\!\!$  1.36 &$\!\!\!$   41.61 &     4.95 &$\!\!\!$   1407. &     0.59 &$\!\!\!$  0.76 &$\!\!\!$ 0.52 \\
  84 &$\!\!\!$  1414  &$\!\!\!\!$ $\gamma_\mathrm{d}\alpha_\mathrm{ce}$ &  0.95 &$\!\!\!$  0.71 &  200 &$\!\!\!\!\!$  188 &    5.9 &  0.43 &     3.51 &$\!\!\!$  3.09 &$\!\!\!$  1.14 &$\!\!\!$   70.81 &     5.99 &$\!\!\!$   358.3 &     0.52 &$\!\!\!$  0.66 &$\!\!\!$ 0.52 \\
  85 &$\!\!\!$  1414  &$\!\!\!\!$ $\gamma_\mathrm{d}\gamma_\mathrm{a}$  &  0.95 &$\!\!\!$  0.99 &  200 &$\!\!\!\!\!$  188 &    5.9 &  0.43 &     3.51 &$\!\!\!$  3.09 &$\!\!\!$  1.14 &$\!\!\!$   70.81 &     5.99 &$\!\!\!$   358.3 &     0.52 &$\!\!\!$  0.66 &$\!\!\!$ 0.52 \\
  86 &$\!\!\!$  1414  &$\!\!\!\!$ $\gamma_\mathrm{d}\gamma_\mathrm{a}$  &  0.90 &$\!\!\!$  1.00 &  200 &$\!\!\!\!\!$  167 &    17. &  0.39 &     3.65 &$\!\!\!$  3.21 &$\!\!\!$  1.14 &$\!\!\!$   75.14 &     5.96 &$\!\!\!$   511.9 &     0.54 &$\!\!\!$  0.69 &$\!\!\!$ 0.52 \\
 \multicolumn{18}{l}{} \\
  87 &$\!\!\!$  1704a &$\!\!\!\!$ $\gamma_\mathrm{s}\alpha_\mathrm{ce}$ &  1.67 &$\!\!\!$  0.60 &  -20 &$\!\!\!\!\!$   52 &   360. &  1.41 &     2.06 &$\!\!\!$  1.88 &$\!\!\!$  1.09 &$\!\!\!$   40.37 &     3.66 &$\!\!\!$   65.66 &     0.51 &$\!\!\!$  0.36 &$\!\!\!$ 0.14 \\
  88 &$\!\!\!$  1704a &$\!\!\!\!$ $\gamma_\mathrm{s}\alpha_\mathrm{ce}$ &  1.88 &$\!\!\!$  0.62 &  -20 &$\!\!\!\!\!$   15 &   175. &  1.17 &     2.19 &$\!\!\!$  2.00 &$\!\!\!$  1.09 &$\!\!\!$   93.52 &     3.79 &$\!\!\!$   66.89 &     0.53 &$\!\!\!$  0.37 &$\!\!\!$ 0.14 \\
  89 &$\!\!\!$  1704a &$\!\!\!\!$ $\gamma_\mathrm{s}\alpha_\mathrm{ce}$ &  2.05 &$\!\!\!$  0.43 &  -20 &$\!\!\!\!\!$    7 &   135. &  1.36 &     2.03 &$\!\!\!$  1.90 &$\!\!\!$  1.07 &$\!\!\!$   252.8 &     3.51 &$\!\!\!$   96.02 &     0.54 &$\!\!\!$  0.38 &$\!\!\!$ 0.14 \\
  90 &$\!\!\!$  1704a &$\!\!\!\!$ $\gamma_\mathrm{s}\gamma_\mathrm{s}$  &  1.67 &$\!\!\!$  1.52 &  -20 &$\!\!\!\!\!$   52 &   360. &  1.41 &     2.06 &$\!\!\!$  1.88 &$\!\!\!$  1.09 &$\!\!\!$   40.37 &     3.66 &$\!\!\!$   65.66 &     0.51 &$\!\!\!$  0.36 &$\!\!\!$ 0.14 \\
  91 &$\!\!\!$  1704a &$\!\!\!\!$ $\gamma_\mathrm{s}\gamma_\mathrm{s}$  &  1.88 &$\!\!\!$  1.50 &  -20 &$\!\!\!\!\!$   15 &   175. &  1.17 &     2.19 &$\!\!\!$  2.00 &$\!\!\!$  1.09 &$\!\!\!$   93.52 &     3.79 &$\!\!\!$   66.89 &     0.53 &$\!\!\!$  0.37 &$\!\!\!$ 0.14 \\
  92 &$\!\!\!$  1704a &$\!\!\!\!$ $\gamma_\mathrm{a}\alpha_\mathrm{ce}$ &  2.64 &$\!\!\!$  0.59 &  -20 &$\!\!\!\!\!$   23 &   215. &  1.22 &     2.16 &$\!\!\!$  1.98 &$\!\!\!$  1.09 &$\!\!\!$   87.04 &     3.74 &$\!\!\!$   69.77 &     0.53 &$\!\!\!$  0.37 &$\!\!\!$ 0.14 \\
  93 &$\!\!\!$  1704a &$\!\!\!\!$ $\gamma_\mathrm{a}\alpha_\mathrm{ce}$ &  1.90 &$\!\!\!$  0.43 &  -20 &$\!\!\!\!\!$  499 &  2596. &  1.36 &     2.43 &$\!\!\!$  1.90 &$\!\!\!$  1.27 &$\!\!\!$   43.81 &     3.51 &$\!\!\!$   96.02 &     0.54 &$\!\!\!$  0.38 &$\!\!\!$ 0.14 \\
  94 &$\!\!\!$  1704a &$\!\!\!\!$ $\gamma_\mathrm{a}\gamma_\mathrm{a}$  &  1.57 &$\!\!\!$  1.09 &  -20 &$\!\!\!\!\!$  884 &  4521. &  1.12 &     3.56 &$\!\!\!$  2.03 &$\!\!\!$  1.75 &$\!\!\!$   43.52 &     3.95 &$\!\!\!$   50.36 &     0.51 &$\!\!\!$  0.36 &$\!\!\!$ 0.14 \\
  95 &$\!\!\!$  1704a &$\!\!\!\!$ $\gamma_\mathrm{d}\alpha_\mathrm{ce}$ &  1.11 &$\!\!\!$  0.60 &  -20 &$\!\!\!\!\!$   52 &   360. &  1.41 &     2.06 &$\!\!\!$  1.88 &$\!\!\!$  1.09 &$\!\!\!$   40.37 &     3.66 &$\!\!\!$   65.66 &     0.51 &$\!\!\!$  0.36 &$\!\!\!$ 0.14 \\
  96 &$\!\!\!$  1704a &$\!\!\!\!$ $\gamma_\mathrm{d}\alpha_\mathrm{ce}$ &  1.24 &$\!\!\!$  0.62 &  -20 &$\!\!\!\!\!$   15 &   175. &  1.17 &     2.19 &$\!\!\!$  2.00 &$\!\!\!$  1.09 &$\!\!\!$   93.52 &     3.79 &$\!\!\!$   66.89 &     0.53 &$\!\!\!$  0.37 &$\!\!\!$ 0.14 \\
  97 &$\!\!\!$  1704a &$\!\!\!\!$ $\gamma_\mathrm{d}\alpha_\mathrm{ce}$ &  1.37 &$\!\!\!$  0.34 &  -20 &$\!\!\!\!\!$    2 &   110. &  1.48 &     1.93 &$\!\!\!$  1.86 &$\!\!\!$  1.04 &$\!\!\!$   294.5 &     3.33 &$\!\!\!$   120.2 &     0.56 &$\!\!\!$  0.39 &$\!\!\!$ 0.14 \\
  98 &$\!\!\!$  1704a &$\!\!\!\!$ $\gamma_\mathrm{d}\gamma_\mathrm{a}$  &  1.11 &$\!\!\!$  1.13 &  -20 &$\!\!\!\!\!$   52 &   360. &  1.41 &     2.06 &$\!\!\!$  1.88 &$\!\!\!$  1.09 &$\!\!\!$   40.37 &     3.66 &$\!\!\!$   65.66 &     0.51 &$\!\!\!$  0.36 &$\!\!\!$ 0.14 \\
  99 &$\!\!\!$  1704a &$\!\!\!\!$ $\gamma_\mathrm{d}\gamma_\mathrm{a}$  &  1.24 &$\!\!\!$  1.11 &  -20 &$\!\!\!\!\!$   15 &   175. &  1.17 &     2.19 &$\!\!\!$  2.00 &$\!\!\!$  1.09 &$\!\!\!$   93.52 &     3.79 &$\!\!\!$   66.89 &     0.53 &$\!\!\!$  0.37 &$\!\!\!$ 0.14 \\
 \end{tabular}
 \caption{{\it (continued)}
 \label{tab:table_app2}  }
 \end{table*}

 \setcounter{table}{5}
 \begin{table*}
 \centering
 \begin{tabular}{rlrrrrrrrrrrrrrrrr}
 \hline \hline
 Nr. & WD & $\!\!$Presc. &  $\gamma_\mathrm{1}$ & $\!\!\gamma_\mathrm{2}$,  & \multicolumn{2}{c}{$\Delta\tau$ (Myr)} & $\!\!\!\!\Delta(\Delta\tau)$ & $\tau_2$  & $M_\mathrm{1i}$ & $M_\mathrm{2i}$ & $q_\mathrm{1i}$ & $P_\mathrm{i}$  &  $q_\mathrm{2m}$ & $P_\mathrm{m}$  &  $M_\mathrm{1f}$ & $M_\mathrm{2f}$  & $P_\mathrm{f}$  \\
 & & & & $\!\!\!\alpha_\mathrm{ce2}$  & Obs & Mdl & \% & Gyr   & $M_\odot$ & $M_\odot$ &  & d   &   & d  &  $M_\odot$ & $M_\odot$ & d   \\
 \hline
 \multicolumn{18}{l}{} \\
 100 &$\!\!\!$  1704b &$\!\!\!\!$ $\gamma_\mathrm{s}\alpha_\mathrm{ce}$ &  1.65 &$\!\!\!$  0.53 &   20 &$\!\!\!\!\!$  292 &  1360. &  0.73 &     2.83 &$\!\!\!$  2.59 &$\!\!\!$  1.09 &$\!\!\!$   47.21 &     6.37 &$\!\!\!$   161.8 &     0.41 &$\!\!\!$  0.58 &$\!\!\!$ 0.14 \\
 101 &$\!\!\!$  1704b &$\!\!\!\!$ $\gamma_\mathrm{s}\alpha_\mathrm{ce}$ &  1.74 &$\!\!\!$  0.76 &   20 &$\!\!\!\!\!$  285 &  1326. &  0.75 &     2.76 &$\!\!\!$  2.55 &$\!\!\!$  1.08 &$\!\!\!$   49.12 &     6.40 &$\!\!\!$   107.3 &     0.40 &$\!\!\!$  0.57 &$\!\!\!$ 0.14 \\
 102 &$\!\!\!$  1704b &$\!\!\!\!$ $\gamma_\mathrm{s}\gamma_\mathrm{s}$  &  1.63 &$\!\!\!$  1.43 &   20 &$\!\!\!\!\!$  205 &   927. &  0.58 &     2.98 &$\!\!\!$  2.79 &$\!\!\!$  1.07 &$\!\!\!$   50.44 &     6.54 &$\!\!\!$   208.9 &     0.43 &$\!\!\!$  0.61 &$\!\!\!$ 0.14 \\
 103 &$\!\!\!$  1704b &$\!\!\!\!$ $\gamma_\mathrm{a}\alpha_\mathrm{ce}$ &  2.36 &$\!\!\!$  0.46 &   20 &$\!\!\!\!\!$  231 &  1054. &  0.58 &     3.05 &$\!\!\!$  2.79 &$\!\!\!$  1.09 &$\!\!\!$   33.85 &     6.54 &$\!\!\!$   208.9 &     0.43 &$\!\!\!$  0.61 &$\!\!\!$ 0.14 \\
 104 &$\!\!\!$  1704b &$\!\!\!\!$ $\gamma_\mathrm{a}\alpha_\mathrm{ce}$ &  1.59 &$\!\!\!$  0.47 &   20 &$\!\!\!\!\!$ 1017 &  4985. &  1.45 &     2.83 &$\!\!\!$  1.95 &$\!\!\!$  1.45 &$\!\!\!$   20.80 &     5.08 &$\!\!\!$   158.7 &     0.38 &$\!\!\!$  0.55 &$\!\!\!$ 0.14 \\
 105 &$\!\!\!$  1704b &$\!\!\!\!$ $\gamma_\mathrm{a}\gamma_\mathrm{a}$  &  1.63 &$\!\!\!$  1.01 &   20 &$\!\!\!\!\!$  498 &  2391. &  0.86 &     3.02 &$\!\!\!$  2.43 &$\!\!\!$  1.24 &$\!\!\!$   31.86 &     5.78 &$\!\!\!$   527.9 &     0.42 &$\!\!\!$  0.60 &$\!\!\!$ 0.14 \\
 106 &$\!\!\!$  1704b &$\!\!\!\!$ $\gamma_\mathrm{a}\gamma_\mathrm{a}$  &  2.56 &$\!\!\!$  0.99 &   20 &$\!\!\!\!\!$  205 &   927. &  0.58 &     2.98 &$\!\!\!$  2.79 &$\!\!\!$  1.07 &$\!\!\!$   50.44 &     6.54 &$\!\!\!$   208.9 &     0.43 &$\!\!\!$  0.61 &$\!\!\!$ 0.14 \\
 107 &$\!\!\!$  1704b &$\!\!\!\!$ $\gamma_\mathrm{d}\alpha_\mathrm{ce}$ &  1.03 &$\!\!\!$  0.15 &   20 &$\!\!\!\!\!$  182 &   810. &  2.23 &     1.68 &$\!\!\!$  1.65 &$\!\!\!$  1.01 &$\!\!\!$   212.1 &     4.08 &$\!\!\!$   478.6 &     0.41 &$\!\!\!$  0.58 &$\!\!\!$ 0.14 \\
 108 &$\!\!\!$  1704b &$\!\!\!\!$ $\gamma_\mathrm{d}\alpha_\mathrm{ce}$ &  1.00 &$\!\!\!$  0.76 &   20 &$\!\!\!\!\!$  332 &  1562. &  0.75 &     2.87 &$\!\!\!$  2.55 &$\!\!\!$  1.12 &$\!\!\!$   33.29 &     6.40 &$\!\!\!$   107.3 &     0.40 &$\!\!\!$  0.57 &$\!\!\!$ 0.14 \\
 109 &$\!\!\!$  1704b &$\!\!\!\!$ $\gamma_\mathrm{d}\gamma_\mathrm{a}$  &  0.97 &$\!\!\!$  0.99 &   20 &$\!\!\!\!\!$  205 &   927. &  0.58 &     2.98 &$\!\!\!$  2.79 &$\!\!\!$  1.07 &$\!\!\!$   50.44 &     6.54 &$\!\!\!$   208.9 &     0.43 &$\!\!\!$  0.61 &$\!\!\!$ 0.14 \\
 110 &$\!\!\!$  1704b &$\!\!\!\!$ $\gamma_\mathrm{d}\gamma_\mathrm{a}$  &  0.99 &$\!\!\!$  0.99 &   20 &$\!\!\!\!\!$  292 &  1360. &  0.73 &     2.83 &$\!\!\!$  2.59 &$\!\!\!$  1.09 &$\!\!\!$   47.21 &     6.37 &$\!\!\!$   161.8 &     0.41 &$\!\!\!$  0.58 &$\!\!\!$ 0.14 \\
 \multicolumn{18}{l}{} \\
 111 &$\!\!\!$  2209  &$\!\!\!\!$ $\gamma_\mathrm{s}\alpha_\mathrm{ce}$ &  1.69 &$\!\!\!$  0.54 &  500 &$\!\!\!\!\!$  517 &    3.3 &  1.45 &     2.37 &$\!\!\!$  1.95 &$\!\!\!$  1.21 &$\!\!\!$   148.5 &     3.55 &$\!\!\!$   168.2 &     0.55 &$\!\!\!$  0.55 &$\!\!\!$ 0.28 \\
 112 &$\!\!\!$  2209  &$\!\!\!\!$ $\gamma_\mathrm{s}\alpha_\mathrm{ce}$ &  1.56 &$\!\!\!$  0.88 &  500 &$\!\!\!\!\!$  552 &    10. &  0.75 &     3.79 &$\!\!\!$  2.55 &$\!\!\!$  1.49 &$\!\!\!$   87.01 &     4.48 &$\!\!\!$   113.8 &     0.57 &$\!\!\!$  0.57 &$\!\!\!$ 0.28 \\
 113 &$\!\!\!$  2209  &$\!\!\!\!$ $\gamma_\mathrm{s}\gamma_\mathrm{s}$  &  1.62 &$\!\!\!$  1.69 &  500 &$\!\!\!\!\!$  477 &    4.5 &  1.37 &     2.40 &$\!\!\!$  2.00 &$\!\!\!$  1.20 &$\!\!\!$   165.4 &     3.64 &$\!\!\!$   258.7 &     0.55 &$\!\!\!$  0.55 &$\!\!\!$ 0.28 \\
 114 &$\!\!\!$  2209  &$\!\!\!\!$ $\gamma_\mathrm{s}\gamma_\mathrm{s}$  &  1.62 &$\!\!\!$  1.63 &  500 &$\!\!\!\!\!$  262 &    48. &  1.20 &     2.37 &$\!\!\!$  2.16 &$\!\!\!$  1.09 &$\!\!\!$   150.0 &     3.93 &$\!\!\!$   304.6 &     0.55 &$\!\!\!$  0.55 &$\!\!\!$ 0.28 \\
 115 &$\!\!\!$  2209  &$\!\!\!\!$ $\gamma_\mathrm{a}\alpha_\mathrm{ce}$ &  1.67 &$\!\!\!$  0.70 &  500 &$\!\!\!\!\!$  690 &    38. &  0.90 &     3.74 &$\!\!\!$  2.40 &$\!\!\!$  1.56 &$\!\!\!$   58.62 &     4.36 &$\!\!\!$   138.0 &     0.55 &$\!\!\!$  0.55 &$\!\!\!$ 0.28 \\
 116 &$\!\!\!$  2209  &$\!\!\!\!$ $\gamma_\mathrm{a}\alpha_\mathrm{ce}$ &  1.30 &$\!\!\!$  0.54 &  500 &$\!\!\!\!\!$ 1245 &   149. &  1.45 &     3.74 &$\!\!\!$  1.95 &$\!\!\!$  1.92 &$\!\!\!$   57.09 &     3.55 &$\!\!\!$   168.2 &     0.55 &$\!\!\!$  0.55 &$\!\!\!$ 0.28 \\
 117 &$\!\!\!$  2209  &$\!\!\!\!$ $\gamma_\mathrm{a}\gamma_\mathrm{a}$  &  1.38 &$\!\!\!$  1.10 &  500 &$\!\!\!\!\!$  542 &    8.4 &  0.70 &     4.15 &$\!\!\!$  2.62 &$\!\!\!$  1.58 &$\!\!\!$   97.10 &     4.16 &$\!\!\!$   677.7 &     0.63 &$\!\!\!$  0.63 &$\!\!\!$ 0.28 \\
 118 &$\!\!\!$  2209  &$\!\!\!\!$ $\gamma_\mathrm{a}\gamma_\mathrm{a}$  &  1.14 &$\!\!\!$  1.16 &  500 &$\!\!\!\!\!$  846 &    69. &  1.00 &     4.15 &$\!\!\!$  2.31 &$\!\!\!$  1.80 &$\!\!\!$   95.54 &     3.66 &$\!\!\!$   821.2 &     0.63 &$\!\!\!$  0.63 &$\!\!\!$ 0.28 \\
 119 &$\!\!\!$  2209  &$\!\!\!\!$ $\gamma_\mathrm{d}\alpha_\mathrm{ce}$ &  1.06 &$\!\!\!$  0.53 &  500 &$\!\!\!\!\!$  347 &    31. &  1.35 &     2.31 &$\!\!\!$  2.03 &$\!\!\!$  1.14 &$\!\!\!$   80.61 &     3.76 &$\!\!\!$   169.5 &     0.54 &$\!\!\!$  0.54 &$\!\!\!$ 0.28 \\
 120 &$\!\!\!$  2209  &$\!\!\!\!$ $\gamma_\mathrm{d}\alpha_\mathrm{ce}$ &  1.12 &$\!\!\!$  0.88 &  500 &$\!\!\!\!\!$  559 &    12. &  0.75 &     3.84 &$\!\!\!$  2.55 &$\!\!\!$  1.50 &$\!\!\!$   71.51 &     4.48 &$\!\!\!$   113.8 &     0.57 &$\!\!\!$  0.57 &$\!\!\!$ 0.28 \\
 121 &$\!\!\!$  2209  &$\!\!\!\!$ $\gamma_\mathrm{d}\gamma_\mathrm{a}$  &  1.06 &$\!\!\!$  1.05 &  500 &$\!\!\!\!\!$  494 &    1.3 &  0.67 &     3.94 &$\!\!\!$  2.65 &$\!\!\!$  1.48 &$\!\!\!$   83.11 &     4.50 &$\!\!\!$   229.2 &     0.59 &$\!\!\!$  0.59 &$\!\!\!$ 0.28 \\
 122 &$\!\!\!$  2209  &$\!\!\!\!$ $\gamma_\mathrm{d}\gamma_\mathrm{a}$  &  1.04 &$\!\!\!$  1.05 &  500 &$\!\!\!\!\!$  340 &    32. &  0.50 &     4.15 &$\!\!\!$  2.94 &$\!\!\!$  1.41 &$\!\!\!$   98.45 &     4.67 &$\!\!\!$   294.3 &     0.63 &$\!\!\!$  0.63 &$\!\!\!$ 0.28 \\
 \end{tabular}
 \caption{{\it (continued)}
 \label{tab:table_label}  }
 \end{table*}

\end{document}